\documentclass[amssymb,twocolumn,floats,amsmath,superscriptaddress]{revtex4-2}
\usepackage[english]{babel}
\usepackage[utf8]{inputenc}
\usepackage[colorinlistoftodos, color=green!40, prependcaption]{todonotes}
\usepackage{amsthm}
\usepackage{mathtools}
\usepackage{physics}
\usepackage{xcolor}
\usepackage{graphicx}
\usepackage[left=23mm,right=13mm,top=35mm,columnsep=15pt]{geometry} 
\usepackage{adjustbox}
\usepackage{placeins}
\usepackage[T1]{fontenc}
\usepackage{csquotes}
\def \IFT{Institute of Theoretical Physics, Faculty of Physics, University of Warsaw, Pasteura St. 5, 02-093 Warsaw, Poland}
\def \PW{Department of Semiconductor Materials Engineering Faculty of Fundamental Problems of Technology Wrocław University of Science and Technology Wybrzeże Wyspiańskiego 27, 50-370 Wrocław, Poland}
\def \AC{II. Institute of Physics B and JARA-FIT, RWTH-Aachen University, 52074 Aachen, Germany}
\def \Haifa{Schulich Faculty of Chemistry, Solid State Institute, Russell Berrie Nanotechnology insistute and Helen Diller Quantum Center, Technion – Israel Institute of Technology, Haifa 3200003, Israel}
\def \HaifaMat{Department of Materials Science and Engineering, Technion – Israel Institute of Technology, Haifa 3200003, Israel}
\def \FZJ{Forschungszentrum Jülich,
Peter Grünberg Institute (PGI-6),
52428 Jülich, Germany}
\def \SOLEIL{Synchrotron-SOLEIL, Université Paris-Saclay, Saint-Aubin, BP48, Paris, F91192, Gif sur Yvette, France}
\def \MAGTOP{International Research Centre MagTop, Institute of Physics, Polish
Academy of Sciences,~\\
 Aleja Lotnik\'ow 32/46, PL-02668 Warsaw, Poland}
\def \ALS{Advanced Light Source, Lawrence Berkeley National Laboratory, One Cyclotron Road, Berkeley, California 94720, USA}
\def\MM{\textcolor{black}}
\def\MMM{\textcolor{black}}

\begin{document}
\title{Electronic band structure changes across the antiferromagnetic phase transition of exfoliated MnPS$_3$ flakes probed by $\mu$-ARPES}

\author{Jeff Strasdas}\affiliation{\AC}
\author{Benjamin Pestka}\affiliation{\AC}
\author{Miłosz Rybak}\affiliation{\PW}
\author{Adam K. Budniak}\affiliation{\Haifa}
\author{Niklas Leuth}\affiliation{\AC}
\author{Honey Boban}\affiliation{\FZJ}
\author{Vitaliy Feyer}\affiliation{\FZJ}
\author{Iulia Cojocariu}\affiliation{\FZJ}
\author{Daniel Baranowski}\affiliation{\FZJ}
\author{José Avila}\affiliation{\SOLEIL}
\author{Pavel Dudin}\affiliation{\SOLEIL}
\author{Aaron Bostwick}\affiliation{\ALS} 
\author{Chris Jozwiak}\affiliation{\ALS}
\author{Eli Rotenberg}\affiliation{\ALS}
\author{Carmine Autieri}\affiliation{\MAGTOP}
\author{Yaron Amouyal}\affiliation{\HaifaMat}
\author{Lukasz Plucinski}\affiliation{\FZJ}
\author{Efrat Lifshitz}\affiliation{\Haifa} 
\author{Magdalena Birowska}\affiliation{\IFT}
\author{Markus Morgenstern}\affiliation{\AC}
\date{\today} 

\begin{abstract}
Exfoliated magnetic 2D materials \MM{enable} versatile tuning of magnetization, e.g., by gating or \MM{providing} proximity-induced exchange interaction. However, \MM{their} electronic band structure after exfoliation has not been probed, \MM{most likely} 
due to their photochemical sensitivity. Here, we provide micron-scale angle-resolved photoelectron spectroscopy of the exfoliated intralayer antiferromagnet MnPS$_3$ above and below the N\'{e}el temperature down to one monolayer. The favorable comparison with density functional theory calculations enables to identify the orbital character of the observed bands. Consistently, we find pronounced changes \MM{across the Nèel temperature for bands} that consist of Mn 3d and 3p levels of adjacent S atoms. The deduced \MM{orbital mixture} indicates that the superexchange is \MM{relevant for} the magnetic \MM{interaction}. There are only minor changes between monolayer and thicker films \MM{demonstrating} the predominant 2D character of MnPS$_3$.
The novel access \MM{is} transferable to other MPX$_3$ materials (M: transition metal, P: phosphorus, X: chalcogenide) providing a multitude of antiferromagnetic arrangements. 
\end{abstract}
\keywords{Magnetic 2D materials, $\mu$-ARPES, density functional theory}

\maketitle
\noindent {{Corresponding author: } 
M.~Morgenstern, email: mmorgens@physik.rwth-aachen.de } 

\section{Introduction}
The first successful exfoliation of CrI$_3$ \cite{Huang2017}, Cr$_2$Ge$_2$Te$_6$ \cite{Gong2017} and other layered magnetic materials \cite{Huang2020,Gibertini2019,Gong2019,Burch2018,Fei2018} added ferro- and antiferromagnets to the toolbox of exfoliation-based heterostructures \cite{Novoselov2016}. This enabled several magnetoelectric effects such as deliberate gate tuning of the spin-flop transition \cite{Huang2018,Jiang2018b} as well as from antiferro- to ferrimagnetism \cite{Jiang2018} or from ferro- to paramagnetism \cite{Deng2018}. Moreover, the exchange coupling of the ferromagnet could be successfully transferred via proximity to neighboring 2D materials \cite{Sierra2021} such as graphene \cite{Wu2020,Ghiasi2021,Tang2020}, WSe$_2$ \cite{Zhong2020}, the topological insulator WTe$_2$ \cite{Zhao2020} and the superconductor NbSe$_2$ \cite{kang2021,Hamill2021,Kezilebieke2020}.
Recently, also twisted layers of 2D antiferromagnets have been prepared exhibiting distinct magnetic properties from their regularly stacked counterparts \cite{Xie2021,Song2021,Xu2021}. 

Nevertheless, the basic electronic band structure of exfoliated 2D magnetic materials has never been probed. This is most likely due to the extreme photochemical sensitivity of most of the flakes \cite{Shcherbakov2018}. Previous angularly resolved photoelectron spectroscopy (ARPES) data have only been obtained on in-situ cleaved bulk materials \cite{Watson2020,Yilmaz2021,Xu2020,Liu2020,Li2018,Suzuki2019,Zhang2018,Kong2019,Jiang2020,Xie2021} or in-situ after molecular beam epitaxy \cite{Zhang2021}. For this reason, the interesting class of intralayer antiferromagnetic insulators \MM{M}PS$_3$ (\MM{M} = Mn, Fe, Co, Ni) \cite{Joy1992,Susner2017,Wang2018} that feature rather strong magnetoelastic \cite{Vaclavkova2020,Liu2021}, magnetoelectric \cite{Ressouche2010,Lai2019,Chu2020,Kim2018,Kang2020} and magnetically induced electron-phonon \cite{Ergeen2022} couplings have not been examined, since their large band gaps prohibit ARPES of cleaved crystals below the Néel temperature \cite{Kamata1997,Bianchi2023}. 

\begin{figure*}[thb]
\centering
\includegraphics[width=\textwidth]{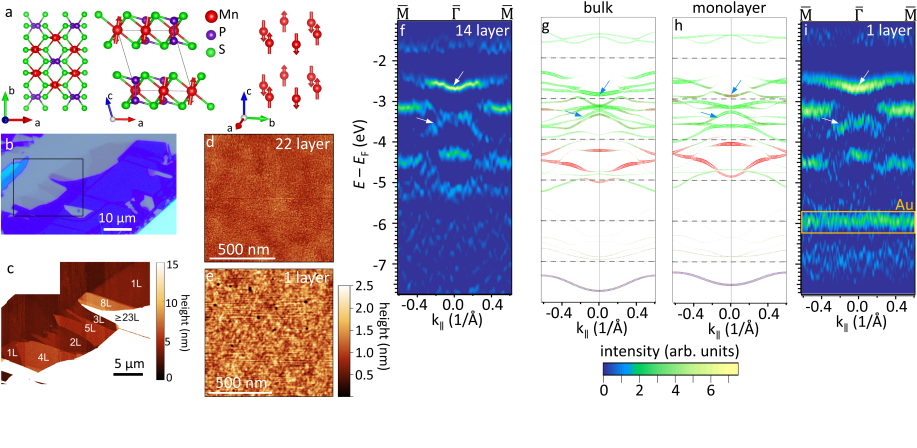}
\vspace{-1.6cm}
\caption{{\bf \MM{Exfoliation and comparison of monolayer and bulk.}}
(a) Atomic and magnetic structure of MnPS$_3$ with Mn magnetic moments depicted as arrows. Left: top view, middle: side view of two layers, right: perspective view with Mn atoms only (Graphics are produced with the VESTA program \cite{VESTA}).
(b) Optical microscope image of a thick flake (violet) with several few-layer flakes at the upper left rim (darker grey areas) exfoliated on Au/Ti/SiO$_2$/Si.
(c) Tapping-mode atomic foce microscopy (AFMi) image of the highlighted area in b with layer thicknesses marked.
(d)--(e) Tapping mode AFMi images of areas probed by ARPES in f, i, rms roughness:  $0.24$\,nm (d), $0.30$\,nm (e). 
\MM{(f) Curvature plot of ARPES data for 14 layers MnPS$_3$, $T=7$\,K, photon energy $h\nu=100$\,eV. 
(g) Selected bands of the band structure from DFT+U for bulk MnPS$_3$ along the same $k_\parallel$ direction as in f, $k_z=0.46$\,\AA, $U=1.8$\,eV. Band selection is according to the matrix elements in ARPES (see text). 
(h) Same as g, but for a monolayer. The colors in g, h mark the dominating orbital with symbol size indicating its strength (color code, Fig.~\ref{Fig_2}). 
(i) Same as f, but for an area with thickness of 1 layer. The band marked Au is due to a Au 5d band of the substrate \MMM{(Supplementary Section S1.G)}   
Arrows in f--i mark bands discussed in the text. } 
}
\label{Fig_1}
\end{figure*}

Here, we investigate exfoliated MnPS$_3$, an intralayer antiferromagnet (AFM) with a honeycomb-type N\'{e}el order of Mn$^{2+}$ ions \MM{(Fig.~\ref{Fig_1}a) \cite{Kurosawa1983,Ressouche2010,Brec1986,Autieri2022}}. 
The AFM order persists down to a monolayer 
\cite{Lim2021,Vaclavkova2020,Kim2019,Long2020}. 
The interlayer exchange interaction is weakly ferromagnetic \cite{Babuka2020} breaking inversion symmetry and, hence, enabling magnetoelectric coupling \cite{Chu2020} \MM{as indeed} found by neutron diffraction 
\cite{Ressouche2010}. \MM{This has been exploited to image the AFM domains ($\sim 100$\,$\mu$m) \cite{Ni2021} and to determine the critical exponent $\beta=0.37$ using second harmonic generation \cite{Chu2020,Shan2021}}.
The N\'{e}el temperature of the bulk material is $T_{\rm N}=78$\,K \MM{\cite{Brec1986,Wildes2006}}. 
The optical band gap 
\MM{is} 2.94\,eV at $T=5$\,K \cite{Grasso1991,Du2015} with additional localized inter-d-level transitions at lower excitation energies \cite{Grasso1991}.
Weak interlayer van-der-Waals binding with energy density $0.25$\,J/m$^2$ (graphene: 0.38\,J/m$^2$) enables exfoliation \cite{Du2015}.
 
Here, we employ such exfoliation down to single layers onto a Au/Ti-covered Si/SiO$_2$ substrate. We probe the insulating material by $\mu$-ARPES above and below $T_{\rm N}$. We find excellent agreement with density functional theory (DFT+U) calculations enabling us to deduce an adequate $U=1.8$\,eV and to assign the orbital character of the observed bands.
The most pronounced changes across $T_{\rm N}$ appear for an occupied band that is dominated by Mn 3d orbitals with additional contributions from the adjacent S 3p orbitals. Weaker changes in band structure are found for less strongly bound bands that are dominated by S 3p orbitals with contributions from Mn 3d levels. These changes are qualitatively reproduced by comparing \MM{with} DFT+U calculations \MM{of the} AFM Néel and the paramagnetic (PM) \MM{phase}. 
 \MM{The changes} are attributed to an influence of, both, direct Mn-Mn exchange and a superexchange that, in turn, lead to a \MM{modification} of the orbital mixture and the dispersion of the bands. The band structure \MM{remains similar} down to the thickness of 1 layer evidencing the strong 2D character of the material. 

\section{Results and discussion}
\subsection{Exfoliation of MnPS$_3$.}
\MM{The exfoliation of MnPS$_3$} typically leads to relatively thick flakes with thin layers down to a monolayer at their rim (Fig.~\ref{Fig_1}b-c) \cite{Du2015}. We found that direct exfoliation after plasma ashing of the Au substrate provides a higher yield of thin areas, e.g.  an  area with a thickness of two layers and a size of $150\times300\,\mu$m$^2$  \MMM{(Supplementary Section S1.B)}.

\subsection{Comparison of bulk and monolayer.}
\MM{The areas of different thickness at the same flake (Fig.~\ref{Fig_1}c) are advantageous \MM{for $\mu$-ARPES with a focus } down to 5 $\mu$m \cite{Wiemann2011}. Hence,  we could probe multiple thicknesess down to the monolayer partially at the same flake. Figure \ref{Fig_1}f--i displays the curvature plots of the recorded ARPES data along $\overline{\rm M}\overline{\Gamma}\overline{\rm M}$ for a thick bulk-type area and a monolayer area, both below $T_{\rm N}$ and recorded with the same ARPES parameters. For comparison, the corresponding bands from the DFT+U calculation \MMM{(Supplementary Section S2)} are displayed. The visible bands are quite similar for the different thicknesses in, both, experiment and calculations.    
This reveals} that the mutual interaction of van der Waals (vdW)-type \MM{between the layers} has \MM{a minor} impact on the electronic properties of MnPS$_3$. \MM{Hence, also the bulk material is predominantly a 2D-type magnet.}  
Minor changes in the band structure are interestingly reproduced by the calculations (arrows in Fig.~\ref{Fig_1}f--i). For example, the nearly parabolic part of the band at $E-E_{\rm F}\simeq -2.5$\,eV \MM{($E_{\rm F}$: Fermi level)} is more curved \MM{around $\overline{\Gamma}$} for one layer in experiment and calculation.
Moreover, the bands propagating  downwards from $\overline{\Gamma}$ at $E-E_{\rm F}\simeq -3.5$\,eV are steeper for the thick film in both panels. This indicates \MM{a detailed agreement between ARPES and DFT+U data or, in turn, an adequate choice of the parameters $U$ and  $k_z$ (wave vector perpendicular to the surface) deduced by detailed comparison \MMM{(Supplementary Section S4)}. Eventually, it demonstrates the appropriateness of DFT+U to describe MnPS$_3$ adequately as discussed in more detail below.}

 \begin{figure*}[htb]
\centering
\includegraphics[width=\textwidth]{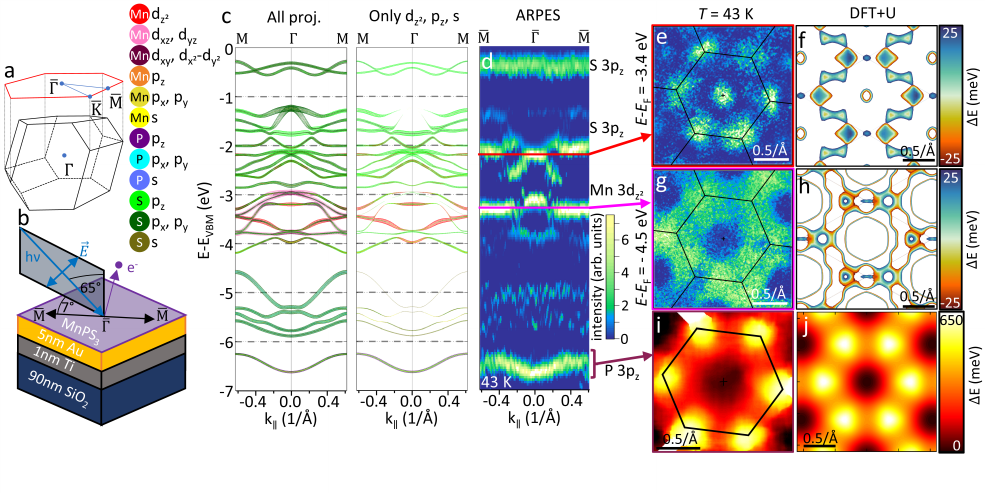}
\vspace{-1.3cm}
\caption{{\bf Electronic structure of \MM{MnPS$_3$ with AFM-Nèel phase}}. (a) \MM{The} first Brillouin zone (BZ) of \MM{the} bulk (black) and \MM{of the} monolayer (red).  (b) 
\MM{Sample} sequence (on top of a Si substrate) and photon beam geometry 
\MM{with} the polarization direction $\vec{E}$ 
\MM{in} the plane of incidence. (c) Band structure obtained 
 within DFT+U approach ($U=1.8$\,eV, $k_z=0.0/$\AA).
The electronic states are color-coded
by their atomic orbital contribution with size giving the contribution strength (color code on the left).
Left: all bands, right: only bands with s, p$_{\rm z}$ and d$_{\rm z^2}$ contribution as required by the \MM{ARPES} selection rules \cite{Moser2017} \MMM{(Supplementary Section S1.J)}.
(d) Curvature plot of ARPES \MM{data below $T_{\rm N}$}, 
$T=43$\,K, $h\nu=50$\,eV,  62 layers. Dominating orbitals of the \MM{visible} bands are \MM{marked.} 
The \MM{intensity scale} applies to d, e, g.
(e) Constant energy ($k_x$, $k_y$) plot of ARPES curvature at $E-E_{\rm F}=-3.4\pm 0.05$\,eV, $T=43$\,K, $h\nu=50$\,eV, 62\,layers. The 2D BZ boundaries 
\MM{are} overlaid (black lines). (f) 
Calculated band energies at
$E-E_{\rm F}=-3.4\pm 0.025$\,eV. 
(g)--(h) same as e--f for $E-E_{\rm F}=-4.5$\,eV.
(i) Color plot of the binding energies for the lowest energy band \MM{in} d, $T=43$\,K, $h\nu=53.5$\,eV,   62 layers. 
The energies are the maximum of the curvature of $I(E)$ for each ($k_x$, $k_y$) within the energy interval $E-E_{\rm F}\in \left[-5,-9\right]$\,eV. (j) Same as i, but deduced from the DFT+U calculations. The arrows from d to e,g,h connect the bands to its ($k_x$, $k_y$) plots. \MM{The horizontal lines mark} the central energies of e, g.}
\label{Fig_2}
\end{figure*}
%
\subsection{Identification of the bands below T$_{\rm N}$.}
Due to the \MM{small} differences between monolayer and bulk, \MM{we concentrate on the thicker layers in the following}.
In order to identify the set of probed bands below \MM{$T_{\rm N}$}, a comparison between the 
\MM{ARPES data at $T=43$\,K and the bands from} the DFT+U approach with the AFM-Néel magnetic phase is discussed.
Figure~\ref{Fig_2}d shows the measured band structure of a thick MnPS$_3$ (62 layers) along the $\overline{\rm M}\overline{\Gamma}\overline{\rm M}$ direction below $T_{\rm N}=78$\,K as recorded at photon energy $h\nu=50$\,eV. This is chosen \MM{to hit} the Mn 3p\,$\rightarrow$\,3d resonance \MM{for} a better visibility of Mn related bands \MMM{(Supplementary Section S1.I}). For comparison, the  \MM{AFM-Néel} band structure calculated by DFT+U is displayed in Fig.~\ref{Fig_2}c with marked orbital contributions. A detailed analysis of characteristic features, that monotonously change with $U$, enabled us to select the proper $U=1.8$\,eV \MMM{(Supplementary Section S4.A)}. Changes of the band structure with $k_z$ are \MM{also observed, but} more subtle. The best agreement \MM{for} $h\nu=50$\,eV is found \MM{at} $k_z\simeq 0.0$/{\AA} \MMM{(Supplementary Section S4.B)}, also leading to a reasonable match of the $k_z$ dispersion deduced from photon energy dependent ARPES \MM{with} DFT+U \MMM{(Supplementary Section S5)}. 
Since matrix elements in our experimental geometry 
(Fig.~\ref{Fig_2}b) select s-type, p$_z$-type and d$_{\rm z^2}$-type states \MMM{(Supplementary Section S1.J)}, we \MM{also} plot the DFT+U results including only bands with these contributions (Fig.~\ref{Fig_2}c, right). Such band selection provides rather good agreement to the experimental data. Moreover, the comparison identifies the \MM{visible} orbital character of the four strongest groups of bands \MM{as labeled on the right of Fig.~\ref{Fig_2}d}. All these bands have additional contributions from other orbitals. The \MM{band} labeled P 3p$_{\rm z}$ has significant contributions from \MM{S 3p orbitals},
the bands labeled Mn 3d$_{\rm z^2}$ consist of all Mn 3d levels and \MM{weaker} contributions from all S 3p orbitals, and the \MM{two blocks of} bands 
labeled S 3$p_{\rm z}$  have  \MM{contributions} from \MM{ the other S 3p} orbitals and weaker contributions from Mn 3d levels.
\MMM{(Supplementary Section S6)}. This demonstrates that all these orbitals are significantly involved in the hybridization, respectively, in the bonding within MnPS$_3$. In particular, the strong mixing of the Mn 3d and S 3p levels \MM{implies their combined involvement as required for the superexchange interaction}. 

To corroborate the good agreement between ARPES and DFT+U data, Fig.~\ref{Fig_2}e--j display the comparison of $(k_x,k_y)$ \MM{cuts.} 
For the low energy band labeled P 3$p_z$, we plot the \MM{band energy deduced by  a maximum finding routine exhibiting} excellent agreement (Fig.~\ref{Fig_2}i--j). Since more bands are involved at higher energy, we plot the ARPES intensity at constant $E-E_{\rm F}$  (Fig.~\ref{Fig_2}e,g) compared with DFT states at \MM{the same energy, but} within a window as given by the ARPES energy resolution (Fig.~\ref{Fig_2}f,h). The states are additionally color coded by its energy. 
Again, one finds a very convincing quantitative agreement between experiment and theory. The ($k_x$, $k_y$) plots also nicely show the \MM{hexagonal} symmetry of the surface projected BZ. \MM{This corroborates that our parameters are adequately chosen and that DFT+U is a reasonable approach to describe MnPS$_3$.}

\subsection{Magnetic mechanism.}
\MM{Next, we relate our data to the knowledge on the magnetic interactions in MnPS$_3$.}
The \MM{monolayer} possesses a D$_{\rm 3d}$ point group symmetry with the Mn atom in a trigonal anti-prismatic environment of S atoms \MM{(Fig.~\ref{Fig_1}a)}. Due to the crystal field splitting, the 3d states of Mn  (3d$^5$, $S=5/2$) \MM{split} into two disentangled subsets \MM{that} exhibit either odd (d$_{\rm xz}$ d$_{\rm yz}$) or even (d$_{\rm xy}$, d$_{\rm x^2-y^2}$, d$_{\rm z^2}$) mirror symmetry  with respect to the monolayer plane. The d$_{\rm z^2}$ orbital further splits from the even subset becoming the lowest energy state.

Let us discuss now the competition between the 
direct exchange or superexchange mechanisms 
by considering the exchange couplings J$_i$, which are are crucial to understand  the origin of the magnetic exchange mechanism underneath.
Both, inelastic neutron scattering \MM{via} the fitted magnon dispersion \cite{Wildes1998} and DFT+U studies \cite{Autieri2022} reveal that the dominant exchange 
\MM{between} the Mn atoms is the nearest neighbor exchange $J_1$ inducing AFM coupling. The second-nearest neighbor coupling $J_2$ is negligible and the third-nearest neighbor coupling $J_3$ is about 1/3 of $J_1$ also leading to AFM coupling. The main contribution to $J_1$ originates from the direct exchange mechanism as revealed, e.g., by model
\MM{H}amiltonians with \textit{ab-initio} parametrization \cite{Autieri2022}. It is mainly caused by the direct overlap between the even orbitals 3d$_{\rm xy}$, 3d$_{\rm z^2}$ and 3d$_{\rm x^2-y^2}$ of neighboring Mn atoms and, to a weaker extent (factor of 5.7 smaller), by  overlapping odd orbitals (d$_{\rm xz}$, d$_{\rm yz}$) \cite{Autieri2022}. Although the effective direct exchange \MM{dominates the} coupling strength 
\MM{\cite{Autieri2022}}, an additional contribution to \MM{$J_1$} originates from \MM{the} superexchange path Mn(d)-S(p)-Mn(p) \MM{as similarly} 
reported for chromium sulfides \MM{(exchange path  in Fig. 5(c) of \cite{Ushakov2013})}. \MM{The} strong hybridization between the S \MM{3p} and neighboring Mn \MM{3d} states is \MM{nicely} visible in the orbital-projected band structure within the energy range -4 to -3 eV (\MM{Fig.~\ref{Fig_2}c,} \MMM{Fig.~S11}). \MM{Interestingly, the} superexchange process \MM{consists of} a strong AFM contribution due to hybridization of the \MM{S 3p$_z$} orbital with t$_{\rm 2g}$ orbitals (neglecting the trigonal distortion, and assuming octahedral coordination \cite{Autieri2022}) and a rather weak FM \MM{part via} the e$_{\rm g}$ orbitals. Overall, the direct and indirect \MM{(superexchange)} mechanism  lead to strong AFM \MM{$J_1$} \cite{Autieri2022}. 

\begin{figure}[htb]
\centering
\includegraphics[width=0.485\textwidth]{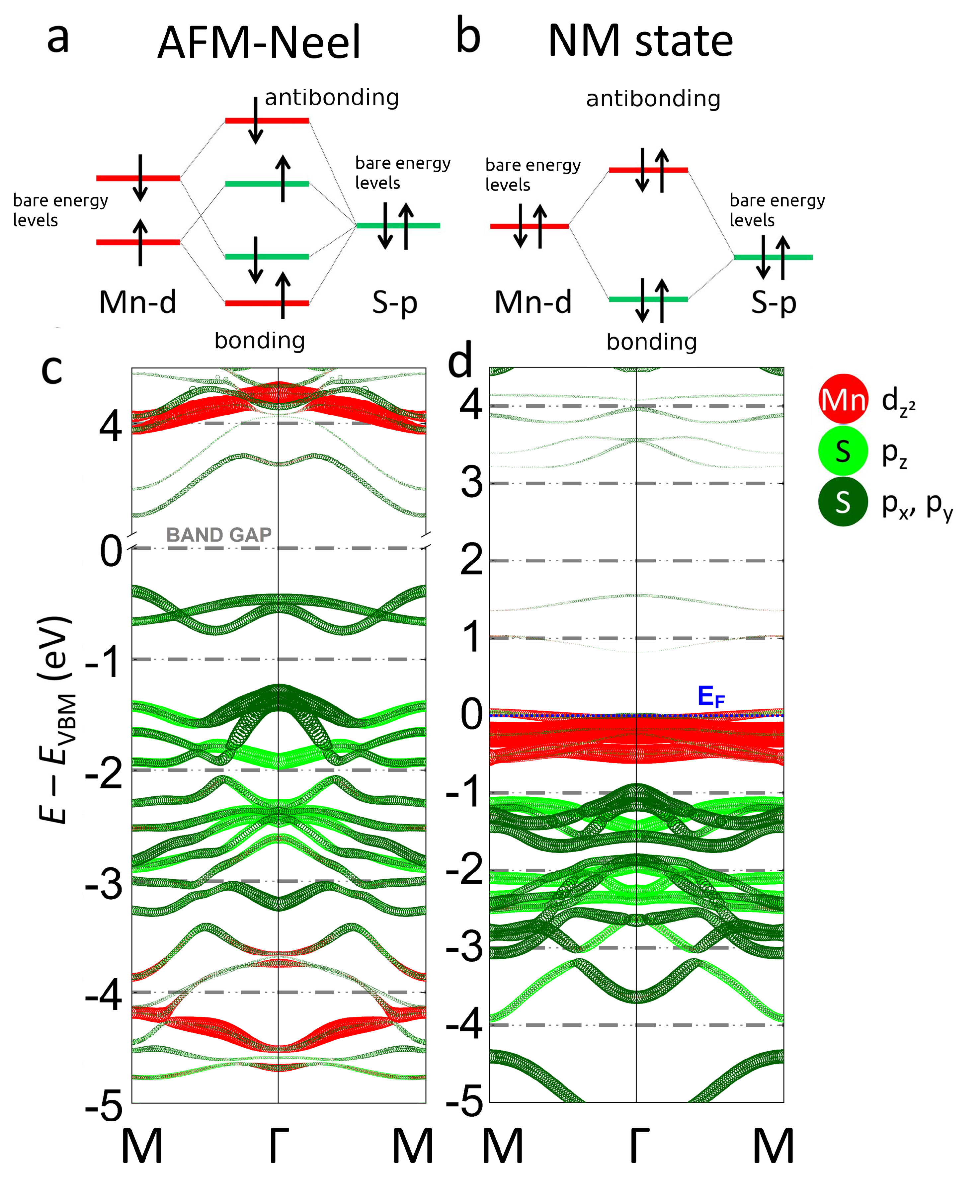}
\caption{{\bf Comparison of magnetic and nonmagnetic (NM) calculations.}
(a), \MM{(b)} \MM{Schematics} of the bare and \MM{the} hybridized atomic orbitals for \MM{(a)} AFM Néel and (b) NM phase. The atomic Mn 3d levels (red) are either spin-split or not and subsequently hybridize with the S 3p levels (green).  Up and down arrows represent electronic states of the majority and the minority spin-channel in (a), respectively.  \MM{(c), (d)} \MM{Orbital-projected Bulk} band structure for \MM{(c)} AFM Néel phase within  DFT+U ($U = 1.8$\,eV, \MM{same as Fig.~\ref{Fig_2}c)} and \MM{(d)} \MM{NM} phase within  DFT calculation ignoring spin degrees of freedom. }
\label{Fig_5}
\end{figure}
%
\subsection{Band changes across the magnetic transition.}

To illustrate the effect of magnetism on the energy
bands across the Néel temperature, we firstly present an
oversimplified model assuming energy levels within magnetic and non-magnetic cases. This established model
provides qualitative understanding of the band alignments and hybridization. 
\MM{Afterwards}, we discuss the \MM{band} changes \MM{across $T_{\rm N}$ observed in \MM{the} ARPES measurements,  comparing them} with the AFM Néel and PM phase \MM{as modeled by DFT+U}.


\MM{The atomic model is} presented in  Fig. \ref{Fig_5}\MM{a--b}. \MM{The} bare energy levels of Mn \MM{3}d and S \MM{3}p \MM{are} shown in the outer parts. 
\MM{With hybridization}, we obtain the order of energetic levels \MM{as shown} in the \MM{center.} 
In \MM{the} NM phase, the bare energy of the Mn \MM{3d$_{z^2}$} orbitals is above the energy of the S \MM{3}p orbitals, therefore, the \MM{pd}-hybridization produces a bonding state that is mainly S \MM{3}p and an antibonding state that is mainly Mn \MM{3}d  \MM{(Fig. \ref{Fig_5}b)}. \MM{This} picture is reflect\MM{ed} in the \MM{calculated} band structure of \MM{the} NM phase,  where spin degrees of freedom and Hubbard \MM{$U$} parameters \MM{are} neglected  (\MM{Fig.~\ref{Fig_5}d}). The NM phase
exhibits bands of pure 3d$_{\rm {z^2}}$ character at $E-E_{\rm VBM}\simeq -0.5$\,eV ($E_{\rm VBM}$: valence band maximum). At slightly higher energies, one finds bands purely from the other even 3d orbitals as expected for the trigonal anti-prismatic crystal field (not shown). 
\MM{The} Mn 3d bands are all at higher energies than the occupied S 3p bands.
In \MM{the} AFM Néel case, the spin split\MM{ting} pushes the majority spins of the Mn \MM{3d$_{z^2}$} below the S \MM{3}p orbitals\MM{. Therefore,} the hybridization \MM{of the majority spins} produces a bonding state that is mainly Mn \MM{3d$_{z^2}$} and an antibonding state that is mainly S \MM{3}p (\MM{Fig.~\ref{Fig_5}a}). This, in turn, is visible in \MM{the calculated} AFM Néel band structure  (\MM{Fig. \ref{Fig_5}c}), where the Mn 3d$_{\rm z^2}$ bands \MM{appear} now $4$\,eV  below  and above \MM{$E_{\rm F}$}, falling into valence and conduction band, respectively. \MM{The level order of the} majority spin-channel\MM{, thus,} is opposite \MM{to the order in} the NM case. 
\MM{One can also see that the orbital character of the bands is significantly  mixed in the AFM case as expected for hybridization. This is even more evident in Fig.~\ref{Fig_2}c. The participation of S 3p and Mn 3d orbitals in the same band, moreover, strongly indicates the involvement of superexchange processes in the AFM ordering \MMM{(Supplementary Section S7)}}. Finally, the confinement of the Mn-d bands at $E_{\rm F}$
strongly promotes interaction effects between the electrons, i.e. the strong Coulomb interactions through the narrow d bands
and the half-filled condition (valence electrons of \MM{the} Mn ion
are in the d$^5$ configuration) \MM{points} to the insulating
Mott state \MM{ even at} high temperature.

\begin{figure}[htb]
\centering
\includegraphics[width=0.485\textwidth]{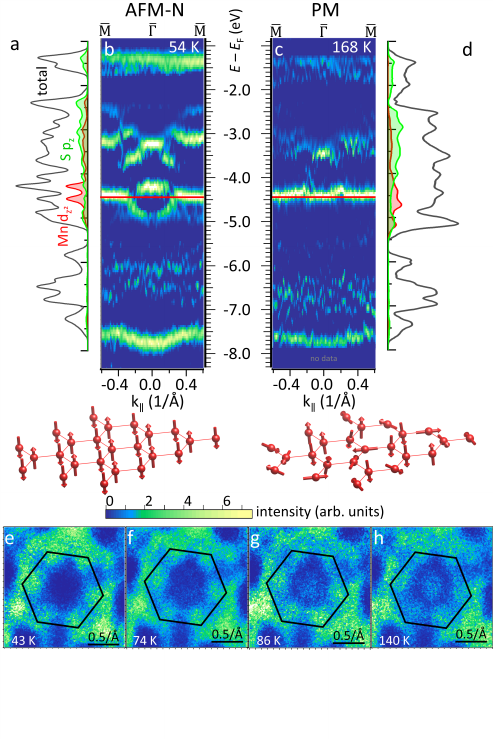}
\vspace{-2.1cm}
\caption{{\bf Comparison between \MM{AFM-Neel and PM phase}.}  \MM{(a)} Density of states (black) and orbital contributions from Mn 3d$_{\rm z^2}$ (red) and S 3p$_{\rm z}$ (green) orbitals for the calculated AFM Néel phase sketched below.  (b)--(c) Curvature plots of ARPES data measured at the same sample spot for the temperatures marked ($T_{\rm N}=78$\,K), $h\nu=50\,eV$, 62\,layers. \MM{(d) Same as (a) for} the PM phase \MM{with spin configuration sketched below}.  (e)--(h) Constant energy ($k_x$, $k_y$) plots of ARPES curvature recorded at the same sample spot at varying $T$ as marked, $E-E_{\rm F}=-4.5\pm 0.05$\,eV \MM{as indicated in b}, $h\nu=50$\,eV, 62\,layers. The 2D BZ of MnPS$_3$ is overlaid (black hexagon).}
\label{Fig_3}
\end{figure}

%
\MM{The experimental changes of the electronic structure with temperature $T$ are shown in Fig.~\ref{Fig_3}, i.e. ARPES data above and below $T_{\rm N}$  recorded at the identical spot of the sample.} They are displayed along the $\overline{\rm M}\overline{\Gamma}\overline{\rm M}$ direction (Fig.~\ref{Fig_3}b,c) and  as ($k_x$,$k_y$) cuts through the band dominated by the Mn 3$d_{\rm z^2}$ orbitals (Fig.~\ref{Fig_3}e--h). ARPES data at additional $T$ \MM{are} presented in \MM{Supplementary Section S3.} Most striking is the apparent splitting of the Mn 3$d_{\rm z^2}$ band in the central region around $\overline{\Gamma}$, that disappears above $T_{\rm N}$. Accordingly, the energy maximum of the upper band shifts from $\overline{\Gamma}$ towards \MM{$|k_\parallel|\simeq 0.25/$\,{\AA}}.
This band change is corroborated by the ($k_x$,$k_y$) plots (Fig.~\ref{Fig_3}e--g) displayed at an energy slightly below the \MM{maxima.} 
\MM{It exhibits a circle of} missing intensity in the BZ center below $T_{\rm N}=78$\,K (Fig.~\ref{Fig_3}e,f) \MM{and}  a ring of reduced intensity above $T_{\rm N}$ (Fig.~\ref{Fig_3}g,h). \MM{The hexagonal symmetry 
is} preserved below $T_{\rm N}$ as expected for \MM{the} AFM N\'{e}el phase \MM{with} primitive magnetic cell commensurate with  the structural cell \cite{Kurosawa1983}. There are additional more subtle changes between Fig.~\ref{Fig_3}b and c. Firstly, the curvature of the P 3p$_z$ band at lowest energy changes. Secondly, the two branches declining with $|k_\parallel|$ at $E-E_{\rm F} \simeq -3.6$\,eV  largely disappear and the central plateau above the declining bands shifts downwards above $T_{\rm N}$. These changes are reproduced in several cool-downs for distinct flakes with different thicknesses 
\MM{and} are also observed in 
\MM{the} raw data (\MMM{Supplementary Fig. S7e}). Hence, the \MM{S 3p} bands \MM{also} change across $T_{\rm N}$, but it remains difficult to pinpoint \MM{their} changes \MM{precisely.} \MM{Nevertheless, this again points to the involvement of superexchange.} 

To \MM{model} the ARPES data above $T_{\rm N}$, we \MM{calculate} the \MM{PM phase} within the DFT+U approach \MMM{(Supplementary Section S2)}. \MM{It} is a disorder\MM{ed} magnetic state, where each spin fluctuates randomly and rapidly without any correlation to \MM{its} neighbors.  The PM \MM{calculation shows}  d\MM{$_{z^2}$ levels} over a \MM{wider energy} range, whereas the AFM Néel \MM{phase exhibits} two \MM{distinct} d\MM{$_{z^2}$} peaks (\MM{Fig.~\ref{Fig_3}a,d)}. \MM{Thus, the leading change of the Mn 3d bands across $T_{\rm N}$ is reproduced, while not the details. Note that also the distribution of S 3p levels changes across $T_{\rm N}$ corroborating their involvement in the magnetic structure.}   

\section{Conclusions}
We have demonstrated the first successful ARPES study of exfoliated magnetic 2D materials using the rather inert MnPS$_3$. We find good agreement with DFT+U calculations, if the selection rules imposed by the photon beam geometry are adequately taken into account. This, in turn, enables to adapt the $U$ parameter to $U=1.8$\,eV and to identify the orbital character of the observed bands.
Cooling below the N\'{e}el temperature, we find  pronounced changes of bands that are dominated by Mn 3d orbitals, but hybridized with S 3p orbitals, and smaller changes for bands with dominating S 3p character, but hybridized with Mn 3d levels. Comparing DFT+U data of the antiferromagnetic and \MM{paramagnetic} phase reproduces the changes qualitatively. 
A very similar band structure is observed down to the monolayer limit with minor changes that appear both in experiment and calculation. 

The large variety of MPX$_3$ materials (M = Ni, Fe, Co, Zn; X = S, Se) as well as corresponding alloys with multiple magnetic and magnetoelectric configurations \cite{Grasso2002} opens a vast field for studying band structure changes across the critical temperature.
This will enable a more concise understanding of the interrelation between magnetism and electronic structure for these novel types of true 2D magnets.


\section*{Acknowledgement}
We acknowledge financial support from the German Research Foundation (DFG) via the project Mo 858/19-1, from the European Union's Horizon 2020 research and innovation program under grant agreement number 881603 (Graphene Flagship, Core 3) and from the German Ministry of Education and Research (Project 05K2022 -ioARPES). M.B. acknowledges support from the University of Warsaw within the project “Excellence Initiative-Research University” program. Access to computing facilities of PL-Grid Polish Infrastructure for Supporting Computational
Science in the European Research Space and of the Interdisciplinary Center of Modeling
(ICM), University of Warsaw are gratefully acknowledged.
We acknowledge access to the NanoESCA beamline of Elettra, the Italian synchrotron facility.
E.L. is supported by the Israel Science Foundation (project no. 2528/19) and by the Deutsche - Israel Program (project no. NA1223/2-1).
A.K.B and E.F. were supported by the European Commission via the Marie-Sklodowska Curie action Phonsi (H2020-MSCA-ITN-642656). C. A. is supported by the Foundation for Polish Science through
the International Research Agendas program co-financed
by the European Union within the Smart Growth Operational Programme (Grant No. MAB/2017/1).


\begin{thebibliography}{66}
\expandafter\ifx\csname natexlab\endcsname\relax\def\natexlab#1{#1}\fi
\expandafter\ifx\csname bibnamefont\endcsname\relax
  \def\bibnamefont#1{#1}\fi
\expandafter\ifx\csname bibfnamefont\endcsname\relax
  \def\bibfnamefont#1{#1}\fi
\expandafter\ifx\csname citenamefont\endcsname\relax
  \def\citenamefont#1{#1}\fi
\expandafter\ifx\csname url\endcsname\relax
  \def\url#1{\texttt{#1}}\fi
\expandafter\ifx\csname urlprefix\endcsname\relax\def\urlprefix{URL }\fi
\providecommand{\bibinfo}[2]{#2}
\providecommand{\eprint}[2][]{\url{#2}}

\bibitem[{\citenamefont{Huang et~al.}(2017)\citenamefont{Huang, Clark,
  Navarro-Moratalla, Klein, Cheng, Seyler, Zhong, Schmidgall, McGuire, Cobden
  et~al.}}]{Huang2017}
\bibinfo{author}{\bibfnamefont{B.}~\bibnamefont{Huang}},
  \bibinfo{author}{\bibfnamefont{G.}~\bibnamefont{Clark}},
  \bibinfo{author}{\bibfnamefont{E.}~\bibnamefont{Navarro-Moratalla}},
  \bibinfo{author}{\bibfnamefont{D.~R.} \bibnamefont{Klein}},
  \bibinfo{author}{\bibfnamefont{R.}~\bibnamefont{Cheng}},
  \bibinfo{author}{\bibfnamefont{K.~L.} \bibnamefont{Seyler}},
  \bibinfo{author}{\bibfnamefont{D.}~\bibnamefont{Zhong}},
  \bibinfo{author}{\bibfnamefont{E.}~\bibnamefont{Schmidgall}},
  \bibinfo{author}{\bibfnamefont{M.~A.} \bibnamefont{McGuire}},
  \bibinfo{author}{\bibfnamefont{D.~H.} \bibnamefont{Cobden}},
  \bibinfo{author}{\bibfnamefont{W.}~\bibnamefont{Yao}},
  \bibinfo{author}{\bibfnamefont{D.}~\bibnamefont{Xiao}},
  \bibinfo{author}{\bibfnamefont{P.}~\bibnamefont{Jarillo-Herrero}},
  \bibnamefont{and} \bibinfo{author}{\bibfnamefont{X.}~\bibnamefont{Xu}},
  \emph{\bibinfo{title}{Layer-dependent ferromagnetism in a van der Waals
  crystal down to the monolayer limit}}, \bibinfo{journal}{Nature}
  \textbf{\bibinfo{volume}{546}}, \bibinfo{pages}{270} (\bibinfo{year}{2017}),
  \urlprefix\url{https://doi.org/10.1038/nature22391}.

\bibitem[{\citenamefont{Gong et~al.}(2017)\citenamefont{Gong, Li, Li, Ji,
  Stern, Xia, Cao, Bao, Wang, Wang et~al.}}]{Gong2017}
\bibinfo{author}{\bibfnamefont{C.}~\bibnamefont{Gong}},
  \bibinfo{author}{\bibfnamefont{L.}~\bibnamefont{Li}},
  \bibinfo{author}{\bibfnamefont{Z.}~\bibnamefont{Li}},
  \bibinfo{author}{\bibfnamefont{H.}~\bibnamefont{Ji}},
  \bibinfo{author}{\bibfnamefont{A.}~\bibnamefont{Stern}},
  \bibinfo{author}{\bibfnamefont{Y.}~\bibnamefont{Xia}},
  \bibinfo{author}{\bibfnamefont{T.}~\bibnamefont{Cao}},
  \bibinfo{author}{\bibfnamefont{W.}~\bibnamefont{Bao}},
  \bibinfo{author}{\bibfnamefont{C.}~\bibnamefont{Wang}},
  \bibinfo{author}{\bibfnamefont{Y.}~\bibnamefont{Wang}},
  \bibinfo{author}{\bibfnamefont{Z.~Q.} \bibnamefont{Qiu}},
  \bibinfo{author}{\bibfnamefont{R.~J.} \bibnamefont{Cava}},
  \bibinfo{author}{\bibfnamefont{S.~G.} \bibnamefont{Louie}},
  \bibinfo{author}{\bibfnamefont{J.}~\bibnamefont{Xia}}, \bibnamefont{and}
  \bibinfo{author}{\bibfnamefont{X.}~\bibnamefont{Zhang}},
  \emph{\bibinfo{title}{Discovery of intrinsic ferromagnetism in
  two-dimensional van der Waals crystals}}, \bibinfo{journal}{Nature}
  \textbf{\bibinfo{volume}{546}}, \bibinfo{pages}{265} (\bibinfo{year}{2017}),
  \urlprefix\url{https://doi.org/10.1038/nature22060}.

\bibitem[{\citenamefont{Huang et~al.}(2020)\citenamefont{Huang, McGuire, May,
  Xiao, Jarillo-Herrero, and Xu}}]{Huang2020}
\bibinfo{author}{\bibfnamefont{B.}~\bibnamefont{Huang}},
  \bibinfo{author}{\bibfnamefont{M.~A.} \bibnamefont{McGuire}},
  \bibinfo{author}{\bibfnamefont{A.~F.} \bibnamefont{May}},
  \bibinfo{author}{\bibfnamefont{D.}~\bibnamefont{Xiao}},
  \bibinfo{author}{\bibfnamefont{P.}~\bibnamefont{Jarillo-Herrero}},
  \bibnamefont{and} \bibinfo{author}{\bibfnamefont{X.}~\bibnamefont{Xu}},
  \emph{\bibinfo{title}{Emergent phenomena and proximity effects in
  two-dimensional magnets and heterostructures}}, \bibinfo{journal}{Nat.
  Mater.} \textbf{\bibinfo{volume}{19}}, \bibinfo{pages}{1276}
  (\bibinfo{year}{2020}),
  \urlprefix\url{https://doi.org/10.1038/s41563-020-0791-8}.

\bibitem[{\citenamefont{Gibertini et~al.}(2019)\citenamefont{Gibertini,
  Koperski, Morpurgo, and Novoselov}}]{Gibertini2019}
\bibinfo{author}{\bibfnamefont{M.}~\bibnamefont{Gibertini}},
  \bibinfo{author}{\bibfnamefont{M.}~\bibnamefont{Koperski}},
  \bibinfo{author}{\bibfnamefont{A.~F.} \bibnamefont{Morpurgo}},
  \bibnamefont{and} \bibinfo{author}{\bibfnamefont{K.~S.}
  \bibnamefont{Novoselov}}, \emph{\bibinfo{title}{Magnetic 2D materials and
  heterostructures}}, \bibinfo{journal}{Nat. Nanotechnol.}
  \textbf{\bibinfo{volume}{14}}, \bibinfo{pages}{408} (\bibinfo{year}{2019}),
  \urlprefix\url{https://doi.org/10.1038/s41565-019-0438-6}.

\bibitem[{\citenamefont{Gong and Zhang}(2019)}]{Gong2019}
\bibinfo{author}{\bibfnamefont{C.}~\bibnamefont{Gong}} \bibnamefont{and}
  \bibinfo{author}{\bibfnamefont{X.}~\bibnamefont{Zhang}},
  \emph{\bibinfo{title}{Two-dimensional magnetic crystals and emergent
  heterostructure devices}}, \bibinfo{journal}{Science}
  \textbf{\bibinfo{volume}{363}} (\bibinfo{year}{2019}),
  \urlprefix\url{https://doi.org/10.1126/science.aav4450}.

\bibitem[{\citenamefont{Burch et~al.}(2018)\citenamefont{Burch, Mandrus, and
  Park}}]{Burch2018}
\bibinfo{author}{\bibfnamefont{K.~S.} \bibnamefont{Burch}},
  \bibinfo{author}{\bibfnamefont{D.}~\bibnamefont{Mandrus}}, \bibnamefont{and}
  \bibinfo{author}{\bibfnamefont{J.-G.} \bibnamefont{Park}},
  \emph{\bibinfo{title}{Magnetism in two-dimensional van der Waals materials}},
  \bibinfo{journal}{Nature} \textbf{\bibinfo{volume}{563}}, \bibinfo{pages}{47}
  (\bibinfo{year}{2018}),
  \urlprefix\url{https://doi.org/10.1038/s41586-018-0631-z}.

\bibitem[{\citenamefont{Fei et~al.}(2018)\citenamefont{Fei, Huang, Malinowski,
  Wang, Song, Sanchez, Yao, Xiao, Zhu, May et~al.}}]{Fei2018}
\bibinfo{author}{\bibfnamefont{Z.}~\bibnamefont{Fei}},
  \bibinfo{author}{\bibfnamefont{B.}~\bibnamefont{Huang}},
  \bibinfo{author}{\bibfnamefont{P.}~\bibnamefont{Malinowski}},
  \bibinfo{author}{\bibfnamefont{W.}~\bibnamefont{Wang}},
  \bibinfo{author}{\bibfnamefont{T.}~\bibnamefont{Song}},
  \bibinfo{author}{\bibfnamefont{J.}~\bibnamefont{Sanchez}},
  \bibinfo{author}{\bibfnamefont{W.}~\bibnamefont{Yao}},
  \bibinfo{author}{\bibfnamefont{D.}~\bibnamefont{Xiao}},
  \bibinfo{author}{\bibfnamefont{X.}~\bibnamefont{Zhu}},
  \bibinfo{author}{\bibfnamefont{A.~F.} \bibnamefont{May}},
  \bibinfo{author}{\bibfnamefont{W.}~\bibnamefont{Wu}},
  \bibinfo{author}{\bibfnamefont{D.~H.} \bibnamefont{Cobden}},
  \bibinfo{author}{\bibfnamefont{J.-H.} \bibnamefont{Chu}}, \bibnamefont{and}
  \bibinfo{author}{\bibfnamefont{X.}~\bibnamefont{Xu}},
  \emph{\bibinfo{title}{Two-dimensional itinerant ferromagnetism in atomically
  thin Fe3GeTe2}}, \bibinfo{journal}{Nat. Mater.}
  \textbf{\bibinfo{volume}{17}}, \bibinfo{pages}{778} (\bibinfo{year}{2018}),
  \urlprefix\url{https://doi.org/10.1038/s41563-018-0149-7}.

\bibitem[{\citenamefont{Novoselov et~al.}(2016)\citenamefont{Novoselov,
  Mishchenko, Carvalho, and Neto}}]{Novoselov2016}
\bibinfo{author}{\bibfnamefont{K.~S.} \bibnamefont{Novoselov}},
  \bibinfo{author}{\bibfnamefont{A.}~\bibnamefont{Mishchenko}},
  \bibinfo{author}{\bibfnamefont{A.}~\bibnamefont{Carvalho}}, \bibnamefont{and}
  \bibinfo{author}{\bibfnamefont{A.~H.~C.} \bibnamefont{Neto}},
  \emph{\bibinfo{title}{2D materials and van der Waals heterostructures}},
  \bibinfo{journal}{Science} \textbf{\bibinfo{volume}{353}}
  (\bibinfo{year}{2016}),
  \urlprefix\url{https://doi.org/10.1126/science.aac9439}.

\bibitem[{\citenamefont{Huang et~al.}(2018)\citenamefont{Huang, Clark, Klein,
  MacNeill, Navarro-Moratalla, Seyler, Wilson, McGuire, Cobden, Xiao
  et~al.}}]{Huang2018}
\bibinfo{author}{\bibfnamefont{B.}~\bibnamefont{Huang}},
  \bibinfo{author}{\bibfnamefont{G.}~\bibnamefont{Clark}},
  \bibinfo{author}{\bibfnamefont{D.~R.} \bibnamefont{Klein}},
  \bibinfo{author}{\bibfnamefont{D.}~\bibnamefont{MacNeill}},
  \bibinfo{author}{\bibfnamefont{E.}~\bibnamefont{Navarro-Moratalla}},
  \bibinfo{author}{\bibfnamefont{K.~L.} \bibnamefont{Seyler}},
  \bibinfo{author}{\bibfnamefont{N.}~\bibnamefont{Wilson}},
  \bibinfo{author}{\bibfnamefont{M.~A.} \bibnamefont{McGuire}},
  \bibinfo{author}{\bibfnamefont{D.~H.} \bibnamefont{Cobden}},
  \bibinfo{author}{\bibfnamefont{D.}~\bibnamefont{Xiao}},
  \bibinfo{author}{\bibfnamefont{W.}~\bibnamefont{Yao}},
  \bibinfo{author}{\bibfnamefont{P.}~\bibnamefont{Jarillo-Herrero}},
  \bibnamefont{and} \bibinfo{author}{\bibfnamefont{X.}~\bibnamefont{Xu}},
  \emph{\bibinfo{title}{Electrical control of 2D magnetism in bilayer {CrI}3}},
  \bibinfo{journal}{Nat. Nanotechnol.} \textbf{\bibinfo{volume}{13}},
  \bibinfo{pages}{544} (\bibinfo{year}{2018}),
  \urlprefix\url{https://doi.org/10.1038/s41565-018-0121-3}.

\bibitem[{\citenamefont{Jiang et~al.}(2018{\natexlab{a}})\citenamefont{Jiang,
  Shan, and Mak}}]{Jiang2018b}
\bibinfo{author}{\bibfnamefont{S.}~\bibnamefont{Jiang}},
  \bibinfo{author}{\bibfnamefont{J.}~\bibnamefont{Shan}}, \bibnamefont{and}
  \bibinfo{author}{\bibfnamefont{K.~F.} \bibnamefont{Mak}},
  \emph{\bibinfo{title}{Electric-field switching of two-dimensional van der
  Waals magnets}}, \bibinfo{journal}{Nat. Mater.}
  \textbf{\bibinfo{volume}{17}}, \bibinfo{pages}{406}
  (\bibinfo{year}{2018}{\natexlab{a}}),
  \urlprefix\url{https://doi.org/10.1038/s41563-018-0040-6}.

\bibitem[{\citenamefont{Jiang et~al.}(2018{\natexlab{b}})\citenamefont{Jiang,
  Li, Wang, Mak, and Shan}}]{Jiang2018}
\bibinfo{author}{\bibfnamefont{S.}~\bibnamefont{Jiang}},
  \bibinfo{author}{\bibfnamefont{L.}~\bibnamefont{Li}},
  \bibinfo{author}{\bibfnamefont{Z.}~\bibnamefont{Wang}},
  \bibinfo{author}{\bibfnamefont{K.~F.} \bibnamefont{Mak}}, \bibnamefont{and}
  \bibinfo{author}{\bibfnamefont{J.}~\bibnamefont{Shan}},
  \emph{\bibinfo{title}{Controlling magnetism in 2D {CrI}3 by electrostatic
  doping}}, \bibinfo{journal}{Nat. Nanotechnol.} \textbf{\bibinfo{volume}{13}},
  \bibinfo{pages}{549} (\bibinfo{year}{2018}{\natexlab{b}}),
  \urlprefix\url{https://doi.org/10.1038/s41565-018-0135-x}.

\bibitem[{\citenamefont{Deng et~al.}(2018)\citenamefont{Deng, Yu, Song, Zhang,
  Wang, Sun, Yi, Wu, Wu, Zhu et~al.}}]{Deng2018}
\bibinfo{author}{\bibfnamefont{Y.}~\bibnamefont{Deng}},
  \bibinfo{author}{\bibfnamefont{Y.}~\bibnamefont{Yu}},
  \bibinfo{author}{\bibfnamefont{Y.}~\bibnamefont{Song}},
  \bibinfo{author}{\bibfnamefont{J.}~\bibnamefont{Zhang}},
  \bibinfo{author}{\bibfnamefont{N.~Z.} \bibnamefont{Wang}},
  \bibinfo{author}{\bibfnamefont{Z.}~\bibnamefont{Sun}},
  \bibinfo{author}{\bibfnamefont{Y.}~\bibnamefont{Yi}},
  \bibinfo{author}{\bibfnamefont{Y.~Z.} \bibnamefont{Wu}},
  \bibinfo{author}{\bibfnamefont{S.}~\bibnamefont{Wu}},
  \bibinfo{author}{\bibfnamefont{J.}~\bibnamefont{Zhu}},
  \bibinfo{author}{\bibfnamefont{J.}~\bibnamefont{Wang}},
  \bibinfo{author}{\bibfnamefont{X.~H.} \bibnamefont{Chen}}, \bibnamefont{and}
  \bibinfo{author}{\bibfnamefont{Y.}~\bibnamefont{Zhang}},
  \emph{\bibinfo{title}{Gate-tunable room-temperature ferromagnetism in
  two-dimensional Fe3GeTe2}}, \bibinfo{journal}{Nature}
  \textbf{\bibinfo{volume}{563}}, \bibinfo{pages}{94} (\bibinfo{year}{2018}),
  \urlprefix\url{https://doi.org/10.1038/s41586-018-0626-9}.

\bibitem[{\citenamefont{Sierra et~al.}(2021)\citenamefont{Sierra, Fabian,
  Kawakami, Roche, and Valenzuela}}]{Sierra2021}
\bibinfo{author}{\bibfnamefont{J.~F.} \bibnamefont{Sierra}},
  \bibinfo{author}{\bibfnamefont{J.}~\bibnamefont{Fabian}},
  \bibinfo{author}{\bibfnamefont{R.~K.} \bibnamefont{Kawakami}},
  \bibinfo{author}{\bibfnamefont{S.}~\bibnamefont{Roche}}, \bibnamefont{and}
  \bibinfo{author}{\bibfnamefont{S.~O.} \bibnamefont{Valenzuela}},
  \emph{\bibinfo{title}{Van der Waals heterostructures for spintronics and
  opto-spintronics}}, \bibinfo{journal}{Nat. Nanotechnol.}
  \textbf{\bibinfo{volume}{16}}, \bibinfo{pages}{856} (\bibinfo{year}{2021}),
  \urlprefix\url{https://doi.org/10.1038/s41565-021-00936-x}.

\bibitem[{\citenamefont{Wu et~al.}(2020)\citenamefont{Wu, Yin, Pan, Grutter,
  Pan, Lee, Gilbert, Borchers, Ratcliff, Li et~al.}}]{Wu2020}
\bibinfo{author}{\bibfnamefont{Y.}~\bibnamefont{Wu}},
  \bibinfo{author}{\bibfnamefont{G.}~\bibnamefont{Yin}},
  \bibinfo{author}{\bibfnamefont{L.}~\bibnamefont{Pan}},
  \bibinfo{author}{\bibfnamefont{A.~J.} \bibnamefont{Grutter}},
  \bibinfo{author}{\bibfnamefont{Q.}~\bibnamefont{Pan}},
  \bibinfo{author}{\bibfnamefont{A.}~\bibnamefont{Lee}},
  \bibinfo{author}{\bibfnamefont{D.~A.} \bibnamefont{Gilbert}},
  \bibinfo{author}{\bibfnamefont{J.~A.} \bibnamefont{Borchers}},
  \bibinfo{author}{\bibfnamefont{W.}~\bibnamefont{Ratcliff}},
  \bibinfo{author}{\bibfnamefont{A.}~\bibnamefont{Li}},
  \bibinfo{author}{\bibfnamefont{X.}~\bibnamefont{dong Han}}, \bibnamefont{and}
  \bibinfo{author}{\bibfnamefont{K.~L.} \bibnamefont{Wang}},
  \emph{\bibinfo{title}{Large exchange splitting in monolayer graphene
  magnetized by an antiferromagnet}}, \bibinfo{journal}{Nat. Electron.}
  \textbf{\bibinfo{volume}{3}}, \bibinfo{pages}{604} (\bibinfo{year}{2020}),
  \urlprefix\url{https://doi.org/10.1038/s41928-020-0458-0}.

\bibitem[{\citenamefont{Ghiasi et~al.}(2021)\citenamefont{Ghiasi, Kaverzin,
  Dismukes, de~Wal, Roy, and van Wees}}]{Ghiasi2021}
\bibinfo{author}{\bibfnamefont{T.~S.} \bibnamefont{Ghiasi}},
  \bibinfo{author}{\bibfnamefont{A.~A.} \bibnamefont{Kaverzin}},
  \bibinfo{author}{\bibfnamefont{A.~H.} \bibnamefont{Dismukes}},
  \bibinfo{author}{\bibfnamefont{D.~K.} \bibnamefont{de~Wal}},
  \bibinfo{author}{\bibfnamefont{X.}~\bibnamefont{Roy}}, \bibnamefont{and}
  \bibinfo{author}{\bibfnamefont{B.~J.} \bibnamefont{van Wees}},
  \emph{\bibinfo{title}{Electrical and thermal generation of spin currents by
  magnetic bilayer graphene}}, \bibinfo{journal}{Nat. Nanotechnol.}
  \textbf{\bibinfo{volume}{16}}, \bibinfo{pages}{788} (\bibinfo{year}{2021}),
  \urlprefix\url{https://doi.org/10.1038/s41565-021-00887-3}.

\bibitem[{\citenamefont{Tang et~al.}(2020)\citenamefont{Tang, Zhang, Lai, Tan,
  and bo~Gao}}]{Tang2020}
\bibinfo{author}{\bibfnamefont{C.}~\bibnamefont{Tang}},
  \bibinfo{author}{\bibfnamefont{Z.}~\bibnamefont{Zhang}},
  \bibinfo{author}{\bibfnamefont{S.}~\bibnamefont{Lai}},
  \bibinfo{author}{\bibfnamefont{Q.}~\bibnamefont{Tan}}, \bibnamefont{and}
  \bibinfo{author}{\bibfnamefont{W.}~\bibnamefont{bo~Gao}},
  \emph{\bibinfo{title}{Magnetic Proximity Effect in Graphene/{CrBr} 3 van der
  Waals Heterostructures}}, \bibinfo{journal}{Adv. Mater.}
  \textbf{\bibinfo{volume}{32}}, \bibinfo{pages}{1908498}
  (\bibinfo{year}{2020}),
  \urlprefix\url{https://doi.org/10.1002/adma.201908498}.

\bibitem[{\citenamefont{Zhong et~al.}(2020)\citenamefont{Zhong, Seyler,
  Linpeng, Wilson, Taniguchi, Watanabe, McGuire, Fu, Xiao, Yao
  et~al.}}]{Zhong2020}
\bibinfo{author}{\bibfnamefont{D.}~\bibnamefont{Zhong}},
  \bibinfo{author}{\bibfnamefont{K.~L.} \bibnamefont{Seyler}},
  \bibinfo{author}{\bibfnamefont{X.}~\bibnamefont{Linpeng}},
  \bibinfo{author}{\bibfnamefont{N.~P.} \bibnamefont{Wilson}},
  \bibinfo{author}{\bibfnamefont{T.}~\bibnamefont{Taniguchi}},
  \bibinfo{author}{\bibfnamefont{K.}~\bibnamefont{Watanabe}},
  \bibinfo{author}{\bibfnamefont{M.~A.} \bibnamefont{McGuire}},
  \bibinfo{author}{\bibfnamefont{K.-M.~C.} \bibnamefont{Fu}},
  \bibinfo{author}{\bibfnamefont{D.}~\bibnamefont{Xiao}},
  \bibinfo{author}{\bibfnamefont{W.}~\bibnamefont{Yao}}, \bibnamefont{and}
  \bibinfo{author}{\bibfnamefont{X.}~\bibnamefont{Xu}},
  \emph{\bibinfo{title}{Layer-resolved magnetic proximity effect in van der
  Waals heterostructures}}, \bibinfo{journal}{Nat. Nanotechnol.}
  \textbf{\bibinfo{volume}{15}}, \bibinfo{pages}{187} (\bibinfo{year}{2020}),
  \urlprefix\url{https://doi.org/10.1038/s41565-019-0629-1}.

\bibitem[{\citenamefont{Zhao et~al.}(2020)\citenamefont{Zhao, Fei, Song, Choi,
  Palomaki, Sun, Malinowski, McGuire, Chu, Xu et~al.}}]{Zhao2020}
\bibinfo{author}{\bibfnamefont{W.}~\bibnamefont{Zhao}},
  \bibinfo{author}{\bibfnamefont{Z.}~\bibnamefont{Fei}},
  \bibinfo{author}{\bibfnamefont{T.}~\bibnamefont{Song}},
  \bibinfo{author}{\bibfnamefont{H.~K.} \bibnamefont{Choi}},
  \bibinfo{author}{\bibfnamefont{T.}~\bibnamefont{Palomaki}},
  \bibinfo{author}{\bibfnamefont{B.}~\bibnamefont{Sun}},
  \bibinfo{author}{\bibfnamefont{P.}~\bibnamefont{Malinowski}},
  \bibinfo{author}{\bibfnamefont{M.~A.} \bibnamefont{McGuire}},
  \bibinfo{author}{\bibfnamefont{J.-H.} \bibnamefont{Chu}},
  \bibinfo{author}{\bibfnamefont{X.}~\bibnamefont{Xu}}, \bibnamefont{and}
  \bibinfo{author}{\bibfnamefont{D.~H.} \bibnamefont{Cobden}},
  \emph{\bibinfo{title}{Magnetic proximity and nonreciprocal current switching
  in a monolayer {WTe}2 helical edge}}, \bibinfo{journal}{Nat. Mater.}
  \textbf{\bibinfo{volume}{19}}, \bibinfo{pages}{503} (\bibinfo{year}{2020}),
  \urlprefix\url{https://doi.org/10.1038/s41563-020-0620-0}.

\bibitem[{\citenamefont{Kang et~al.}(2021)\citenamefont{Kang, Jiang, Berger,
  Watanabe, Taniguchi, Forró, Shan, and Mak}}]{kang2021}
\bibinfo{author}{\bibfnamefont{K.}~\bibnamefont{Kang}},
  \bibinfo{author}{\bibfnamefont{S.}~\bibnamefont{Jiang}},
  \bibinfo{author}{\bibfnamefont{H.}~\bibnamefont{Berger}},
  \bibinfo{author}{\bibfnamefont{K.}~\bibnamefont{Watanabe}},
  \bibinfo{author}{\bibfnamefont{T.}~\bibnamefont{Taniguchi}},
  \bibinfo{author}{\bibfnamefont{L.}~\bibnamefont{Forró}},
  \bibinfo{author}{\bibfnamefont{J.}~\bibnamefont{Shan}}, \bibnamefont{and}
  \bibinfo{author}{\bibfnamefont{K.~F.} \bibnamefont{Mak}},
  \emph{\bibinfo{title}{Giant anisotropic magnetoresistance in Ising
  superconductor-magnetic insulator tunnel junctions}},
  \bibinfo{journal}{arXiv:} p. \bibinfo{pages}{.01327} (\bibinfo{year}{2021}),
  \urlprefix\url{https://arxiv.org/abs/2101.01327}.

\bibitem[{\citenamefont{Hamill et~al.}(2021)\citenamefont{Hamill, Heischmidt,
  Sohn, Shaffer, Tsai, Zhang, Xi, Suslov, Berger, Forr{\'{o}}
  et~al.}}]{Hamill2021}
\bibinfo{author}{\bibfnamefont{A.}~\bibnamefont{Hamill}},
  \bibinfo{author}{\bibfnamefont{B.}~\bibnamefont{Heischmidt}},
  \bibinfo{author}{\bibfnamefont{E.}~\bibnamefont{Sohn}},
  \bibinfo{author}{\bibfnamefont{D.}~\bibnamefont{Shaffer}},
  \bibinfo{author}{\bibfnamefont{K.-T.} \bibnamefont{Tsai}},
  \bibinfo{author}{\bibfnamefont{X.}~\bibnamefont{Zhang}},
  \bibinfo{author}{\bibfnamefont{X.}~\bibnamefont{Xi}},
  \bibinfo{author}{\bibfnamefont{A.}~\bibnamefont{Suslov}},
  \bibinfo{author}{\bibfnamefont{H.}~\bibnamefont{Berger}},
  \bibinfo{author}{\bibfnamefont{L.}~\bibnamefont{Forr{\'{o}}}},
  \bibinfo{author}{\bibfnamefont{F.~J.} \bibnamefont{Burnell}},
  \bibinfo{author}{\bibfnamefont{J.}~\bibnamefont{Shan}},
  \bibinfo{author}{\bibfnamefont{K.~F.} \bibnamefont{Mak}},
  \bibinfo{author}{\bibfnamefont{R.~M.} \bibnamefont{Fernandes}},
  \bibinfo{author}{\bibfnamefont{K.}~\bibnamefont{Wang}}, \bibnamefont{and}
  \bibinfo{author}{\bibfnamefont{V.~S.} \bibnamefont{Pribiag}},
  \emph{\bibinfo{title}{Two-fold symmetric superconductivity in few-layer
  {NbSe}2}}, \bibinfo{journal}{Nat. Phys.} \textbf{\bibinfo{volume}{17}},
  \bibinfo{pages}{949} (\bibinfo{year}{2021}),
  \urlprefix\url{https://doi.org/10.1038/s41567-021-01219-x}.

\bibitem[{\citenamefont{Kezilebieke et~al.}(2020)\citenamefont{Kezilebieke,
  Huda, Va{\v{n}}o, Aapro, Ganguli, Silveira, G{\l}odzik, Foster, Ojanen, and
  Liljeroth}}]{Kezilebieke2020}
\bibinfo{author}{\bibfnamefont{S.}~\bibnamefont{Kezilebieke}},
  \bibinfo{author}{\bibfnamefont{M.~N.} \bibnamefont{Huda}},
  \bibinfo{author}{\bibfnamefont{V.}~\bibnamefont{Va{\v{n}}o}},
  \bibinfo{author}{\bibfnamefont{M.}~\bibnamefont{Aapro}},
  \bibinfo{author}{\bibfnamefont{S.~C.} \bibnamefont{Ganguli}},
  \bibinfo{author}{\bibfnamefont{O.~J.} \bibnamefont{Silveira}},
  \bibinfo{author}{\bibfnamefont{S.}~\bibnamefont{G{\l}odzik}},
  \bibinfo{author}{\bibfnamefont{A.~S.} \bibnamefont{Foster}},
  \bibinfo{author}{\bibfnamefont{T.}~\bibnamefont{Ojanen}}, \bibnamefont{and}
  \bibinfo{author}{\bibfnamefont{P.}~\bibnamefont{Liljeroth}},
  \emph{\bibinfo{title}{Topological superconductivity in a van der Waals
  heterostructure}}, \bibinfo{journal}{Nature} \textbf{\bibinfo{volume}{588}},
  \bibinfo{pages}{424} (\bibinfo{year}{2020}),
  \urlprefix\url{https://doi.org/10.1038/s41586-020-2989-y}.

\bibitem[{\citenamefont{Xie et~al.}(2021)\citenamefont{Xie, Luo, Ye, Ye, Ge,
  Sung, Rennich, Yan, Fu, Tian et~al.}}]{Xie2021}
\bibinfo{author}{\bibfnamefont{H.}~\bibnamefont{Xie}},
  \bibinfo{author}{\bibfnamefont{X.}~\bibnamefont{Luo}},
  \bibinfo{author}{\bibfnamefont{G.}~\bibnamefont{Ye}},
  \bibinfo{author}{\bibfnamefont{Z.}~\bibnamefont{Ye}},
  \bibinfo{author}{\bibfnamefont{H.}~\bibnamefont{Ge}},
  \bibinfo{author}{\bibfnamefont{S.~H.} \bibnamefont{Sung}},
  \bibinfo{author}{\bibfnamefont{E.}~\bibnamefont{Rennich}},
  \bibinfo{author}{\bibfnamefont{S.}~\bibnamefont{Yan}},
  \bibinfo{author}{\bibfnamefont{Y.}~\bibnamefont{Fu}},
  \bibinfo{author}{\bibfnamefont{S.}~\bibnamefont{Tian}},
  \bibinfo{author}{\bibfnamefont{H.}~\bibnamefont{Lei}},
  \bibinfo{author}{\bibfnamefont{R.}~\bibnamefont{Hovden}},
  \bibinfo{author}{\bibfnamefont{K.}~\bibnamefont{Sun}},
  \bibinfo{author}{\bibfnamefont{R.}~\bibnamefont{He}}, \bibnamefont{and}
  \bibinfo{author}{\bibfnamefont{L.}~\bibnamefont{Zhao}},
  \emph{\bibinfo{title}{Twist engineering of the two-dimensional magnetism in
  double bilayer chromium triiodide homostructures}}, \bibinfo{journal}{Nat.
  Phys.} \textbf{\bibinfo{volume}{18}}, \bibinfo{pages}{30}
  (\bibinfo{year}{2021}),
  \urlprefix\url{https://doi.org/10.1038/s41567-021-01408-8}.

\bibitem[{\citenamefont{Song et~al.}(2021)\citenamefont{Song, Sun, Anderson,
  Wang, Qian, Taniguchi, Watanabe, McGuire, St\"{o}hr, Xiao et~al.}}]{Song2021}
\bibinfo{author}{\bibfnamefont{T.}~\bibnamefont{Song}},
  \bibinfo{author}{\bibfnamefont{Q.-C.} \bibnamefont{Sun}},
  \bibinfo{author}{\bibfnamefont{E.}~\bibnamefont{Anderson}},
  \bibinfo{author}{\bibfnamefont{C.}~\bibnamefont{Wang}},
  \bibinfo{author}{\bibfnamefont{J.}~\bibnamefont{Qian}},
  \bibinfo{author}{\bibfnamefont{T.}~\bibnamefont{Taniguchi}},
  \bibinfo{author}{\bibfnamefont{K.}~\bibnamefont{Watanabe}},
  \bibinfo{author}{\bibfnamefont{M.~A.} \bibnamefont{McGuire}},
  \bibinfo{author}{\bibfnamefont{R.}~\bibnamefont{St\"{o}hr}},
  \bibinfo{author}{\bibfnamefont{D.}~\bibnamefont{Xiao}},
  \bibinfo{author}{\bibfnamefont{T.}~\bibnamefont{Cao}},
  \bibinfo{author}{\bibfnamefont{J.}~\bibnamefont{Wrachtrup}},
  \bibnamefont{and} \bibinfo{author}{\bibfnamefont{X.}~\bibnamefont{Xu}},
  \emph{\bibinfo{title}{Direct visualization of magnetic domains and
  moir{\'{e}} magnetism in twisted 2D magnets}}, \bibinfo{journal}{Science}
  \textbf{\bibinfo{volume}{374}}, \bibinfo{pages}{1140} (\bibinfo{year}{2021}),
  \urlprefix\url{https://doi.org/10.1126/science.abj7478}.

\bibitem[{\citenamefont{Xu et~al.}(2021)\citenamefont{Xu, Ray, Shao, Jiang,
  Lee, Weber, Goldberger, Watanabe, Taniguchi, Muller et~al.}}]{Xu2021}
\bibinfo{author}{\bibfnamefont{Y.}~\bibnamefont{Xu}},
  \bibinfo{author}{\bibfnamefont{A.}~\bibnamefont{Ray}},
  \bibinfo{author}{\bibfnamefont{Y.-T.} \bibnamefont{Shao}},
  \bibinfo{author}{\bibfnamefont{S.}~\bibnamefont{Jiang}},
  \bibinfo{author}{\bibfnamefont{K.}~\bibnamefont{Lee}},
  \bibinfo{author}{\bibfnamefont{D.}~\bibnamefont{Weber}},
  \bibinfo{author}{\bibfnamefont{J.~E.} \bibnamefont{Goldberger}},
  \bibinfo{author}{\bibfnamefont{K.}~\bibnamefont{Watanabe}},
  \bibinfo{author}{\bibfnamefont{T.}~\bibnamefont{Taniguchi}},
  \bibinfo{author}{\bibfnamefont{D.~A.} \bibnamefont{Muller}},
  \bibinfo{author}{\bibfnamefont{K.~F.} \bibnamefont{Mak}}, \bibnamefont{and}
  \bibinfo{author}{\bibfnamefont{J.}~\bibnamefont{Shan}},
  \emph{\bibinfo{title}{Coexisting ferromagnetic{\textendash}antiferromagnetic
  state in twisted bilayer {CrI}3}}, \bibinfo{journal}{Nat. Nanotechnol.}
  \textbf{\bibinfo{volume}{17}}, \bibinfo{pages}{143} (\bibinfo{year}{2021}),
  \urlprefix\url{https://doi.org/10.1038/s41565-021-01014-y}.

\bibitem[{\citenamefont{Shcherbakov et~al.}(2018)\citenamefont{Shcherbakov,
  Stepanov, Weber, Wang, Hu, Zhu, Watanabe, Taniguchi, Mao, Windl
  et~al.}}]{Shcherbakov2018}
\bibinfo{author}{\bibfnamefont{D.}~\bibnamefont{Shcherbakov}},
  \bibinfo{author}{\bibfnamefont{P.}~\bibnamefont{Stepanov}},
  \bibinfo{author}{\bibfnamefont{D.}~\bibnamefont{Weber}},
  \bibinfo{author}{\bibfnamefont{Y.}~\bibnamefont{Wang}},
  \bibinfo{author}{\bibfnamefont{J.}~\bibnamefont{Hu}},
  \bibinfo{author}{\bibfnamefont{Y.}~\bibnamefont{Zhu}},
  \bibinfo{author}{\bibfnamefont{K.}~\bibnamefont{Watanabe}},
  \bibinfo{author}{\bibfnamefont{T.}~\bibnamefont{Taniguchi}},
  \bibinfo{author}{\bibfnamefont{Z.}~\bibnamefont{Mao}},
  \bibinfo{author}{\bibfnamefont{W.}~\bibnamefont{Windl}},
  \bibinfo{author}{\bibfnamefont{J.}~\bibnamefont{Goldberger}},
  \bibinfo{author}{\bibfnamefont{M.}~\bibnamefont{Bockrath}}, \bibnamefont{and}
  \bibinfo{author}{\bibfnamefont{C.~N.} \bibnamefont{Lau}},
  \emph{\bibinfo{title}{Raman Spectroscopy, Photocatalytic Degradation, and
  Stabilization of Atomically Thin Chromium Tri-iodide}},
  \bibinfo{journal}{Nano Lett.} \textbf{\bibinfo{volume}{18}},
  \bibinfo{pages}{4214} (\bibinfo{year}{2018}),
  \urlprefix\url{https://doi.org/10.1021/acs.nanolett.8b01131}.

\bibitem[{\citenamefont{Watson et~al.}(2020)\citenamefont{Watson,
  Markovi{\'{c}}, Mazzola, Rajan, Morales, Burn, Hesjedal, van~der Laan,
  Mukherjee, Kim et~al.}}]{Watson2020}
\bibinfo{author}{\bibfnamefont{M.~D.} \bibnamefont{Watson}},
  \bibinfo{author}{\bibfnamefont{I.}~\bibnamefont{Markovi{\'{c}}}},
  \bibinfo{author}{\bibfnamefont{F.}~\bibnamefont{Mazzola}},
  \bibinfo{author}{\bibfnamefont{A.}~\bibnamefont{Rajan}},
  \bibinfo{author}{\bibfnamefont{E.~A.} \bibnamefont{Morales}},
  \bibinfo{author}{\bibfnamefont{D.~M.} \bibnamefont{Burn}},
  \bibinfo{author}{\bibfnamefont{T.}~\bibnamefont{Hesjedal}},
  \bibinfo{author}{\bibfnamefont{G.}~\bibnamefont{van~der Laan}},
  \bibinfo{author}{\bibfnamefont{S.}~\bibnamefont{Mukherjee}},
  \bibinfo{author}{\bibfnamefont{T.~K.} \bibnamefont{Kim}},
  \bibinfo{author}{\bibfnamefont{C.}~\bibnamefont{Bigi}},
  \bibinfo{author}{\bibfnamefont{I.}~\bibnamefont{Vobornik}},
  \bibinfo{author}{\bibfnamefont{M.~C.} \bibnamefont{Hatnean}},
  \bibinfo{author}{\bibfnamefont{G.}~\bibnamefont{Balakrishnan}},
  \bibnamefont{and} \bibinfo{author}{\bibfnamefont{P.~D.~C.}
  \bibnamefont{King}}, \emph{\bibinfo{title}{Direct observation of the energy
  gain underpinning ferromagnetic superexchange in the electronic structure of
  {CrGeTe}$_3$}}, \bibinfo{journal}{Phys. Rev. B}
  \textbf{\bibinfo{volume}{101}}, \bibinfo{pages}{205125}
  (\bibinfo{year}{2020}),
  \urlprefix\url{https://doi.org/10.1103/physrevb.101.205125}.

\bibitem[{\citenamefont{Yilmaz et~al.}(2021)\citenamefont{Yilmaz, Geilhufe,
  Pletikosi{\'{c}}, Fernando, Cava, Valla, Vescovo, and Sinkovic}}]{Yilmaz2021}
\bibinfo{author}{\bibfnamefont{T.}~\bibnamefont{Yilmaz}},
  \bibinfo{author}{\bibfnamefont{R.~M.} \bibnamefont{Geilhufe}},
  \bibinfo{author}{\bibfnamefont{I.}~\bibnamefont{Pletikosi{\'{c}}}},
  \bibinfo{author}{\bibfnamefont{G.~W.} \bibnamefont{Fernando}},
  \bibinfo{author}{\bibfnamefont{R.~J.} \bibnamefont{Cava}},
  \bibinfo{author}{\bibfnamefont{T.}~\bibnamefont{Valla}},
  \bibinfo{author}{\bibfnamefont{E.}~\bibnamefont{Vescovo}}, \bibnamefont{and}
  \bibinfo{author}{\bibfnamefont{B.}~\bibnamefont{Sinkovic}},
  \emph{\bibinfo{title}{Multi-hole bands and
  quasi{\textendash}two-dimensionality in Cr$_2$Ge$_2$Te$_6$ studied by
  angle-resolved photoemission spectroscopy}}, \bibinfo{journal}{Europhys.
  Lett.} \textbf{\bibinfo{volume}{133}}, \bibinfo{pages}{27002}
  (\bibinfo{year}{2021}),
  \urlprefix\url{https://doi.org/10.1209/0295-5075/133/27002}.

\bibitem[{\citenamefont{Xu et~al.}(2020)\citenamefont{Xu, Li, Duan, Zhang,
  Chen, Kang, Liang, Chen, Xia, Xu et~al.}}]{Xu2020}
\bibinfo{author}{\bibfnamefont{X.}~\bibnamefont{Xu}},
  \bibinfo{author}{\bibfnamefont{Y.~W.} \bibnamefont{Li}},
  \bibinfo{author}{\bibfnamefont{S.~R.} \bibnamefont{Duan}},
  \bibinfo{author}{\bibfnamefont{S.~L.} \bibnamefont{Zhang}},
  \bibinfo{author}{\bibfnamefont{Y.~J.} \bibnamefont{Chen}},
  \bibinfo{author}{\bibfnamefont{L.}~\bibnamefont{Kang}},
  \bibinfo{author}{\bibfnamefont{A.~J.} \bibnamefont{Liang}},
  \bibinfo{author}{\bibfnamefont{C.}~\bibnamefont{Chen}},
  \bibinfo{author}{\bibfnamefont{W.}~\bibnamefont{Xia}},
  \bibinfo{author}{\bibfnamefont{Y.}~\bibnamefont{Xu}},
  \bibinfo{author}{\bibfnamefont{P.}~\bibnamefont{Malinowski}},
  \bibinfo{author}{\bibfnamefont{X.~D.} \bibnamefont{Xu}},
  \bibinfo{author}{\bibfnamefont{J.-H.} \bibnamefont{Chu}},
  \bibinfo{author}{\bibfnamefont{G.}~\bibnamefont{Li}},
  \bibinfo{author}{\bibfnamefont{Y.~F.} \bibnamefont{Guo}},
  \bibinfo{author}{\bibfnamefont{Z.~K.} \bibnamefont{Liu}},
  \bibinfo{author}{\bibfnamefont{L.~X.} \bibnamefont{Yang}}, \bibnamefont{and}
  \bibinfo{author}{\bibfnamefont{Y.~L.} \bibnamefont{Chen}},
  \emph{\bibinfo{title}{Signature for non-Stoner ferromagnetism in the van der
  Waals ferromagnet Fe3GeTe2}}, \bibinfo{journal}{Phys. Rev. B}
  \textbf{\bibinfo{volume}{101}}, \bibinfo{pages}{201104}
  (\bibinfo{year}{2020}),
  \urlprefix\url{https://doi.org/10.1103/physrevb.101.201104}.

\bibitem[{\citenamefont{Liu et~al.}(2020)\citenamefont{Liu, Huan, Liu, Liu,
  Liu, Lu, Huang, Jiang, Wang, Yu et~al.}}]{Liu2020}
\bibinfo{author}{\bibfnamefont{J.~S.} \bibnamefont{Liu}},
  \bibinfo{author}{\bibfnamefont{S.~C.} \bibnamefont{Huan}},
  \bibinfo{author}{\bibfnamefont{Z.~H.} \bibnamefont{Liu}},
  \bibinfo{author}{\bibfnamefont{W.~L.} \bibnamefont{Liu}},
  \bibinfo{author}{\bibfnamefont{Z.~T.} \bibnamefont{Liu}},
  \bibinfo{author}{\bibfnamefont{X.~L.} \bibnamefont{Lu}},
  \bibinfo{author}{\bibfnamefont{Z.}~\bibnamefont{Huang}},
  \bibinfo{author}{\bibfnamefont{Z.~C.} \bibnamefont{Jiang}},
  \bibinfo{author}{\bibfnamefont{X.}~\bibnamefont{Wang}},
  \bibinfo{author}{\bibfnamefont{N.}~\bibnamefont{Yu}},
  \bibinfo{author}{\bibfnamefont{Z.~Q.} \bibnamefont{Zou}},
  \bibinfo{author}{\bibfnamefont{Y.~F.} \bibnamefont{Guo}}, \bibnamefont{and}
  \bibinfo{author}{\bibfnamefont{D.~W.} \bibnamefont{Shen}},
  \emph{\bibinfo{title}{Electronic structure of the high-mobility
  two-dimensional antiferromagnetic metal {GdTe}$_3$}}, \bibinfo{journal}{Phys.
  Rev. Mat.} \textbf{\bibinfo{volume}{4}}, \bibinfo{pages}{114005}
  (\bibinfo{year}{2020}),
  \urlprefix\url{https://doi.org/10.1103/physrevmaterials.4.114005}.

\bibitem[{\citenamefont{Li et~al.}(2018)\citenamefont{Li, Wang, Guo, Gu, Sun,
  He, Zhou, Gu, Nie, and Pan}}]{Li2018}
\bibinfo{author}{\bibfnamefont{Y.~F.} \bibnamefont{Li}},
  \bibinfo{author}{\bibfnamefont{W.}~\bibnamefont{Wang}},
  \bibinfo{author}{\bibfnamefont{W.}~\bibnamefont{Guo}},
  \bibinfo{author}{\bibfnamefont{C.~Y.} \bibnamefont{Gu}},
  \bibinfo{author}{\bibfnamefont{H.~Y.} \bibnamefont{Sun}},
  \bibinfo{author}{\bibfnamefont{L.}~\bibnamefont{He}},
  \bibinfo{author}{\bibfnamefont{J.}~\bibnamefont{Zhou}},
  \bibinfo{author}{\bibfnamefont{Z.~B.} \bibnamefont{Gu}},
  \bibinfo{author}{\bibfnamefont{Y.~F.} \bibnamefont{Nie}}, \bibnamefont{and}
  \bibinfo{author}{\bibfnamefont{X.~Q.} \bibnamefont{Pan}},
  \emph{\bibinfo{title}{Electronic structure of ferromagnetic semiconductor
  {CrGeTe}$_3$ by angle-resolved photoemission spectroscopy}},
  \bibinfo{journal}{Phys. Rev. B} \textbf{\bibinfo{volume}{98}},
  \bibinfo{pages}{125127} (\bibinfo{year}{2018}),
  \urlprefix\url{https://doi.org/10.1103/physrevb.98.125127}.

\bibitem[{\citenamefont{Suzuki et~al.}(2019)\citenamefont{Suzuki, Gao,
  Koshiishi, Nakata, Hagiwara, Lin, Wan, Kumigashira, Ono, Kang
  et~al.}}]{Suzuki2019}
\bibinfo{author}{\bibfnamefont{M.}~\bibnamefont{Suzuki}},
  \bibinfo{author}{\bibfnamefont{B.}~\bibnamefont{Gao}},
  \bibinfo{author}{\bibfnamefont{K.}~\bibnamefont{Koshiishi}},
  \bibinfo{author}{\bibfnamefont{S.}~\bibnamefont{Nakata}},
  \bibinfo{author}{\bibfnamefont{K.}~\bibnamefont{Hagiwara}},
  \bibinfo{author}{\bibfnamefont{C.}~\bibnamefont{Lin}},
  \bibinfo{author}{\bibfnamefont{Y.~X.} \bibnamefont{Wan}},
  \bibinfo{author}{\bibfnamefont{H.}~\bibnamefont{Kumigashira}},
  \bibinfo{author}{\bibfnamefont{K.}~\bibnamefont{Ono}},
  \bibinfo{author}{\bibfnamefont{S.}~\bibnamefont{Kang}},
  \bibinfo{author}{\bibfnamefont{S.}~\bibnamefont{Kang}},
  \bibinfo{author}{\bibfnamefont{J.}~\bibnamefont{Yu}},
  \bibinfo{author}{\bibfnamefont{M.}~\bibnamefont{Kobayashi}},
  \bibinfo{author}{\bibfnamefont{S.-W.} \bibnamefont{Cheong}},
  \bibnamefont{and} \bibinfo{author}{\bibfnamefont{A.}~\bibnamefont{Fujimori}},
  \emph{\bibinfo{title}{Coulomb-interaction effect on the two-dimensional
  electronic structure of the van der Waals ferromagnet Cr2Ge2Te6}},
  \bibinfo{journal}{Phys. Rev. B} \textbf{\bibinfo{volume}{99}},
  \bibinfo{pages}{161401} (\bibinfo{year}{2019}),
  \urlprefix\url{https://doi.org/10.1103/physrevb.99.161401}.

\bibitem[{\citenamefont{Zhang et~al.}(2018)\citenamefont{Zhang, Lu, Zhu, Tan,
  Feng, Liu, Zhang, Chen, Liu, Luo et~al.}}]{Zhang2018}
\bibinfo{author}{\bibfnamefont{Y.}~\bibnamefont{Zhang}},
  \bibinfo{author}{\bibfnamefont{H.}~\bibnamefont{Lu}},
  \bibinfo{author}{\bibfnamefont{X.}~\bibnamefont{Zhu}},
  \bibinfo{author}{\bibfnamefont{S.}~\bibnamefont{Tan}},
  \bibinfo{author}{\bibfnamefont{W.}~\bibnamefont{Feng}},
  \bibinfo{author}{\bibfnamefont{Q.}~\bibnamefont{Liu}},
  \bibinfo{author}{\bibfnamefont{W.}~\bibnamefont{Zhang}},
  \bibinfo{author}{\bibfnamefont{Q.}~\bibnamefont{Chen}},
  \bibinfo{author}{\bibfnamefont{Y.}~\bibnamefont{Liu}},
  \bibinfo{author}{\bibfnamefont{X.}~\bibnamefont{Luo}},
  \bibinfo{author}{\bibfnamefont{D.}~\bibnamefont{Xie}},
  \bibinfo{author}{\bibfnamefont{L.}~\bibnamefont{Luo}},
  \bibinfo{author}{\bibfnamefont{Z.}~\bibnamefont{Zhang}}, \bibnamefont{and}
  \bibinfo{author}{\bibfnamefont{X.}~\bibnamefont{Lai}},
  \emph{\bibinfo{title}{Emergence of Kondo lattice behavior in a van der Waals
  itinerant ferromagnet, Fe$_3${GeTe}$_2$}}, \bibinfo{journal}{Sci. Adv.}
  \textbf{\bibinfo{volume}{4}}, \bibinfo{pages}{aao6791}
  (\bibinfo{year}{2018}),
  \urlprefix\url{https://doi.org/10.1126/sciadv.aao6791}.

\bibitem[{\citenamefont{Kong et~al.}(2019)\citenamefont{Kong, Stolze, Timmons,
  Tao, Ni, Guo, Yang, Prozorov, and Cava}}]{Kong2019}
\bibinfo{author}{\bibfnamefont{T.}~\bibnamefont{Kong}},
  \bibinfo{author}{\bibfnamefont{K.}~\bibnamefont{Stolze}},
  \bibinfo{author}{\bibfnamefont{E.~I.} \bibnamefont{Timmons}},
  \bibinfo{author}{\bibfnamefont{J.}~\bibnamefont{Tao}},
  \bibinfo{author}{\bibfnamefont{D.}~\bibnamefont{Ni}},
  \bibinfo{author}{\bibfnamefont{S.}~\bibnamefont{Guo}},
  \bibinfo{author}{\bibfnamefont{Z.}~\bibnamefont{Yang}},
  \bibinfo{author}{\bibfnamefont{R.}~\bibnamefont{Prozorov}}, \bibnamefont{and}
  \bibinfo{author}{\bibfnamefont{R.~J.} \bibnamefont{Cava}},
  \emph{\bibinfo{title}{{VI}$_3$- a New Layered Ferromagnetic Semiconductor}},
  \bibinfo{journal}{Adv. Mater.} \textbf{\bibinfo{volume}{31}},
  \bibinfo{pages}{1808074} (\bibinfo{year}{2019}),
  \urlprefix\url{https://doi.org/10.1002/adma.201808074}.

\bibitem[{\citenamefont{Jiang et~al.}(2020)\citenamefont{Jiang, Yang, Li, Wang,
  Jing, Guan, Ma, Zhang, and Qian}}]{Jiang2020}
\bibinfo{author}{\bibfnamefont{W.}~\bibnamefont{Jiang}},
  \bibinfo{author}{\bibfnamefont{Z.}~\bibnamefont{Yang}},
  \bibinfo{author}{\bibfnamefont{Y.}~\bibnamefont{Li}},
  \bibinfo{author}{\bibfnamefont{G.}~\bibnamefont{Wang}},
  \bibinfo{author}{\bibfnamefont{Q.}~\bibnamefont{Jing}},
  \bibinfo{author}{\bibfnamefont{D.}~\bibnamefont{Guan}},
  \bibinfo{author}{\bibfnamefont{J.}~\bibnamefont{Ma}},
  \bibinfo{author}{\bibfnamefont{W.}~\bibnamefont{Zhang}}, \bibnamefont{and}
  \bibinfo{author}{\bibfnamefont{D.}~\bibnamefont{Qian}},
  \emph{\bibinfo{title}{Spin-split valence bands of the ferromagnetic insulator
  Cr2Ge2Te6 studied by angle-resolved photoemission spectroscopy}},
  \bibinfo{journal}{J. Appl. Phys.} \textbf{\bibinfo{volume}{127}},
  \bibinfo{pages}{023901} (\bibinfo{year}{2020}),
  \urlprefix\url{https://doi.org/10.1063/1.5135759}.

\bibitem[{\citenamefont{Zhang et~al.}(2021)\citenamefont{Zhang, Lu, Liu, Niu,
  Sun, Cook, Vaninger, Miceli, Singh, Lian et~al.}}]{Zhang2021}
\bibinfo{author}{\bibfnamefont{X.}~\bibnamefont{Zhang}},
  \bibinfo{author}{\bibfnamefont{Q.}~\bibnamefont{Lu}},
  \bibinfo{author}{\bibfnamefont{W.}~\bibnamefont{Liu}},
  \bibinfo{author}{\bibfnamefont{W.}~\bibnamefont{Niu}},
  \bibinfo{author}{\bibfnamefont{J.}~\bibnamefont{Sun}},
  \bibinfo{author}{\bibfnamefont{J.}~\bibnamefont{Cook}},
  \bibinfo{author}{\bibfnamefont{M.}~\bibnamefont{Vaninger}},
  \bibinfo{author}{\bibfnamefont{P.~F.} \bibnamefont{Miceli}},
  \bibinfo{author}{\bibfnamefont{D.~J.} \bibnamefont{Singh}},
  \bibinfo{author}{\bibfnamefont{S.-W.} \bibnamefont{Lian}},
  \bibinfo{author}{\bibfnamefont{T.-R.} \bibnamefont{Chang}},
  \bibinfo{author}{\bibfnamefont{X.}~\bibnamefont{He}},
  \bibinfo{author}{\bibfnamefont{J.}~\bibnamefont{Du}},
  \bibinfo{author}{\bibfnamefont{L.}~\bibnamefont{He}},
  \bibinfo{author}{\bibfnamefont{R.}~\bibnamefont{Zhang}},
  \bibinfo{author}{\bibfnamefont{G.}~\bibnamefont{Bian}}, \bibnamefont{and}
  \bibinfo{author}{\bibfnamefont{Y.}~\bibnamefont{Xu}},
  \emph{\bibinfo{title}{Room-temperature intrinsic ferromagnetism in epitaxial
  {CrTe}$_2$ ultrathin films}}, \bibinfo{journal}{Nat. Commun.}
  \textbf{\bibinfo{volume}{12}}, \bibinfo{pages}{2492} (\bibinfo{year}{2021}),
  \urlprefix\url{https://doi.org/10.1038/s41467-021-22777-x}.

\bibitem[{\citenamefont{Joy and Vasudevan}(1992)}]{Joy1992}
\bibinfo{author}{\bibfnamefont{P.~A.} \bibnamefont{Joy}} \bibnamefont{and}
  \bibinfo{author}{\bibfnamefont{S.}~\bibnamefont{Vasudevan}},
  \emph{\bibinfo{title}{Magnetism in the layered transition-metal
  {thiophosphates MPS}$_3$ (M=Mn, Fe, and Ni)}}, \bibinfo{journal}{Phys. Rev.
  B} \textbf{\bibinfo{volume}{46}}, \bibinfo{pages}{5425}
  (\bibinfo{year}{1992}),
  \urlprefix\url{https://doi.org/10.1103/physrevb.46.5425}.

\bibitem[{\citenamefont{Susner et~al.}(2017)\citenamefont{Susner,
  Chyasnavichyus, McGuire, Ganesh, and Maksymovych}}]{Susner2017}
\bibinfo{author}{\bibfnamefont{M.~A.} \bibnamefont{Susner}},
  \bibinfo{author}{\bibfnamefont{M.}~\bibnamefont{Chyasnavichyus}},
  \bibinfo{author}{\bibfnamefont{M.~A.} \bibnamefont{McGuire}},
  \bibinfo{author}{\bibfnamefont{P.}~\bibnamefont{Ganesh}}, \bibnamefont{and}
  \bibinfo{author}{\bibfnamefont{P.}~\bibnamefont{Maksymovych}},
  \emph{\bibinfo{title}{Metal Thio- and Selenophosphates as Multifunctional van
  der Waals Layered Materials}}, \bibinfo{journal}{Adv. Mater.}
  \textbf{\bibinfo{volume}{29}}, \bibinfo{pages}{1602852}
  (\bibinfo{year}{2017}),
  \urlprefix\url{https://doi.org/10.1002/adma.201602852}.

\bibitem[{\citenamefont{Wang et~al.}(2018)\citenamefont{Wang, Shifa, Yu, He,
  Liu, Wang, Wang, Zhan, Lou, Xia et~al.}}]{Wang2018}
\bibinfo{author}{\bibfnamefont{F.}~\bibnamefont{Wang}},
  \bibinfo{author}{\bibfnamefont{T.~A.} \bibnamefont{Shifa}},
  \bibinfo{author}{\bibfnamefont{P.}~\bibnamefont{Yu}},
  \bibinfo{author}{\bibfnamefont{P.}~\bibnamefont{He}},
  \bibinfo{author}{\bibfnamefont{Y.}~\bibnamefont{Liu}},
  \bibinfo{author}{\bibfnamefont{F.}~\bibnamefont{Wang}},
  \bibinfo{author}{\bibfnamefont{Z.}~\bibnamefont{Wang}},
  \bibinfo{author}{\bibfnamefont{X.}~\bibnamefont{Zhan}},
  \bibinfo{author}{\bibfnamefont{X.}~\bibnamefont{Lou}},
  \bibinfo{author}{\bibfnamefont{F.}~\bibnamefont{Xia}}, \bibnamefont{and}
  \bibinfo{author}{\bibfnamefont{J.}~\bibnamefont{He}},
  \emph{\bibinfo{title}{New Frontiers on van der Waals Layered Metal
  Phosphorous Trichalcogenides}}, \bibinfo{journal}{Adv. Funct. Mater.}
  \textbf{\bibinfo{volume}{28}}, \bibinfo{pages}{1802151}
  (\bibinfo{year}{2018}),
  \urlprefix\url{https://doi.org/10.1002/adfm.201802151}.

\bibitem[{\citenamefont{Vaclavkova et~al.}(2020)\citenamefont{Vaclavkova,
  Delhomme, Faugeras, Potemski, Bogucki, Suffczy{\'{n}}ski, Kossacki, Wildes,
  Gr{\'{e}}maud, and Sa{\'{u}}l}}]{Vaclavkova2020}
\bibinfo{author}{\bibfnamefont{D.}~\bibnamefont{Vaclavkova}},
  \bibinfo{author}{\bibfnamefont{A.}~\bibnamefont{Delhomme}},
  \bibinfo{author}{\bibfnamefont{C.}~\bibnamefont{Faugeras}},
  \bibinfo{author}{\bibfnamefont{M.}~\bibnamefont{Potemski}},
  \bibinfo{author}{\bibfnamefont{A.}~\bibnamefont{Bogucki}},
  \bibinfo{author}{\bibfnamefont{J.}~\bibnamefont{Suffczy{\'{n}}ski}},
  \bibinfo{author}{\bibfnamefont{P.}~\bibnamefont{Kossacki}},
  \bibinfo{author}{\bibfnamefont{A.~R.} \bibnamefont{Wildes}},
  \bibinfo{author}{\bibfnamefont{B.}~\bibnamefont{Gr{\'{e}}maud}},
  \bibnamefont{and}
  \bibinfo{author}{\bibfnamefont{A.}~\bibnamefont{Sa{\'{u}}l}},
  \emph{\bibinfo{title}{Magnetoelastic interaction in the two-dimensional
  magnetic material {MnPS}$_3$ studied by first principles calculations and
  Raman experiments}}, \bibinfo{journal}{2D Mater.}
  \textbf{\bibinfo{volume}{7}}, \bibinfo{pages}{035030} (\bibinfo{year}{2020}),
  \urlprefix\url{https://doi.org/10.1088/2053-1583/ab93e3}.

\bibitem[{\citenamefont{Liu et~al.}(2021)\citenamefont{Liu, del {\'{A}}guila,
  Bhowmick, Gan, Do, Prosnikov, Sedmidubsk{\'{y}}, Sofer, Christianen, Sengupta
  et~al.}}]{Liu2021}
\bibinfo{author}{\bibfnamefont{S.}~\bibnamefont{Liu}},
  \bibinfo{author}{\bibfnamefont{A.~G.} \bibnamefont{del {\'{A}}guila}},
  \bibinfo{author}{\bibfnamefont{D.}~\bibnamefont{Bhowmick}},
  \bibinfo{author}{\bibfnamefont{C.~K.} \bibnamefont{Gan}},
  \bibinfo{author}{\bibfnamefont{T.~T.~H.} \bibnamefont{Do}},
  \bibinfo{author}{\bibfnamefont{M.}~\bibnamefont{Prosnikov}},
  \bibinfo{author}{\bibfnamefont{D.}~\bibnamefont{Sedmidubsk{\'{y}}}},
  \bibinfo{author}{\bibfnamefont{Z.}~\bibnamefont{Sofer}},
  \bibinfo{author}{\bibfnamefont{P.~C.} \bibnamefont{Christianen}},
  \bibinfo{author}{\bibfnamefont{P.}~\bibnamefont{Sengupta}}, \bibnamefont{and}
  \bibinfo{author}{\bibfnamefont{Q.}~\bibnamefont{Xiong}},
  \emph{\bibinfo{title}{Direct Observation of Magnon-Phonon Strong Coupling in
  Two-Dimensional Antiferromagnet at High Magnetic Fields}},
  \bibinfo{journal}{Phys. Rev. Lett.} \textbf{\bibinfo{volume}{127}},
  \bibinfo{pages}{097401} (\bibinfo{year}{2021}),
  \urlprefix\url{https://doi.org/10.1103/physrevlett.127.097401}.

\bibitem[{\citenamefont{Ressouche et~al.}(2010)\citenamefont{Ressouche, Loire,
  Simonet, Ballou, Stunault, and Wildes}}]{Ressouche2010}
\bibinfo{author}{\bibfnamefont{E.}~\bibnamefont{Ressouche}},
  \bibinfo{author}{\bibfnamefont{M.}~\bibnamefont{Loire}},
  \bibinfo{author}{\bibfnamefont{V.}~\bibnamefont{Simonet}},
  \bibinfo{author}{\bibfnamefont{R.}~\bibnamefont{Ballou}},
  \bibinfo{author}{\bibfnamefont{A.}~\bibnamefont{Stunault}}, \bibnamefont{and}
  \bibinfo{author}{\bibfnamefont{A.}~\bibnamefont{Wildes}},
  \emph{\bibinfo{title}{{Magnetoelectric MnPS}$_3$ as a candidate for
  ferrotoroidicity}}, \bibinfo{journal}{Phys. Rev. B}
  \textbf{\bibinfo{volume}{82}}, \bibinfo{pages}{100408}
  (\bibinfo{year}{2010}),
  \urlprefix\url{https://doi.org/10.1103/physrevb.82.100408}.

\bibitem[{\citenamefont{Lai et~al.}(2019)\citenamefont{Lai, Song, Wan, Xue,
  Wang, Ye, Dai, Zhang, Yang, Du et~al.}}]{Lai2019}
\bibinfo{author}{\bibfnamefont{Y.}~\bibnamefont{Lai}},
  \bibinfo{author}{\bibfnamefont{Z.}~\bibnamefont{Song}},
  \bibinfo{author}{\bibfnamefont{Y.}~\bibnamefont{Wan}},
  \bibinfo{author}{\bibfnamefont{M.}~\bibnamefont{Xue}},
  \bibinfo{author}{\bibfnamefont{C.}~\bibnamefont{Wang}},
  \bibinfo{author}{\bibfnamefont{Y.}~\bibnamefont{Ye}},
  \bibinfo{author}{\bibfnamefont{L.}~\bibnamefont{Dai}},
  \bibinfo{author}{\bibfnamefont{Z.}~\bibnamefont{Zhang}},
  \bibinfo{author}{\bibfnamefont{W.}~\bibnamefont{Yang}},
  \bibinfo{author}{\bibfnamefont{H.}~\bibnamefont{Du}}, \bibnamefont{and}
  \bibinfo{author}{\bibfnamefont{J.}~\bibnamefont{Yang}},
  \emph{\bibinfo{title}{Two-dimensional ferromagnetism and driven
  ferroelectricity in van der Waals {CuCrP}$_2$S$_6$}},
  \bibinfo{journal}{Nanoscale} \textbf{\bibinfo{volume}{11}},
  \bibinfo{pages}{5163} (\bibinfo{year}{2019}),
  \urlprefix\url{https://doi.org/10.1039/c9nr00738e}.

\bibitem[{\citenamefont{Chu et~al.}(2020)\citenamefont{Chu, Roh, Island, Li,
  Lee, Chen, Park, Young, Lee, and Hsieh}}]{Chu2020}
\bibinfo{author}{\bibfnamefont{H.}~\bibnamefont{Chu}},
  \bibinfo{author}{\bibfnamefont{C.~J.} \bibnamefont{Roh}},
  \bibinfo{author}{\bibfnamefont{J.~O.} \bibnamefont{Island}},
  \bibinfo{author}{\bibfnamefont{C.}~\bibnamefont{Li}},
  \bibinfo{author}{\bibfnamefont{S.}~\bibnamefont{Lee}},
  \bibinfo{author}{\bibfnamefont{J.}~\bibnamefont{Chen}},
  \bibinfo{author}{\bibfnamefont{J.-G.} \bibnamefont{Park}},
  \bibinfo{author}{\bibfnamefont{A.~F.} \bibnamefont{Young}},
  \bibinfo{author}{\bibfnamefont{J.~S.} \bibnamefont{Lee}}, \bibnamefont{and}
  \bibinfo{author}{\bibfnamefont{D.}~\bibnamefont{Hsieh}},
  \emph{\bibinfo{title}{Linear Magnetoelectric Phase in Ultrathin {MnPS}$_3$
  Probed by Optical Second Harmonic Generation}}, \bibinfo{journal}{Phys. Rev.
  Lett.} \textbf{\bibinfo{volume}{124}}, \bibinfo{pages}{027601}
  (\bibinfo{year}{2020}),
  \urlprefix\url{https://doi.org/10.1103/physrevlett.124.027601}.

\bibitem[{\citenamefont{Kim et~al.}(2018)\citenamefont{Kim, Kim, Sandilands,
  Sinn, Lee, Son, Lee, Choi, Kim, Park et~al.}}]{Kim2018}
\bibinfo{author}{\bibfnamefont{S.~Y.} \bibnamefont{Kim}},
  \bibinfo{author}{\bibfnamefont{T.~Y.} \bibnamefont{Kim}},
  \bibinfo{author}{\bibfnamefont{L.~J.} \bibnamefont{Sandilands}},
  \bibinfo{author}{\bibfnamefont{S.}~\bibnamefont{Sinn}},
  \bibinfo{author}{\bibfnamefont{M.-C.} \bibnamefont{Lee}},
  \bibinfo{author}{\bibfnamefont{J.}~\bibnamefont{Son}},
  \bibinfo{author}{\bibfnamefont{S.}~\bibnamefont{Lee}},
  \bibinfo{author}{\bibfnamefont{K.-Y.} \bibnamefont{Choi}},
  \bibinfo{author}{\bibfnamefont{W.}~\bibnamefont{Kim}},
  \bibinfo{author}{\bibfnamefont{B.-G.} \bibnamefont{Park}},
  \bibinfo{author}{\bibfnamefont{C.}~\bibnamefont{Jeon}},
  \bibinfo{author}{\bibfnamefont{H.-D.} \bibnamefont{Kim}},
  \bibinfo{author}{\bibfnamefont{C.-H.} \bibnamefont{Park}},
  \bibinfo{author}{\bibfnamefont{J.-G.} \bibnamefont{Park}},
  \bibinfo{author}{\bibfnamefont{S.}~\bibnamefont{Moon}}, \bibnamefont{and}
  \bibinfo{author}{\bibfnamefont{T.}~\bibnamefont{Noh}},
  \emph{\bibinfo{title}{Charge-Spin Correlation in van der Waals
  Antiferromagnet {NiPS}$_3$}}, \bibinfo{journal}{Phys. Rev. Lett.}
  \textbf{\bibinfo{volume}{120}}, \bibinfo{pages}{136402}
  (\bibinfo{year}{2018}),
  \urlprefix\url{https://doi.org/10.1103/physrevlett.120.136402}.

\bibitem[{\citenamefont{Kang et~al.}(2020)\citenamefont{Kang, Kim, Kim, Kim,
  Sim, Lee, Lee, Park, Yun, Kim et~al.}}]{Kang2020}
\bibinfo{author}{\bibfnamefont{S.}~\bibnamefont{Kang}},
  \bibinfo{author}{\bibfnamefont{K.}~\bibnamefont{Kim}},
  \bibinfo{author}{\bibfnamefont{B.~H.} \bibnamefont{Kim}},
  \bibinfo{author}{\bibfnamefont{J.}~\bibnamefont{Kim}},
  \bibinfo{author}{\bibfnamefont{K.~I.} \bibnamefont{Sim}},
  \bibinfo{author}{\bibfnamefont{J.-U.} \bibnamefont{Lee}},
  \bibinfo{author}{\bibfnamefont{S.}~\bibnamefont{Lee}},
  \bibinfo{author}{\bibfnamefont{K.}~\bibnamefont{Park}},
  \bibinfo{author}{\bibfnamefont{S.}~\bibnamefont{Yun}},
  \bibinfo{author}{\bibfnamefont{T.}~\bibnamefont{Kim}},
  \bibinfo{author}{\bibfnamefont{A.}~\bibnamefont{Nag}},
  \bibinfo{author}{\bibfnamefont{A.}~\bibnamefont{Walters}},
  \bibinfo{author}{\bibfnamefont{M.}~\bibnamefont{Garcia-Fernandez}},
  \bibinfo{author}{\bibfnamefont{J.}~\bibnamefont{Li}},
  \bibinfo{author}{\bibfnamefont{L.}~\bibnamefont{Chapon}},
  \bibinfo{author}{\bibfnamefont{K.-J.} \bibnamefont{Zhou}},
  \bibinfo{author}{\bibfnamefont{Y.-W.} \bibnamefont{Son}},
  \bibinfo{author}{\bibfnamefont{J.~H.} \bibnamefont{Kim}},
  \bibinfo{author}{\bibfnamefont{H.}~\bibnamefont{Cheong}}, \bibnamefont{and}
  \bibinfo{author}{\bibfnamefont{J.-G.} \bibnamefont{Park}},
  \emph{\bibinfo{title}{Coherent many-body exciton in van der Waals
  antiferromagnet {NiPS}$_3$}}, \bibinfo{journal}{Nature}
  \textbf{\bibinfo{volume}{583}}, \bibinfo{pages}{785} (\bibinfo{year}{2020}),
  \urlprefix\url{https://doi.org/10.1038/s41586-020-2520-5}.

\bibitem[{\citenamefont{Erge{\c{c}}en et~al.}(2022)\citenamefont{Erge{\c{c}}en,
  Ilyas, Mao, Po, Yilmaz, Kim, Park, Senthil, and Gedik}}]{Ergeen2022}
\bibinfo{author}{\bibfnamefont{E.}~\bibnamefont{Erge{\c{c}}en}},
  \bibinfo{author}{\bibfnamefont{B.}~\bibnamefont{Ilyas}},
  \bibinfo{author}{\bibfnamefont{D.}~\bibnamefont{Mao}},
  \bibinfo{author}{\bibfnamefont{H.~C.} \bibnamefont{Po}},
  \bibinfo{author}{\bibfnamefont{M.~B.} \bibnamefont{Yilmaz}},
  \bibinfo{author}{\bibfnamefont{J.}~\bibnamefont{Kim}},
  \bibinfo{author}{\bibfnamefont{J.-G.} \bibnamefont{Park}},
  \bibinfo{author}{\bibfnamefont{T.}~\bibnamefont{Senthil}}, \bibnamefont{and}
  \bibinfo{author}{\bibfnamefont{N.}~\bibnamefont{Gedik}},
  \emph{\bibinfo{title}{Magnetically brightened dark electron-phonon bound
  states in a van der Waals antiferromagnet}}, \bibinfo{journal}{Nature
  Communications} \textbf{\bibinfo{volume}{13}} (\bibinfo{year}{2022}),
  \urlprefix\url{https://doi.org/10.1038/s41467-021-27741-3}.

\bibitem[{\citenamefont{Kamata et~al.}(1997)\citenamefont{Kamata, Noguchi,
  Suzuki, Tezuka, Kashiwakura, Ohno, and ichi Nakai}}]{Kamata1997}
\bibinfo{author}{\bibfnamefont{A.}~\bibnamefont{Kamata}},
  \bibinfo{author}{\bibfnamefont{K.}~\bibnamefont{Noguchi}},
  \bibinfo{author}{\bibfnamefont{K.}~\bibnamefont{Suzuki}},
  \bibinfo{author}{\bibfnamefont{H.}~\bibnamefont{Tezuka}},
  \bibinfo{author}{\bibfnamefont{T.}~\bibnamefont{Kashiwakura}},
  \bibinfo{author}{\bibfnamefont{Y.}~\bibnamefont{Ohno}}, \bibnamefont{and}
  \bibinfo{author}{\bibfnamefont{S.}~\bibnamefont{ichi Nakai}},
  \emph{\bibinfo{title}{Resonant 2p $\rightarrow$ 3d Photoemission Measurement
  of {MPS}$_3$ (M=Mn, Fe, Ni)}}, \bibinfo{journal}{J. Phys. Soc. Jp.}
  \textbf{\bibinfo{volume}{66}}, \bibinfo{pages}{401} (\bibinfo{year}{1997}),
  \urlprefix\url{https://doi.org/10.1143/jpsj.66.401}.

\bibitem[{\citenamefont{Bianchi et~al.}(2023)\citenamefont{Bianchi, Acharya,
  Dirnberger, Klein, Pashov, Mosina, Sofer, Rudenko, Katsnelson, van
  Schilfgaarde et~al.}}]{Bianchi2023}
\bibinfo{author}{\bibfnamefont{M.}~\bibnamefont{Bianchi}},
  \bibinfo{author}{\bibfnamefont{S.}~\bibnamefont{Acharya}},
  \bibinfo{author}{\bibfnamefont{F.}~\bibnamefont{Dirnberger}},
  \bibinfo{author}{\bibfnamefont{J.}~\bibnamefont{Klein}},
  \bibinfo{author}{\bibfnamefont{D.}~\bibnamefont{Pashov}},
  \bibinfo{author}{\bibfnamefont{K.}~\bibnamefont{Mosina}},
  \bibinfo{author}{\bibfnamefont{Z.}~\bibnamefont{Sofer}},
  \bibinfo{author}{\bibfnamefont{A.~N.} \bibnamefont{Rudenko}},
  \bibinfo{author}{\bibfnamefont{M.~I.} \bibnamefont{Katsnelson}},
  \bibinfo{author}{\bibfnamefont{M.}~\bibnamefont{van Schilfgaarde}},
  \bibinfo{author}{\bibfnamefont{M.}~\bibnamefont{Rösner}}, \bibnamefont{and}
  \bibinfo{author}{\bibfnamefont{P.}~\bibnamefont{Hofmann}},
  \emph{\bibinfo{title}{Paramagnetic Electronic Structure of CrSBr: Comparison
  between Ab Initio GW Theory and Angle-Resolved Photoemission Spectroscopy}},
  \bibinfo{journal}{arXiv:} p. \bibinfo{pages}{2303.01292}
  (\bibinfo{year}{2023}), \urlprefix\url{https://arxiv.org/abs/2303.01292}.

\bibitem[{\citenamefont{Kurosawa et~al.}(1983)\citenamefont{Kurosawa, Saito,
  and Yamaguchi}}]{Kurosawa1983}
\bibinfo{author}{\bibfnamefont{K.}~\bibnamefont{Kurosawa}},
  \bibinfo{author}{\bibfnamefont{S.}~\bibnamefont{Saito}}, \bibnamefont{and}
  \bibinfo{author}{\bibfnamefont{Y.}~\bibnamefont{Yamaguchi}},
  \emph{\bibinfo{title}{Neutron Diffraction Study on {MnPS}$_3$ and
  {FePS}$_3$}}, \bibinfo{journal}{J. Phys. Soc. Jpn.}
  \textbf{\bibinfo{volume}{52}}, \bibinfo{pages}{3919} (\bibinfo{year}{1983}),
  \urlprefix\url{https://doi.org/10.1143/jpsj.52.3919}.

\bibitem[{\citenamefont{Brec}(1986)}]{Brec1986}
\bibinfo{author}{\bibfnamefont{R.}~\bibnamefont{Brec}},
  \emph{\bibinfo{title}{Review on structural and chemical properties of
  transition metal phosphorous trisulfides MPS$_3$}}, \bibinfo{journal}{Solid
  State Ionics} \textbf{\bibinfo{volume}{22}}, \bibinfo{pages}{3 }
  (\bibinfo{year}{1986}), ISSN \bibinfo{issn}{0167-2738}.

\bibitem[{\citenamefont{Autieri et~al.}(2022)\citenamefont{Autieri, Cuono,
  Noce, Rybak, Kotur, Agrapidis, Wohlfeld, and Birowska}}]{Autieri2022}
\bibinfo{author}{\bibfnamefont{C.}~\bibnamefont{Autieri}},
  \bibinfo{author}{\bibfnamefont{G.}~\bibnamefont{Cuono}},
  \bibinfo{author}{\bibfnamefont{C.}~\bibnamefont{Noce}},
  \bibinfo{author}{\bibfnamefont{M.}~\bibnamefont{Rybak}},
  \bibinfo{author}{\bibfnamefont{K.~M.} \bibnamefont{Kotur}},
  \bibinfo{author}{\bibfnamefont{C.~E.} \bibnamefont{Agrapidis}},
  \bibinfo{author}{\bibfnamefont{K.}~\bibnamefont{Wohlfeld}}, \bibnamefont{and}
  \bibinfo{author}{\bibfnamefont{M.}~\bibnamefont{Birowska}},
  \emph{\bibinfo{title}{Limited Ferromagnetic Interactions in Monolayers of
  MPS$_3$ (M = Mn and Ni)}}, \bibinfo{journal}{The Journal of Physical
  Chemistry C} \textbf{\bibinfo{volume}{126}}, \bibinfo{pages}{6791}
  (\bibinfo{year}{2022}), ISSN \bibinfo{issn}{1932-7447},
  \urlprefix\url{https://doi.org/10.1021/acs.jpcc.2c00646}.

\bibitem[{\citenamefont{Lim et~al.}(2021)\citenamefont{Lim, Kim, Lee, Park, and
  Cheong}}]{Lim2021}
\bibinfo{author}{\bibfnamefont{S.~Y.} \bibnamefont{Lim}},
  \bibinfo{author}{\bibfnamefont{K.}~\bibnamefont{Kim}},
  \bibinfo{author}{\bibfnamefont{S.}~\bibnamefont{Lee}},
  \bibinfo{author}{\bibfnamefont{J.-G.} \bibnamefont{Park}}, \bibnamefont{and}
  \bibinfo{author}{\bibfnamefont{H.}~\bibnamefont{Cheong}},
  \emph{\bibinfo{title}{Thickness dependence of antiferromagnetic phase
  transition in Heisenberg-type {MnPS}$_3$}}, \bibinfo{journal}{Current Appl.
  Phys.} \textbf{\bibinfo{volume}{21}}, \bibinfo{pages}{1}
  (\bibinfo{year}{2021}),
  \urlprefix\url{https://doi.org/10.1016/j.cap.2020.09.017}.

\bibitem[{\citenamefont{Kim et~al.}(2019)\citenamefont{Kim, Lim, Kim, Lee, Lee,
  Kim, Park, Son, Park, Park et~al.}}]{Kim2019}
\bibinfo{author}{\bibfnamefont{K.}~\bibnamefont{Kim}},
  \bibinfo{author}{\bibfnamefont{S.~Y.} \bibnamefont{Lim}},
  \bibinfo{author}{\bibfnamefont{J.}~\bibnamefont{Kim}},
  \bibinfo{author}{\bibfnamefont{J.-U.} \bibnamefont{Lee}},
  \bibinfo{author}{\bibfnamefont{S.}~\bibnamefont{Lee}},
  \bibinfo{author}{\bibfnamefont{P.}~\bibnamefont{Kim}},
  \bibinfo{author}{\bibfnamefont{K.}~\bibnamefont{Park}},
  \bibinfo{author}{\bibfnamefont{S.}~\bibnamefont{Son}},
  \bibinfo{author}{\bibfnamefont{C.-H.} \bibnamefont{Park}},
  \bibinfo{author}{\bibfnamefont{J.-G.} \bibnamefont{Park}}, \bibnamefont{and}
  \bibinfo{author}{\bibfnamefont{H.}~\bibnamefont{Cheong}},
  \emph{\bibinfo{title}{Antiferromagnetic ordering in van der Waals 2D magnetic
  material {MnPS}$_3$ probed by Raman spectroscopy}}, \bibinfo{journal}{2D
  Mater.} \textbf{\bibinfo{volume}{6}}, \bibinfo{pages}{041001}
  (\bibinfo{year}{2019}),
  \urlprefix\url{https://doi.org/10.1088/2053-1583/ab27d5}.

\bibitem[{\citenamefont{Long et~al.}(2020)\citenamefont{Long, Henck, Gibertini,
  Dumcenco, Wang, Taniguchi, Watanabe, Giannini, and Morpurgo}}]{Long2020}
\bibinfo{author}{\bibfnamefont{G.}~\bibnamefont{Long}},
  \bibinfo{author}{\bibfnamefont{H.}~\bibnamefont{Henck}},
  \bibinfo{author}{\bibfnamefont{M.}~\bibnamefont{Gibertini}},
  \bibinfo{author}{\bibfnamefont{D.}~\bibnamefont{Dumcenco}},
  \bibinfo{author}{\bibfnamefont{Z.}~\bibnamefont{Wang}},
  \bibinfo{author}{\bibfnamefont{T.}~\bibnamefont{Taniguchi}},
  \bibinfo{author}{\bibfnamefont{K.}~\bibnamefont{Watanabe}},
  \bibinfo{author}{\bibfnamefont{E.}~\bibnamefont{Giannini}}, \bibnamefont{and}
  \bibinfo{author}{\bibfnamefont{A.~F.} \bibnamefont{Morpurgo}},
  \emph{\bibinfo{title}{Persistence of Magnetism in Atomically Thin {MnPS}$_3$
  Crystals}}, \bibinfo{journal}{Nano Lett.} \textbf{\bibinfo{volume}{20}},
  \bibinfo{pages}{2452} (\bibinfo{year}{2020}),
  \urlprefix\url{https://doi.org/10.1021/acs.nanolett.9b05165}.

\bibitem[{\citenamefont{Babuka et~al.}(2020)\citenamefont{Babuka,
  Makowska-Janusik, Peschanskii, Glukhov, Gnatchenko, and
  Vysochanskii}}]{Babuka2020}
\bibinfo{author}{\bibfnamefont{T.}~\bibnamefont{Babuka}},
  \bibinfo{author}{\bibfnamefont{M.}~\bibnamefont{Makowska-Janusik}},
  \bibinfo{author}{\bibfnamefont{A.}~\bibnamefont{Peschanskii}},
  \bibinfo{author}{\bibfnamefont{K.}~\bibnamefont{Glukhov}},
  \bibinfo{author}{\bibfnamefont{S.}~\bibnamefont{Gnatchenko}},
  \bibnamefont{and}
  \bibinfo{author}{\bibfnamefont{Y.}~\bibnamefont{Vysochanskii}},
  \emph{\bibinfo{title}{Electronic and vibrational properties of pure
  {MnPS}$_3$ crystal: Theoretical and experimental investigation}},
  \bibinfo{journal}{Comp. Mater. Sci.} \textbf{\bibinfo{volume}{177}},
  \bibinfo{pages}{109592} (\bibinfo{year}{2020}),
  \urlprefix\url{https://doi.org/10.1016/j.commatsci.2020.109592}.

\bibitem[{\citenamefont{Ni et~al.}(2021)\citenamefont{Ni, Zhang, Hopper,
  Haglund, Huang, Jariwala, Bassett, Mandrus, Mele, Kane et~al.}}]{Ni2021}
\bibinfo{author}{\bibfnamefont{Z.}~\bibnamefont{Ni}},
  \bibinfo{author}{\bibfnamefont{H.}~\bibnamefont{Zhang}},
  \bibinfo{author}{\bibfnamefont{D.~A.} \bibnamefont{Hopper}},
  \bibinfo{author}{\bibfnamefont{A.~V.} \bibnamefont{Haglund}},
  \bibinfo{author}{\bibfnamefont{N.}~\bibnamefont{Huang}},
  \bibinfo{author}{\bibfnamefont{D.}~\bibnamefont{Jariwala}},
  \bibinfo{author}{\bibfnamefont{L.~C.} \bibnamefont{Bassett}},
  \bibinfo{author}{\bibfnamefont{D.~G.} \bibnamefont{Mandrus}},
  \bibinfo{author}{\bibfnamefont{E.~J.} \bibnamefont{Mele}},
  \bibinfo{author}{\bibfnamefont{C.~L.} \bibnamefont{Kane}}, \bibnamefont{and}
  \bibinfo{author}{\bibfnamefont{L.}~\bibnamefont{Wu}},
  \emph{\bibinfo{title}{Direct Imaging of Antiferromagnetic Domains and
  Anomalous Layer-Dependent Mirror Symmetry Breaking in Atomically Thin
  {MnPS}$_3$}}, \bibinfo{journal}{Phys. Rev.Lett.}
  \textbf{\bibinfo{volume}{127}}, \bibinfo{pages}{187201}
  (\bibinfo{year}{2021}),
  \urlprefix\url{https://doi.org/10.1103/physrevlett.127.187201}.

\bibitem[{\citenamefont{Shan et~al.}(2021)\citenamefont{Shan, Ye, Chu, Lee,
  Park, Balents, and Hsieh}}]{Shan2021}
\bibinfo{author}{\bibfnamefont{J.-Y.} \bibnamefont{Shan}},
  \bibinfo{author}{\bibfnamefont{M.}~\bibnamefont{Ye}},
  \bibinfo{author}{\bibfnamefont{H.}~\bibnamefont{Chu}},
  \bibinfo{author}{\bibfnamefont{S.}~\bibnamefont{Lee}},
  \bibinfo{author}{\bibfnamefont{J.-G.} \bibnamefont{Park}},
  \bibinfo{author}{\bibfnamefont{L.}~\bibnamefont{Balents}}, \bibnamefont{and}
  \bibinfo{author}{\bibfnamefont{D.}~\bibnamefont{Hsieh}},
  \emph{\bibinfo{title}{Giant modulation of optical nonlinearity by Floquet
  engineering}}, \bibinfo{journal}{Nature} \textbf{\bibinfo{volume}{600}},
  \bibinfo{pages}{235} (\bibinfo{year}{2021}),
  \urlprefix\url{https://doi.org/10.1038/s41586-021-04051-8}.

\bibitem[{\citenamefont{Wildes et~al.}(2006)\citenamefont{Wildes, R{\o}nnow,
  Roessli, Harris, and Godfrey}}]{Wildes2006}
\bibinfo{author}{\bibfnamefont{A.~R.} \bibnamefont{Wildes}},
  \bibinfo{author}{\bibfnamefont{H.~M.} \bibnamefont{R{\o}nnow}},
  \bibinfo{author}{\bibfnamefont{B.}~\bibnamefont{Roessli}},
  \bibinfo{author}{\bibfnamefont{M.~J.} \bibnamefont{Harris}},
  \bibnamefont{and} \bibinfo{author}{\bibfnamefont{K.~W.}
  \bibnamefont{Godfrey}}, \emph{\bibinfo{title}{Static and dynamic critical
  properties of the quasi-two-dimensional {antiferromagnet MnPS}3}},
  \bibinfo{journal}{Phys. Rev. B} \textbf{\bibinfo{volume}{74}},
  \bibinfo{pages}{094422} (\bibinfo{year}{2006}),
  \urlprefix\url{https://doi.org/10.1103/physrevb.74.094422}.

\bibitem[{\citenamefont{Grasso et~al.}(1991)\citenamefont{Grasso, Neri,
  Perillo, Silipigni, and Piacentini}}]{Grasso1991}
\bibinfo{author}{\bibfnamefont{V.}~\bibnamefont{Grasso}},
  \bibinfo{author}{\bibfnamefont{F.}~\bibnamefont{Neri}},
  \bibinfo{author}{\bibfnamefont{P.}~\bibnamefont{Perillo}},
  \bibinfo{author}{\bibfnamefont{L.}~\bibnamefont{Silipigni}},
  \bibnamefont{and}
  \bibinfo{author}{\bibfnamefont{M.}~\bibnamefont{Piacentini}},
  \emph{\bibinfo{title}{Optical-absorption spectra of crystal-field transitions
  in MnPS$_3$ at low temperatures}}, \bibinfo{journal}{Phys. Rev. B}
  \textbf{\bibinfo{volume}{44}}, \bibinfo{pages}{11060} (\bibinfo{year}{1991}),
  \urlprefix\url{https://doi.org/10.1103/physrevb.44.11060}.

\bibitem[{\citenamefont{Du et~al.}(2016)\citenamefont{Du, Wang, Liu, Hu, Utama,
  Gan, Xiong, and Kloc}}]{Du2015}
\bibinfo{author}{\bibfnamefont{K.-Z.} \bibnamefont{Du}},
  \bibinfo{author}{\bibfnamefont{X.-Z.} \bibnamefont{Wang}},
  \bibinfo{author}{\bibfnamefont{Y.}~\bibnamefont{Liu}},
  \bibinfo{author}{\bibfnamefont{P.}~\bibnamefont{Hu}},
  \bibinfo{author}{\bibfnamefont{M.~I.~B.} \bibnamefont{Utama}},
  \bibinfo{author}{\bibfnamefont{C.~K.} \bibnamefont{Gan}},
  \bibinfo{author}{\bibfnamefont{Q.}~\bibnamefont{Xiong}}, \bibnamefont{and}
  \bibinfo{author}{\bibfnamefont{C.}~\bibnamefont{Kloc}},
  \emph{\bibinfo{title}{Weak Van der Waals Stacking, Wide-Range Band Gap, and
  Raman Study on Ultrathin Layers of Metal Phosphorus Trichalcogenides}},
  \bibinfo{journal}{{ACS} Nano} \textbf{\bibinfo{volume}{10}},
  \bibinfo{pages}{1738} (\bibinfo{year}{2016}),
  \urlprefix\url{https://doi.org/10.1021/acsnano.5b05927}.

\bibitem[{\citenamefont{Momma and Izumi}(2011)}]{VESTA}
\bibinfo{author}{\bibfnamefont{K.}~\bibnamefont{Momma}} \bibnamefont{and}
  \bibinfo{author}{\bibfnamefont{F.}~\bibnamefont{Izumi}},
  \emph{\bibinfo{title}{{{\it VESTA3} for three-dimensional visualization of
  crystal, volumetric and morphology data}}}, \bibinfo{journal}{Journal of
  Applied Crystallography} \textbf{\bibinfo{volume}{44}}, \bibinfo{pages}{1272}
  (\bibinfo{year}{2011}),
  \urlprefix\url{https://doi.org/10.1107/S0021889811038970}.

\bibitem[{\citenamefont{Wiemann et~al.}(2011)\citenamefont{Wiemann, Patt, Krug,
  Weber, Escher, Merkel, and Schneider}}]{Wiemann2011}
\bibinfo{author}{\bibfnamefont{C.}~\bibnamefont{Wiemann}},
  \bibinfo{author}{\bibfnamefont{M.}~\bibnamefont{Patt}},
  \bibinfo{author}{\bibfnamefont{I.~P.} \bibnamefont{Krug}},
  \bibinfo{author}{\bibfnamefont{N.~B.} \bibnamefont{Weber}},
  \bibinfo{author}{\bibfnamefont{M.}~\bibnamefont{Escher}},
  \bibinfo{author}{\bibfnamefont{M.}~\bibnamefont{Merkel}}, \bibnamefont{and}
  \bibinfo{author}{\bibfnamefont{C.~M.} \bibnamefont{Schneider}},
  \emph{\bibinfo{title}{A New Nanospectroscopy Tool with Synchrotron Radiation:
  NanoESCA@Elettra}}, \bibinfo{journal}{e-Journal of Surface Science and
  Nanotechnology} \textbf{\bibinfo{volume}{9}}, \bibinfo{pages}{395}
  (\bibinfo{year}{2011}).

\bibitem[{\citenamefont{Moser}(2017)}]{Moser2017}
\bibinfo{author}{\bibfnamefont{S.}~\bibnamefont{Moser}},
  \emph{\bibinfo{title}{An experimentalist{\textquotesingle}s guide to the
  matrix element in angle resolved photoemission}}, \bibinfo{journal}{J.
  Electr. Spectr. Rel. Phen.} \textbf{\bibinfo{volume}{214}},
  \bibinfo{pages}{29} (\bibinfo{year}{2017}),
  \urlprefix\url{https://doi.org/10.1016/j.elspec.2016.11.007}.

\bibitem[{\citenamefont{Wildes et~al.}(1998)\citenamefont{Wildes, Roessli,
  Lebech, and Godfrey}}]{Wildes1998}
\bibinfo{author}{\bibfnamefont{A.~R.} \bibnamefont{Wildes}},
  \bibinfo{author}{\bibfnamefont{B.}~\bibnamefont{Roessli}},
  \bibinfo{author}{\bibfnamefont{B.}~\bibnamefont{Lebech}}, \bibnamefont{and}
  \bibinfo{author}{\bibfnamefont{K.~W.} \bibnamefont{Godfrey}},
  \emph{\bibinfo{title}{Spin waves and the critical behaviour of the
  magnetization in MnPS$_3$}}, \bibinfo{journal}{J. Phys.: Cond. Mat.}
  \textbf{\bibinfo{volume}{10}}, \bibinfo{pages}{6417} (\bibinfo{year}{1998}),
  \urlprefix\url{https://doi.org/10.1088/0953-8984/10/28/020}.

\bibitem[{\citenamefont{Ushakov et~al.}(2013)\citenamefont{Ushakov, Kukusta,
  Yaresko, and Khomskii}}]{Ushakov2013}
\bibinfo{author}{\bibfnamefont{A.~V.} \bibnamefont{Ushakov}},
  \bibinfo{author}{\bibfnamefont{D.~A.} \bibnamefont{Kukusta}},
  \bibinfo{author}{\bibfnamefont{A.~N.} \bibnamefont{Yaresko}},
  \bibnamefont{and} \bibinfo{author}{\bibfnamefont{D.~I.}
  \bibnamefont{Khomskii}}, \emph{\bibinfo{title}{Magnetism of layered chromium
  sulfides $M$CRS$_2$ ($M=$ Na, K, Ag, and Au): A first-principles study}},
  \bibinfo{journal}{Phys. Rev. B} \textbf{\bibinfo{volume}{87}},
  \bibinfo{pages}{014418} (\bibinfo{year}{2013}),
  \urlprefix\url{https://doi.org/10.1103/physrevb.87.014418}.

\bibitem[{\citenamefont{Grasso and Silipigni}(2002)}]{Grasso2002}
\bibinfo{author}{\bibfnamefont{V.}~\bibnamefont{Grasso}} \bibnamefont{and}
  \bibinfo{author}{\bibfnamefont{L.}~\bibnamefont{Silipigni}},
  \emph{\bibinfo{title}{Low-dimensional materials: The {MPX}$_3$ family,
  physical features and potential future applications}}, \bibinfo{journal}{Riv.
  Nuov. Cim.} \textbf{\bibinfo{volume}{25}}, \bibinfo{pages}{1}
  (\bibinfo{year}{2002}), \urlprefix\url{https://doi.org/10.1007/bf03548909}.

\end{thebibliography}
\end{document}


\title{Supplementary Information: Electronic band structure changes across the antiferromagnetic phase transition of exfoliated MnPS$_3$ flakes probed by $\mu$-ARPES}

\author{Jeff Strasdas}\affiliation{\AC}
\author{Benjamin Pestka}\affiliation{\AC}
\author{Miłosz Rybak}\affiliation{\PW}
\author{Adam K. Budniak}\affiliation{\Haifa}
\author{Niklas Leuth}\affiliation{\AC}
\author{Honey Boban}\affiliation{\FZJ}
\author{Vitaliy Feyer}\affiliation{\FZJ}
\author{Iulia Cojocariu}\affiliation{\FZJ}
\author{Daniel Baranowski}\affiliation{\FZJ}
\author{José Avila}\affiliation{\SOLEIL}
\author{Pavel Dudin}\affiliation{\SOLEIL}
\author{Aaron Bostwick}\affiliation{\ALS} 
\author{Chris Jozwiak}\affiliation{\ALS}
\author{Eli Rotenberg}\affiliation{\ALS}
\author{Carmine Autieri}\affiliation{\MAGTOP}
\author{Yaron Amouyal}\affiliation{\HaifaMat}
\author{Lukasz Plucinski}\affiliation{\FZJ}
\author{Efrat Lifshitz}\affiliation{\Haifa}
\author{Magdalena Birowska}\affiliation{\IFT}
\author{Markus Morgenstern}\affiliation{\AC}
\date{\today} 

\maketitle
\noindent {{Corresponding author: } 
M.~Morgenstern, email: mmorgens@physik.rwth-aachen.de } 
\tableofcontents

%
\section{Experimental details}
\label{sec:exp_det}
\begin{figure*}[t]
\centering
\includegraphics[width=0.8\textwidth]{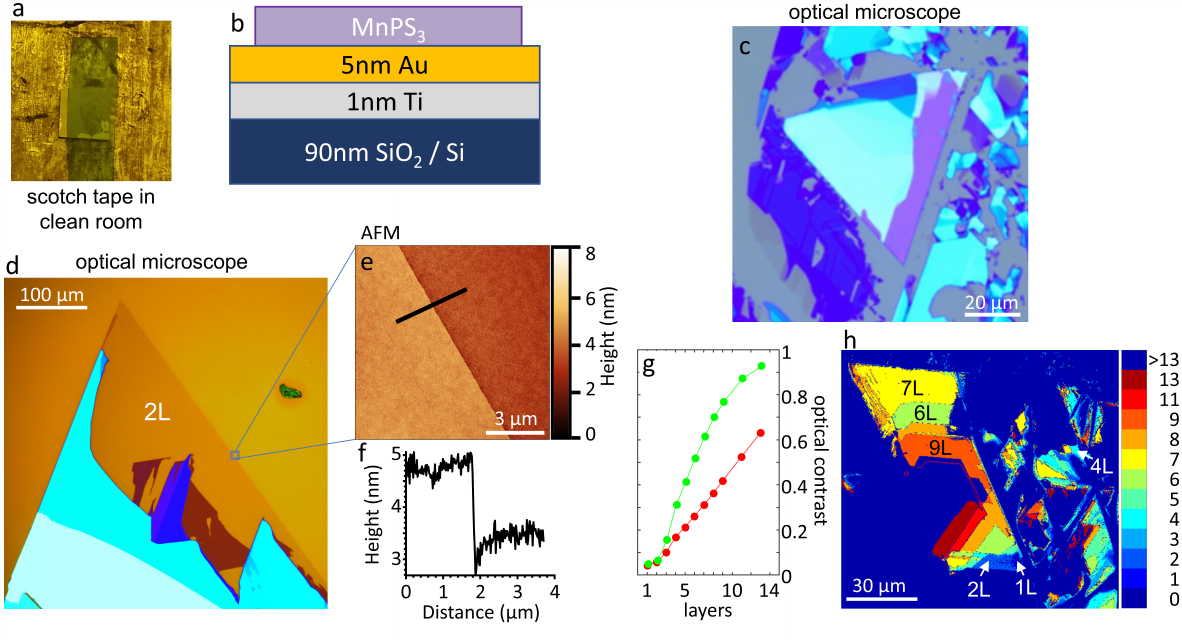}
\vspace{-0.5cm}
\caption{
{\bf Sample preparation}
(a) Photograph of a scotch tape in contact with a chip during exfoliation.
(b) Schematic sketch of the sample structure.
(c) Optical microscope image of MnPS$_3$ flakes of various thickness.
(d) Optical microscope image of a MnPS$_3$ flake with a large part of thickness of two layers and size $\sim 150\times300\,\mu$m$^2$.
(e) Tapping-mode atomic force microscopy (AFMi) image of the highlighted area in d.
(f) Profile line along the path marked in e.
\MM{(g) Optical contrast with respect to the substrate for MnPS$_3$ flakes as a function of thicknesses. The green and red color filters of a Leica DFC 450 camera are used for the symbols of respective colors. The layer thicknesses are determined by AFMi. Error bars are smaller than the symbol size.}
(h) False-color image of optically estimated layer thicknesses of MnPS$_3$ using the relation shown in g.
}
\label{Fig_S1}
\end{figure*}

\subsection{Crystal growth}
MnPS$_3$  was synthesized in a quartz tube via vapor transport synthesis (VTS) – without any transporting agent, i.e. no iodine was used \cite{Taylor1973, Budniak2020}. One gram of an elements mixture of metal manganese powder (Alfa Aesar), red phosphorus (Riedel-de Haën), and sulfur (Sigma-Aldrich) with atomic ratio Mn:P:S = 1:1:3 was grounded in an agate mortar. The mixture was moved into a quartz ampoule, evacuated to high vacuum ($< 4.5\cdot 10^{-5}$\,mbar) with a turbo molecular pump, and then closed by a flamer. The sealed ampoule was placed inside a previously warmed 3-zone furnace, calibrated in the way that the mixture of elements (substrate zone) was kept at 650$^\circ$\,C and the deposition zone was at 600$^\circ$\,C. After 5 days the furnace was turned off and the sample was left to cool down inside. Then the ampoule was opened and only recrystallized MnPS$_3$ from the deposition zone was collected.

\subsection{Flake preparation by exfoliation}

Thin flakes of MnPS$_3$ are prepared by an optimized exfoliation process utilizing the scotch tape method within a clean room evironment.
Figure~\ref{Fig_S1}a shows a photograph of an adhesive tape (Nitto ELP BT-150E-KL) during exfoliation of MnPS$_3$ at 60$^\circ$\,C.
The resulting sample structure with substrate and flake is sketched in Fig.~\ref{Fig_S1}b. As substrate, 
we use Si/SiO$_2$ (SiO$_2$ thickness: 90\,nm) for the purpose of flake visibility  and enhance flake adhesion by a thin Au/Ti film deposited  at a substrate temperature of 370\,K in ultrahigh vacuum (UHV) \MM{\cite{Huang2020b,Velicky2018}.
Before exfoliation, the substrate is exposed to plasma ashing using O$_2$ at 300\,K with a power of 50\,W and a flux of 100\,sccm for 25\,s. Within 10\,min
afterwards, MnPS$_3$ is exfoliated. The procedure reveals
MnPS$_3$ flakes with large areas of various thicknesses down to a monolayer (Fig.~\ref{Fig_S1} c--h).} 
The number of layers can be inferred from the flat-field corrected optical Michelson contrast by selecting the green and red band of a standard camera. \MM{Calibration of the thicknesses is done by atomic force microscopy (AFMi) in tapping mode using a commercial Bruker system (Fig.~\ref{Fig_S1}e--f). A monotonous increase of contrast with thickness is found, at least, up to 14 layers, (Fig.~\ref{Fig_S1}g) as reported previously \cite{Sun2019,Ni2021}. It enables an automatic build-up of thickness maps (Fig.~\ref{Fig_S1}h).}
%
\subsection{ARPES and XPS measurements}

Flakes of various thicknesses with clean surfaces, as cross-checked by AFMi and X-ray photoelectron spectroscopy (XPS) \MM{(Section S1.D)}, are investigated by $\mu$-ARPES at the NanoESCA beamline of Elettra, the Italian synchrotron radiation facility, using a FOCUS NanoESCA photoemission electron microscope (PEEM) in the $k$ space mapping mode operation \cite{Schneider2012}. The PEEM is operated at a background pressure $p<5^{-11}$\,mbar. 
XPS is performed with the same instrument.
The photoelectron signal at various photon energies $h\nu$ is collected from a spot size with diameter $5-10$\,$\mu$m.  The total energy resolution of beamline and analyser is 50\,meV. The geometry of incident light (p-polarized) is sketched in \MM{Fig.~2b}, main text. The sample temperature 
$T$ during $\mu$-ARPES is measured by a Si diode (Lake shore DT-670E-BR) at the sample holder and adjusted by changing the flow rate of liquid He. Prior to $\mu$-ARPES, the samples are annealed at $T=200^\circ$\,C for $60$\,min.  
Additional measurements after the same sample preparation, that compare the monolayer with thicker films of MnPS$_3$ (\MM{Fig.~1f, i}, main text), are acquired from the MAESTRO endstation (7.0.2.1) at the Advanced Light Source at Lawrence Berkeley National Laboratory with a beam spot diameter of $20$\,$\mu$m and an energy resolution of $16$\,meV. 
In both experiments, the Fermi level $E_{\rm F}$ is determined on the Au substrate next to the MnPS$_3$ flakes. The work function of MnPS$_3$ is deduced from the onset energy of secondary electrons 
\MM{(Section S1.E)}. The Brillouin zone (BZ) orientation is adapted by the repetitive features in first and second BZs, while the scale of wave vectors parallel to the surface $k_\parallel$ is deduced from the angular dependence of the  onset of secondary photoelectron emission \MM{(Section S1.F)}. 
For the sake of better visibility, we display the curvature of the measured photoelectron intensity with respect to energy. All discussed band structure features are cross-checked to be visible in the raw ARPES data as well \MM{(Section S1.H)}. 

\subsection{Sample cleanliness}

\begin{figure*}[htb]
\centering
\includegraphics[width=0.9\textwidth]{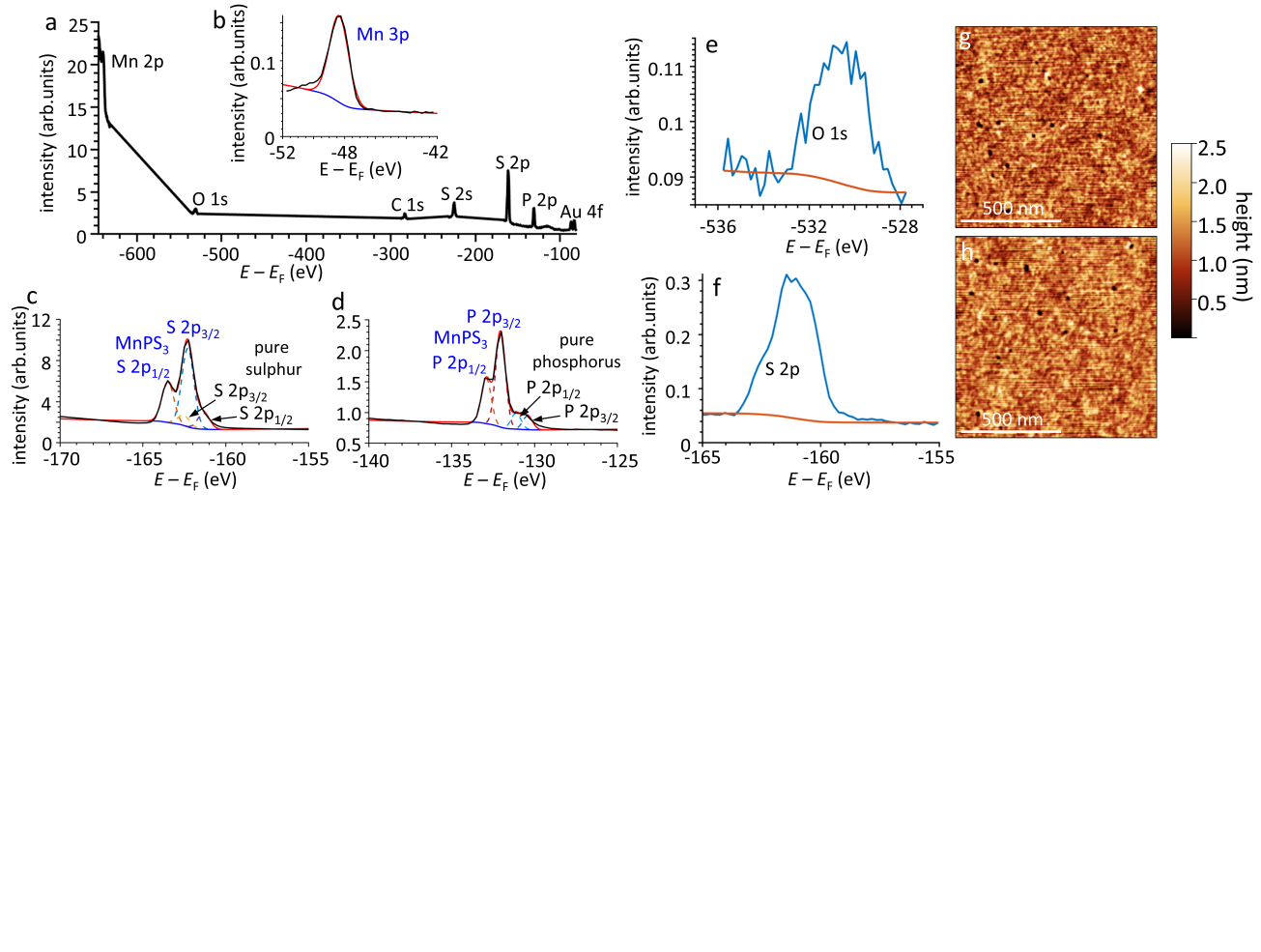}
\vspace{-6.2cm}
\caption{{\bf Analysis of XPS data}
(a) XPS data of 13\,layer MnPS$_3$ recorded after extensive ARPES measurements on that area. Peaks are labeled by chemical element and atomic orbital. 
(b)--(d) Higher resolution XPS data of  62\,layer MnPS$_3$ as studied in \MM{Fig.~2, 4}, main text, displaying the Mn 3p, S 2p and P 2p peak, respectively. Fit lines are displayed in red. They consist of the labeled dashed Lorentzian peaks and the blue Shirley–Proctor–Sherwood background. 
Peaks labeled blue belong to intact MnPS$_3$. Peaks labeled black are likely caused by remaining intercalated pure sulfur and phosphorus from the crystal growth.
(e), (f) Higher resolution XPS data from (a) containing the O 1s and S 2p peaks of a 13\,layer MnPS$_3$ area. A comparison of the O 1s and the S 2p peak reveals that the
oxygen density is less than 8 \% of the sulfur density. (a)--(f) $h\nu=700$\,eV.
(g), (h) AFMi images of the monolayer MnPS$_3$ area that is  studied in \MM{Fig.~1i}, main text before (g, $\sigma_{\rm rms}=0.30$\,nm) and after (h, $\sigma_{\rm rms}=0.29$\,nm) ARPES measurements.
}
\label{Fig_S2}
\end{figure*}

\begin{figure}[htb]
\centering
\includegraphics[width=\textwidth/2]{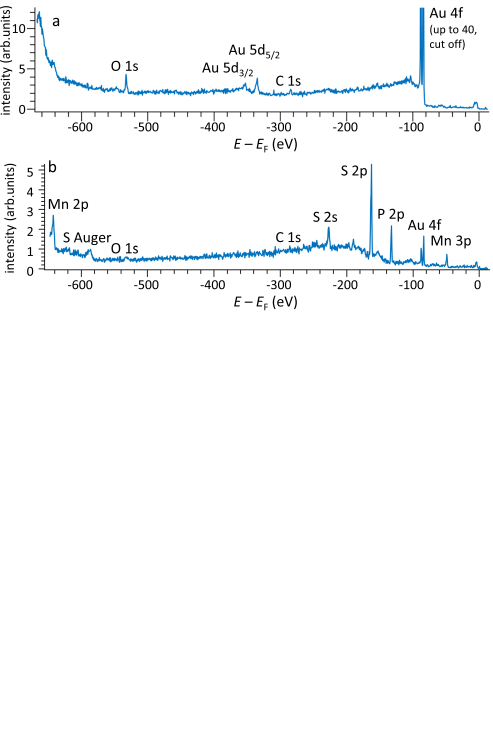}
\vspace{-8.6cm}
\caption{{\bf XPS: Au substrate vs. MnPS$_3$}
(a) XPS data on the Au substrate next to a MnPS$_3$ flake, 
$h\nu=750$\,eV. Peaks are labeled by chemical element and atomic orbital. 
The dominating Au 4f double peak at $E-E_{\rm F}\approx -92$\,eV exhibits a maximum of 40 arb. units.
(b) XPS data of 14\,layer MnPS$_3$ after ARPES measurements, $h\nu = 750$\,eV. The scales (arb. units) can be compared directly, since roughly proportional to photoelectrons/photon.}
\label{Fig_S3}
\end{figure}

The sample quality prior to and after the ARPES measurements is probed by XPS 
and AFMi. Fig.~\ref{Fig_S2}a shows the XPS data for a large energy range as recorded after ARPES measurements. 
The peaks are labeled revealing the presence of O and C contamination. 
To estimate the relative amount of oxygen compared to sulfur, the corresponding peaks are probed at higher resolution (Fig. \ref{Fig_S2}e, f). 
The ratio of peak areas weighted with the atomic sensitivity factors directly yields the relative atomic fractions \cite{moulder1992}.
Using the available atomic sensitivity factors for O 1s and S 2p \cite{Biesinger2009, yeh1993, yeh1985}, we find that the oxygen density is about 8\% of the sulfur density only. 

The relative peak heights of O 1s to S 2p do not change between XPS recorded before and after the ARPES measurements.
This excludes that the O and C contamination originates from ARPES. 
In line, AFMi data recorded on the photon beam spot area exhibit the same roughness $\sigma_{\rm rms}$ of the surface prior and after the ARPES measurements as exemplarily shown for the monolayer in Fig.~\ref{Fig_S2}g--h.

Since the electron extraction area is significantly less focused for XPS ($\sim 15\,\mu$m), we conjecture that the C and O peaks are caused by remaining glue on the substrate. Such glue residuals are partially observed by AFMi as well. Nevertheless, in order to probe the possible oxidation of the Mn, S and P of MnPS$_3$, we fitted the corresponding XPS peaks (Fig.~\ref{Fig_S2}b--d). For the S 2p and P 2p peak, we can exclude oxidation, since it would appear at larger binding energy ($164-170$\,eV for S, $> 134$\,eV for P) \cite{siow2018, franke1991}. In contrast, we observe minority contributions to both peaks at lower binding energy (Fig.~\ref{Fig_S2}c--d). We attribute these contributions to remaining pure S and P that might be intercalated in between the layers during the crystal growth process. 
The Mn 3p peak (Fig.~\ref{Fig_S2}b) could be fitted by a single Lorentzian with an energy that indicates a charge transfer. The energy shift is is compatible with both, S or O bonding to the Mn \cite{Ilton2016, Nelson2000}. Hence, we cannot exclude partial oxidation of the Mn. However, even if, only  8\,\% of the S bonds would be replaced by O bonds. Notice that a replacement of S by O has been observed in electron microscopy, but there it is induced by the high energy electrons that kick off the S atoms prior to oxidation \cite{Zhou2022,Koester2022}.

Figure~\ref{Fig_S3} shows XPS spectra focused to the Au substrate and to a close-by MnPS$_3$ flake for direct comparison. The scale is roughly proportional to the number of photoelectrons per photon. While the C 1s peak is similar in both spectra, i.e. barely above the noise level, the O 1s peak is a factor of ten larger on the Au substrate. This corroborates our suspicion that the oxygen originates from remaining glue on the substrate.

\subsection{Determination of Fermi level and work function}
\label{sub:kz}

\begin{figure}[bt]
\centering
\includegraphics[width=\textwidth/2]{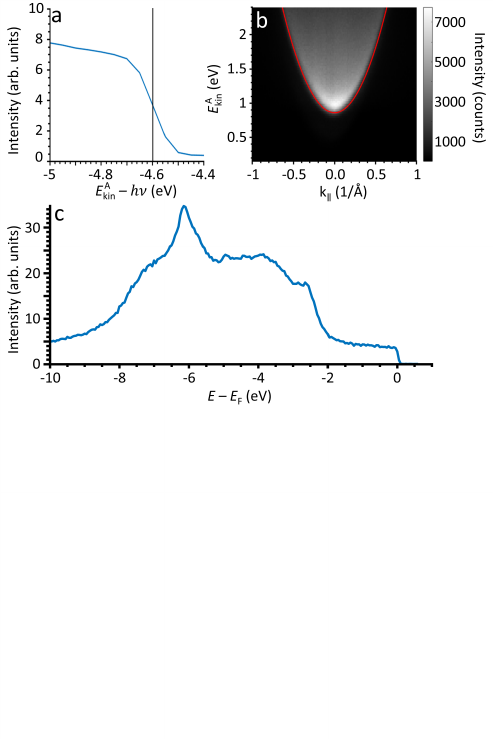} 
\vspace{-6.8cm}
\caption{{\bf Work function and spectrum of Au substrate}
(a) ARPES intensity of gold substrate near the Fermi edge. The black line marks the inflection point of $I(E_{\rm kin}^{\rm A}-h\nu)$ regarded as $E_{\rm F}$. The analyzer work function, thus, reads $\phi^{\rm A} = 4.6$\,eV (see text), $h\nu = 50$\,eV, $\theta =0^\circ$.
(b) Intensity of secondary electron emission as a function of measured kinetic energy \MM{$E_{\rm kin}^{\rm A}$} and momentum parallel to the surface $k_\parallel$. Red line: free electron parabola fitting the onset of secondary photoelectrons to determine the work function of MnPS$_3$ $\phi=5.45$\,eV, $h\nu = 50$\,eV, 13\,layers.
(c) ARPES intensity of gold substrate next to MnPS$_3$. The sharp peak at $E-E_{\rm F} = -6\,$eV is a Au 5d band \cite{Rangel2012} also appearing in \MM{Fig.~1i}, main text, $h\nu = 50$\,eV, $\theta=0^\circ$.}
\label{Fig_S4}
\end{figure}

The Fermi level $E_{\rm F}$ is determined with a precision of about 20\,meV on the Au substrate next to the MnPS$_3$ flakes (Fig.~\ref{Fig_S4}a). The detected photoelectrons from $E_{\rm F}$  are directly related to the work function of the analyzer
$\phi^\text{A}$ via their kinetic energy $E_\text{kin, F}^\text{A}$ reading 
\begin{equation}
    E_\text{kin,F}^\text{A}+\phi^\text{A}=h\nu
\end{equation}
with photon energy $h\nu$. This leads to $\phi^\text{A}=4.6$\,eV (Fig.~\ref{Fig_S4}a).
To find the work function $\phi$ of MnPS$_3$, the onset energy of secondary photoelectrons $E_\text{kin, min}^\text{A}(k_\parallel)$ is employed (Fig.~\ref{Fig_S4}b). We use the free electron parabola in vacuum for $k_z=0/$\AA\, to fit the onset by
\begin{equation}
\label{eq:S2}
    E_\text{kin, min}^\text{A}(k_\parallel) = \frac{\hbar^2}{2m} {k_\parallel}^2 + \phi - \phi^\text{A}
\end{equation}
 ($m$: free electron mass) \cite{Sobota2021}.
Practically, $E_\text{kin, min}^\text{A}$ is taken as the energy with strongest positive slope of $I(E_\text{kin}^\text{A})$ in the region $E_\text{kin}^\text{A}\in\left[0\,{\rm eV}, 3\,{\rm eV}\right]$ for each $k_\parallel$. 
The fitted free electron parabola exhibits a minimum at the kinetic energy $E_{\rm kin, min}^{\rm A} (k_\parallel=0)=0.85$\,eV resulting in $\phi = 5.45$\,eV.

\subsection{Determination of the Brillouin zone orientation}

A parabolic fit of the secondary emission as displayed in Fig.~\ref{Fig_S4}b directly reveals the $k_\parallel=0.0/$\AA\, point as the minimum position. 
Such a fit in two perpendicular directions, thus, reveals the position of
the $\overline{\Gamma}$ point. The curvature of the fitted parabola gives a scale factor for $k_\parallel$ in the two different directions. Hence, the size and center of the hexagonal Brillouin zone projection to the surface is fixed. To determine the angular orientation, we use the P 3p band at low energy that features a clear minimum at $\overline{\Gamma}$ (\MM{Fig.~2d, 4b, c}, main text). The minimum is also visible in adjacent Brillouin zones such that the deduced direction  between such minima is the $\overline{\rm M}\overline{\Gamma}\overline{\rm M}$ direction. Subsequently, we crosscheck at various energies that the  ARPES features are compatible with the deduced Brillouin zone. We estimate the errors in the $k_\parallel$ scaling factor to be $\pm\, 0.9$\,\%, in the $\overline{\Gamma}$ determination to be $\pm\, 0.006/$\AA\, and in the azimuthal orientation of the Brillouin zone to be $\pm\, 0.5^\circ$.

\FloatBarrier
\subsection{Prominent ARPES features of Au substrate}
The energy distribution curve (EDC) of the Au substrate for $\theta=0^\circ$ is shown in Fig.~\ref{Fig_S5}c. The most promiment peak at $E-E_{\rm F} \approx -6\,$eV can be attributed to a Au 5d state \cite{Rangel2012}.
This state is also visible in the monolayer measurements of MnPS$_3$ as a flat band  (\MM{Fig.~1i}, main text).


\subsection{Curvature of ARPES data}

\begin{figure*}[thb]
\centering
\includegraphics[width=0.8\textwidth]{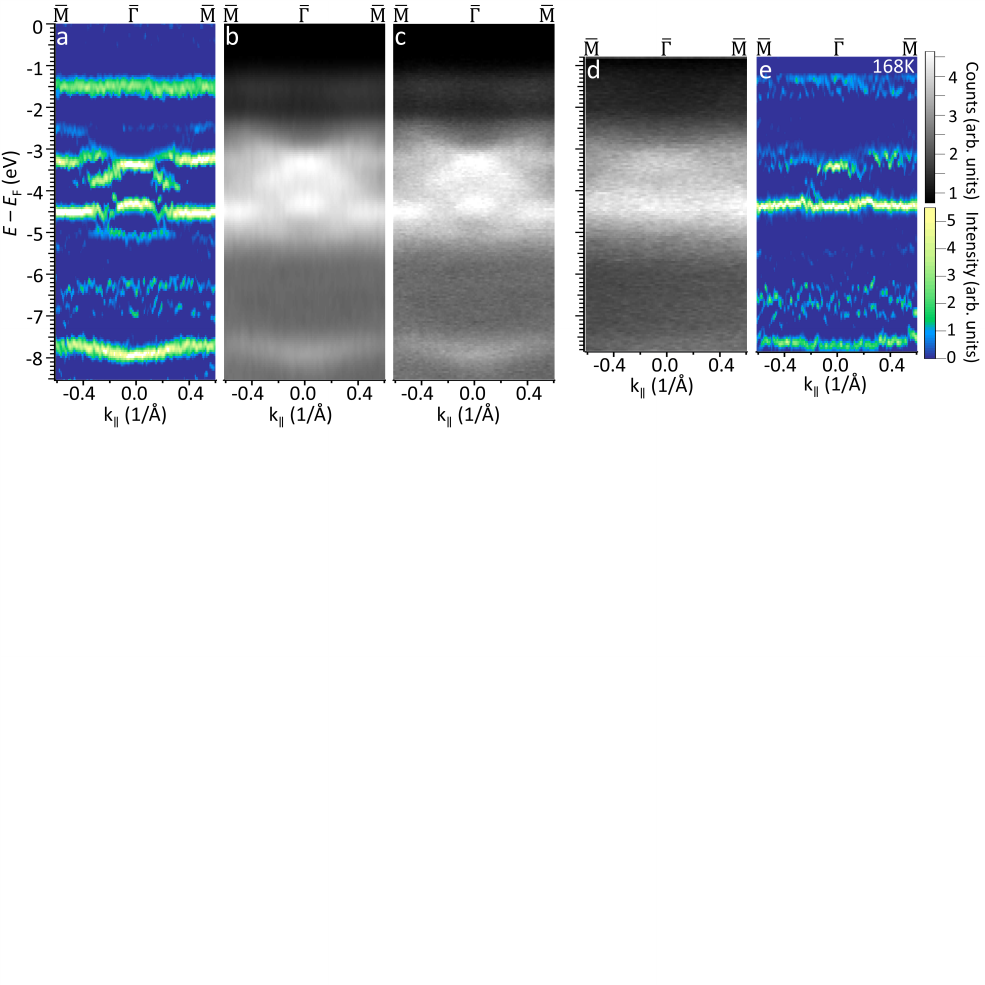}
\vspace{-8.4cm}
\caption{{\bf ARPES data and curvature plots},  MnPS$_3$, 62\,layers, $h\nu=50$\,eV.
(a) Curvature plot of the data in b using $a_0 = 0.05$. (b) Smoothed ARPES intensity of the raw data from c by $N=2$ convolutions with a Gaussian of width $\sigma = 95$\,meV.
(c) Raw ARPES data at $T=43$\,K, i.e below the Néel temperature $T_{\rm N}=78 $\,K.
(d) Raw ARPES data of the same area as in c at $T=168$\,K.
(e) Curvature plot of d using $a_0 = 0.05$ after smoothing with $N = 2$, $\sigma = 95$\,meV.
}
\label{Fig_S5}
\end{figure*}

To visualize the features of the photoemission intensity in ARPES, the curvature with respect to energy is calculated.
As the raw spectrum contains an increasing background for larger binding energy, this simplifies the identification of bands by highlighting local maxima in the energy distribution curves (EDCs) \cite{Peng2020}.
Before calculating the curvature, the measured intensity as displayed in Fig. \ref{Fig_S5}c is smoothed along energy by up to $2$ subsequent convolutions with a Gaussian of the same energy width.
The standard deviation $\sigma$ of this Gaussian and the number of smoothing steps $N$ is  tabulated in table~\ref{Tab_S1}. The chosen $\sigma$ is always below $100\,$meV to avoid flattening of features.
After smoothing, the curvature $C(E)$ with respect to energy $E$ is determined: 
\begin{equation}
    C(E) = \frac{I''(E)}{(C_0 + I'(E)^2)^{3/2}}\,\hspace{3mm}{\rm with}\,\hspace{3mm}  C_0= a_0 |I'|_\text{max}^2
    \label{eq:S1}
\end{equation}  
Here, $I = I(E,k_\parallel)$ is the smoothed intensity of the ARPES data at kinetic energy $E$ and in-plane momentum $k_\parallel$, $I'$ and $I''$ are the first and second partial derivatives with respect to $E$, respectively, and $C_0$  
is a free parameter that equalizes the relative intensity of visible bands by artificially broadening the most sharp ones \cite{Peng2020}.
Technically, we relate  $C_0$ to $|I'|_\text{max}^2$, the largest derivative with respect to $E$ within the smoothed $I(E,k_\parallel)$ plot, where $a_0$ is a tuning parameter as displayed in table~\ref{Tab_S1}.
For adequately discriminating bands from noise, we always compare the curvature plots with the raw and the smoothed ARPES data as exemplarily shown in Fig. \ref{Fig_S5}.
 ARPES spectra that are directly compared are always treated with identical smoothing and curvature parameters.

\begin{table}[b]
\centering
\begin{tabular}{|c c c c|}
 \hline
 Figure & $N$ & ~$\sigma$(meV)~ & $a_0$ \\
 \hline
 \MM{1f,i} & 2 & 75 & 0.05 \\
 \hline
 \MM{2d,e,g} & 2 & 95 & 0.05 \\ 
 \hline
 \MM{4b--c} & 2 & 95 & 0.05 \\
 \hline
 \MM{4e--h} & 1 & 90 & 0.01 \\
 \hline
 \ref{Fig_S5}a,e & 2 & 95 & 0.05 \\
 \hline
 \ref{Fig_S6}a-c & 2 & 95 & 0.05 \\
 \hline
 \ref{Fig_S7}a-d & 1 & 90 & 0.01 \\
 \hline
 \ref{Fig_S10}a-f & 2 & 90 & 0.025 \\
 \hline
 \ref{Fig_S13}a,c & 2 & 95 & 0.05 \\
 \hline
 \ref{Fig_S13}d & 2 & 50 & 0.1 \\
 \hline
\end{tabular}
\caption{Parameters used to obtain the curvature plots from the raw ARPES data. The raw data is firstly smoothed by $N$ convolutions with a Gaussian of standard deviation $\sigma$. The parameter $a_0$ equalizes the intensity of different bands that appear in the curvature plots  using eq.~(\ref{eq:S1}) \cite{Peng2020}. 
}
\label{Tab_S1}
\end{table}

\subsection{Resonant excitation}
\label{sub:res}
 \begin{figure}[t]
\centering
\includegraphics[width=\textwidth/2]{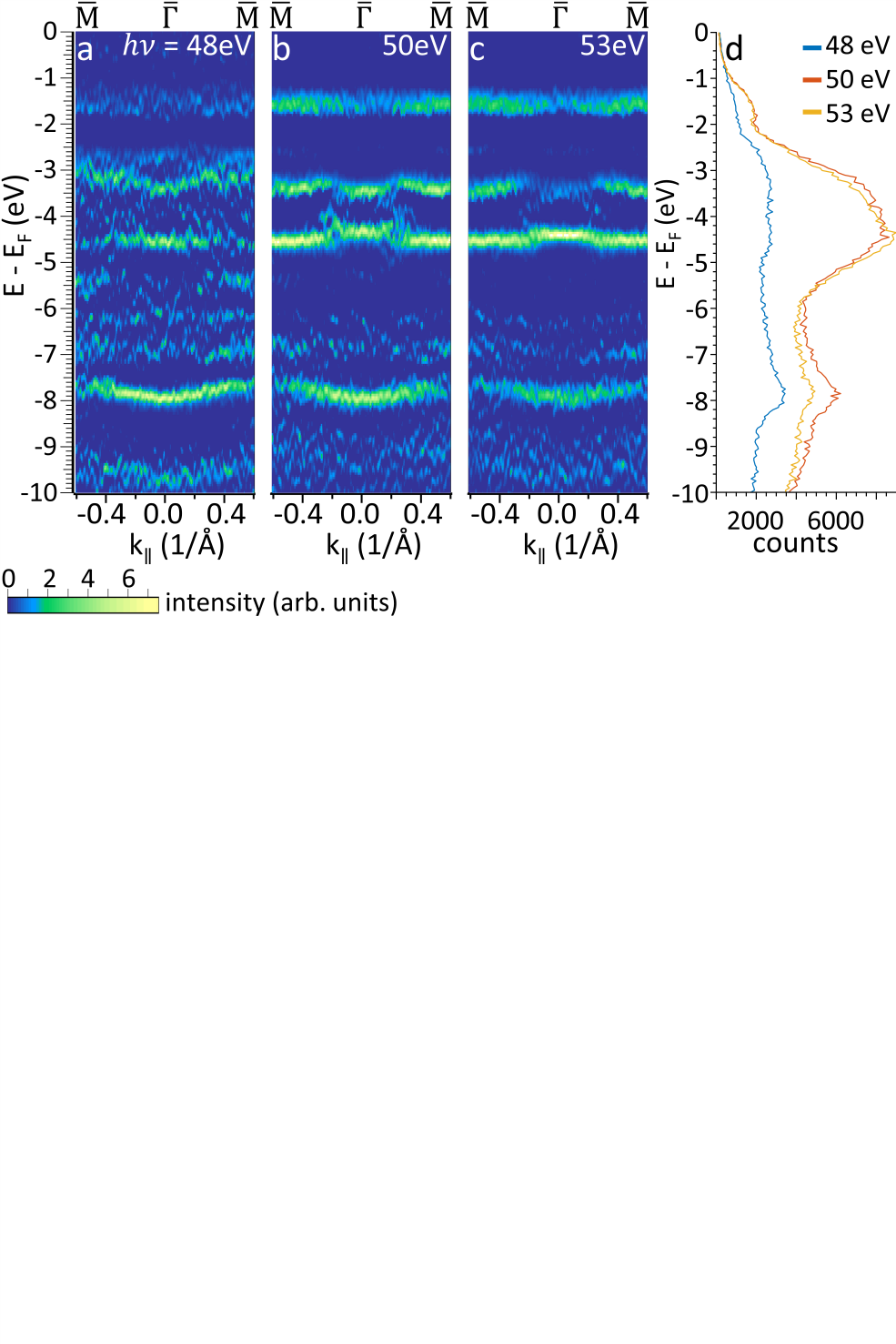}
\vspace{-8.0cm}
\caption{{\bf Resonant excitation},
13\,layers, $T = 42$\,K.
(a)--(c) Curvature plots of ARPES data for the indicated $h\nu$   below (a), at (b) and slightly above (c) resonance. The intensity scale is the same as in Fig.~\ref{Fig_S5}.
(d) EDCs at $\overline{\Gamma}$ for the same $h\nu$ as a--c. A strong enhancement of intensity is observed at resonance for $ E-E_{\rm F} \in \left[ -6\,{\rm eV},-2\,{\rm eV} \right]$. The same integration time has been used for all three $h\nu$.
}
\label{Fig_S6}
\end{figure}

Figure~\ref{Fig_S6} shows a strong enhancement of some band intensities by changing $h\nu$. 
At $h\nu=50-53$\,eV,  the intensity of bands with dominating Mn 3d 
or S 3p character ($E - E_{\rm F} \in \left[-5.5\,\text{eV}, -1\,\text{eV}\right]$) is much larger than at $h\nu=48$\,eV. This does not apply for the P 3p band at $E - E_{\rm F}\approx -7.8$\,eV. The EDCs at $\overline{\Gamma}$ (Fig.~\ref{Fig_S6}d) concomitantly show a strong enhancement of band intensity (peak height above background) for $E - E_{\rm F} \in \left[-6\,\text{eV}, -2\,\text{eV}\right]$.   
We conjecture that the enhancement is caused by a Mn 3p\,$\rightarrow$\,3d resonance. Such resonant enhancement of photoelectron intensity by an induced Auger like process  has previously been observed for a 2p\,$\rightarrow$\,3d transition of MnPS$_3$ \cite{Kamata1997} and for a 3p\,$\rightarrow$\,3d transition of FePS$_3$, exhibiting, both, the initial transitions into a multiplet of empty 3d levels \cite{Choi1994}. The Mn 3p level indeed is at $E-E_{\rm F}=-48.5$\,eV (Fig.~\ref{Fig_S2}b) compatible with resonant excitations into the empty Mn 3d levels at slightly higher energy. 
Moreover, the orbital decomposition of the DFT+U data (Fig.~\ref{Fig_S11}) reveals that all the bands with enhanced intensity have a significant Mn 3d contribution that, in addition, is strongest for the bands at $E-E_{\rm F}\in \left[-5\,{\rm eV}, -4\,{\rm eV} \right]$, where we see the strongest enhancement of the EDCs. Consequently, it is favorable to use $h\nu =50-53$\,eV for probing Mn related bands.



\subsection{Selection rules}
\label{sub:Selection}

The visibility of bands in ARPES depends also on their orbital character via matrix elements of the photoelectron excitation. 
For the sake of simplicity, we assume the sudden dipole approximation, plane waves as final vacuum states without exponential decay  into the crystal, and negligible photon wave vector $\textbf{\textit{k}}_{h\nu} << \textbf{\textit{k}}_{\rm f}$ with wave vector of the emitted photoelectrons $\textbf{\textit{k}}_{\rm f}$. The matrix element M$_{\textbf{\textit{k}}_{\rm f} {\textbf{\textit{k}}}}$ between initial state at wave vector $\textbf{\textit{k}}$ and final state in vacuum then reads (based on eq.~(20) in \cite{Moser2017})

 %
\begin{equation}
    M_{\textbf{\textit{k}}_{\rm f} \textbf{\textit{k}}} \sim \textbf{\textit{E}} \cdot \textbf{\textit{k}}_{\rm f} \braket{e^{i \textbf{\textit{k}}_{\rm f}\textbf{\textit{r}}}}{\psi_{\kappa\,, \textbf{\textit{R}}=\textbf{\textit{0}}}(\textbf{\textit{r}})} ,
\label{eq:S0}
\end{equation}
where $\textbf{\textit{E}}$ is the polarization vector of the incoming light and $\psi_{\kappa\,, \textbf{\textit{R}}=\textbf{\textit{0}}}(\textbf{\textit{r}})$ is the Wannier function of the initial state centered at position $\textbf{\textit{R}}=\textbf{\textit{0}}$ of the band with index $\kappa$. 
The Wannier function can be written as a sum of atomic orbitals revealing selection rules for the orbital contributions to each band \cite{Moser2017}. As shown in \cite{Moser2017}, 
linear polarized light with polarization parallel to the incident plane ($\pi$-polarization) selectively excites s,  p$_{\rm z}$, and d$_{\rm z^2}$ orbitals, if the photoelectrons are emitted close to normal emission. 
This condition applies for the probed photoelectrons from the first Brillouin zone ($k_\parallel < 0.7\,$\AA), since we use a large $h\nu=50$\,eV implying $\left|\textbf{\textit{k}}_{\rm f}\right| \approx 3.6/$\AA.
Consequently, we compare the experimental data with the calculated band structure after projection to these orbitals. The contribution of s orbitals turns out to be largely irrelevant for the probed energy regime such that it is not considered explicitly.   



\section{Computational Details}
The bandstructure calculations are performed within the framework of the Density Functional Theory (DFT) implemented in VASP software \cite{Kresse1993,Kresse1996}. The PAW pseudopotentials \cite{Blochl1994,Kresse1999} and Perdew-Burke-Ernzerhof (PBE) exchange-correlation functional are used \cite{Perdew1996}. 
The kinetic energy cutoff for the plane-wave expansion is set to 500\,eV. A $\mathbf{k}$-mesh of $10\times10\times 9$ ($10\times10\times 2$ and $10\times6\times 2$) is taken to sample an irreducible first BZ of the primitive bulk cell (primitive and rectangular planar cell of a monolayer, respectively). We employ the GGA+U formalism proposed by Dudarev \cite{Dudarev1998} to properly account for on-site Coulomb repulsion between $3d$ electrons of Mn ions using effective Hubbard $U$ parameters. The lattice parameters and position of the atoms are fully optimized within this approach assuming the magnetic state of AFM-N\'{e}el type as found experimentally \cite{Kurosawa1983,Brec1986}. The convergence criteria for the energy and force are set to 10$^{-5}$\,eV and 10$^{-3}$\,eV/\AA, respectively. In the case of the monolayer, we added 20\,{\AA} of vacuum to avoid the spurious interaction between periodic replicas. The non-local nature of dispersive forces, crucial for layered materials 
\cite{Tawfik2018,Birowska2011,Birowska2021b} are accounted for within the semi-empirical Grimme approach \cite{Grimme2006} with a D3 parametrization \cite{Grimme2010}. The paramagnetic phase was mimicked by a $4\times4$ supercell with a random spin orientation resulting in zero magnetization. The nonmagnetic case (NM) was assumed neglecting the spin degrees of freedom performing non-spin polarized DFT calculations. 
\MM{Ultimately, the quantitative comparison with ARPES data should be done at} the level of spectral functions, which can be obtained for instance within LDA+DMFT approach \cite{Nekrasov2021} \MM{employing the one-step model \cite{Pendry1976, Minar2011, Minar2013}.} 
However, such calculations  are computationally demanding. 
Hence, we limit our approach to the 
comparison of energy positions of the electronic bands from the DFT+U approach with the ARPES data.

\section{Evolution of the band structure with temperature}
\label{sec:tempdep}

\begin{figure}[t]
\centering
\includegraphics[width=\textwidth/2]{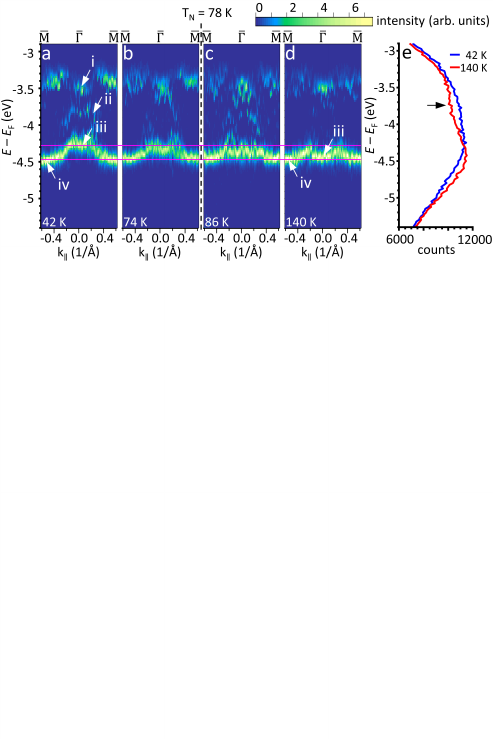}
\vspace{-9.2cm}
\caption{{\bf Evolution of bands across the Neél temperature}, h$\nu = 50$\,eV. Plane of incident light with respect to $\overline{\Gamma}\overline{\rm M}$: $37^\circ$.
(a)--(b) Curvature plots of ARPES data below $T_{\rm N} = 78$\,K, 13\,layers. The marked structures (i--iv) are discussed in the text. 
(c)--(d) Same as a--b, but above $T_{\rm N} = 78$\,K. 
The horizontal lines across a--d are guides to the eye. 
The intensity scale for a--d is the same as in Fig.~\ref{Fig_S5}.
(e) Raw EDCs recorded at $\overline{\Gamma}$ for $T$ as in a, d. The peak shift from $-4.3$\,eV to $-4.5$\,eV with increasing $T$
matches the downwards shift of the maximum of the Mn d$_{\rm z^2}$ band at $\Gamma$ in a--d (feature iii). The additional decrease in intensity at $-3.75\,$eV (arrow) is related to the disappearance of the feature ii.
}
\label{Fig_S7}
\end{figure}

\begin{figure*}[bht]
\centering
\includegraphics[width=\textwidth]{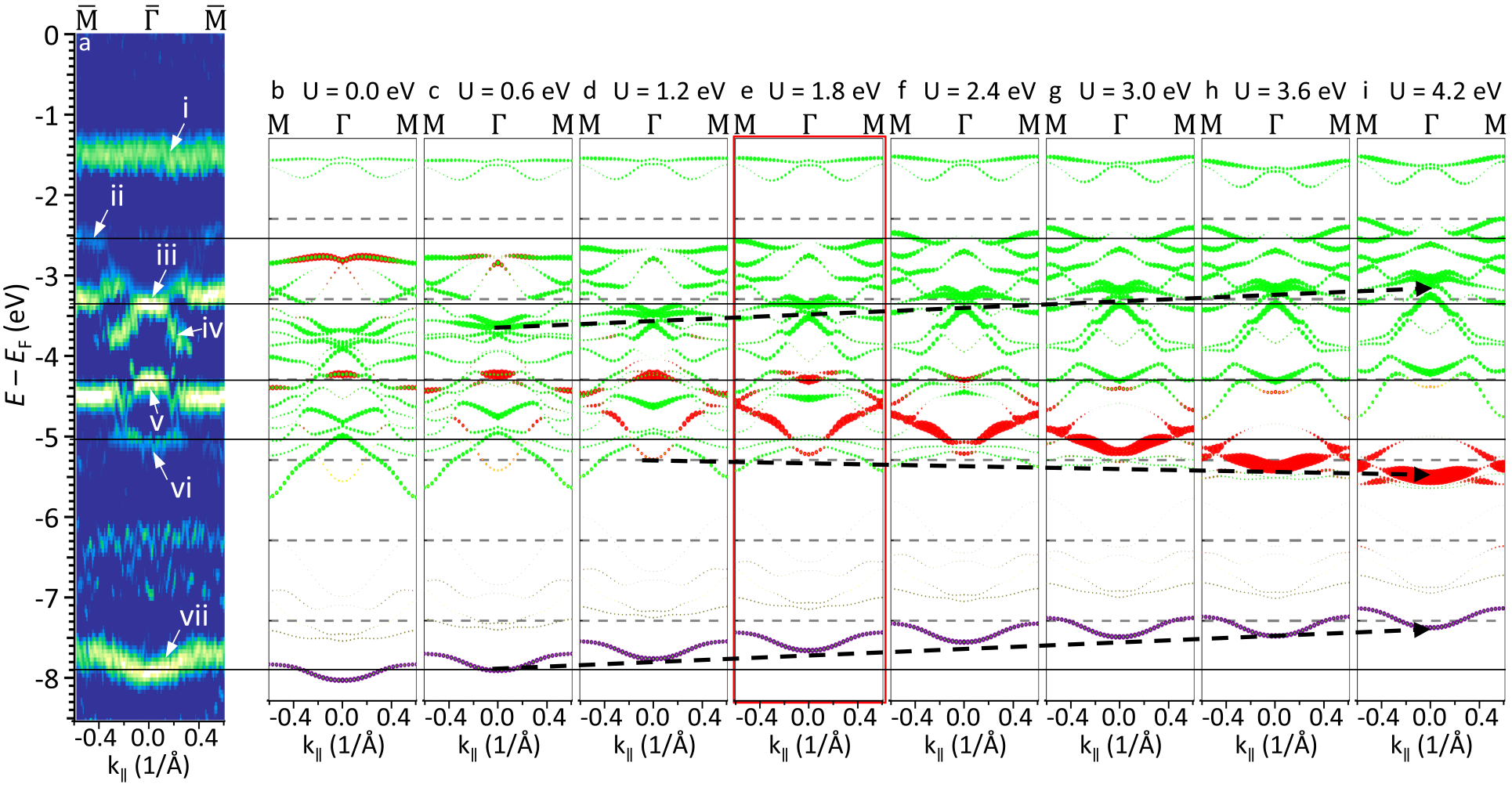}
\vspace{-0.8cm}
\caption{(a) Curvature plot of ARPES data, 62\,layers, $T=43$\,K, $h\nu=50$\,eV (same as \MM{Fig.~2d}, main text). 
The symbols i--vii are discussed in the text. (b)--(i) DFT +U data at various $U$ as marked on top. The \MM{dashed} arrows indicate trends of some features that are used to select $U$.  The five horizontal black \MM{solid} lines mark features in experiment to be matched by theory. The red box highlights the selected most favorable $U=1.8$\,eV, $k_z = 0.48/\,$\AA.
}
\label{Fig_S8}
\end{figure*}

To pinpoint the band changes across the magnetic transition at $T_{\rm N} = 78\,$K, ARPES measurements at various temperatures are compared in Fig.~\ref{Fig_S7}.
We employ h$\nu = 50$\,eV, at resonance of the Mn bands, strengthening the intensity of these bands (Fig.~\ref{Fig_S6}).
For all temperatures, the same integration time is used enabling a quantitative comparison, in particular, of the EDCs (Fig.~\ref{Fig_S7}e).
Some prominent features are marked with arrows labelled i--iv. 
They are attributed to the S 3p$_{\rm z}$ bands (i, ii) and Mn 3d$_{\rm z^2}$ bands (iii, iv) according to the comparison with the DFT+U data in \MM{Fig.~2}, main text.
The  characteristic 'double-hump' structure (i) with tails towards lower energy (ii) of the S 3p$_{\rm z}$ bands only slightly changes across $T_{\rm N}$.
These changes have already been discussed with respect to  \MM{Fig.~4}, main text. The relatively weak intensity of these bands, in particular at larger $T$, makes it difficult to pinpoint the changes. However, a pronounced change is observed in the raw data EDC at $\overline{\Gamma}$ (Fig.~\ref{Fig_S7}e). A dip appears at $E-E_{\rm F}\approx -3.75$\,eV (arrow) above $T_{\rm N}$. This feature is reproducibly observed in several cooldowns for flakes of different thickness \MM{and is in line with the disappearance of the tail structures}. 

In contrast, the Mn 3d band at slightly lower energy (iii, iv) changes unambiguously across $T_{\rm N}$.
While below $T_{\rm N}$, the band maximum appears at $\overline{\Gamma}$ (iii) and the total dispersion amounts to $\sim 0.2$\,eV, the band flattens above $T_{\rm N}$ showing two maxima at $k_\parallel\approx \pm 0.25/$\,\AA. The change of the bands does not appear abruptly at $T_{\rm N}$, but rather continuously across the phase transition, however, with a relatively strong change around $T_{\rm N}$ (Fig.~\ref{Fig_S7}b--c). This is in line with the largely two-dimensional magnetic interactions in MnPS$_3$ as deduced from inelastic neutron diffraction \cite{Wildes1998}. 

The split band observed in Fig.~2b, main text, is not visible in these data, most likely since the angle of incident light with respect to the $\overline{\Gamma}\overline{\rm M}$ direction is relatively rotated by $30^\circ$. 


%
\section{Adjusting $U$ and $k_z$ of the DFT+U calculations to the ARPES data}
\label{sec:U_kz_select}

\subsection{Adjusting $U$}
\label{sub:select_U}

\begin{figure*}[tbh]
\centering
\includegraphics[width=\textwidth]{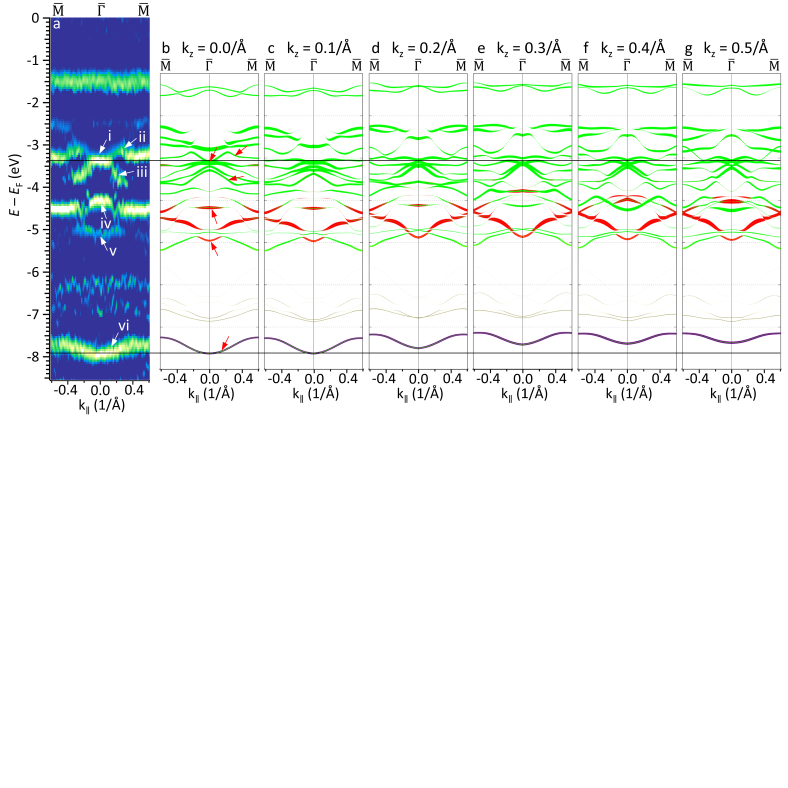}
\vspace{-9.2cm}
\caption{(a) Curvature plot of ARPES data, 62\,layers, $h\nu=50$\,eV, $T = 43$\,K (same as \MM{Fig.~2d}, main text). The intensity scale is the same as in Fig.~\ref{Fig_S5}. The symbols i--vii are discussed in the text. (b)--(g) DFT +U (U = 1.8\,eV) data at various $k_z$ as marked on top.  The arrows in b highlight the features that we identify with the experimental ones marked i-vi. The horizontal black lines mark the features i and iv in a.}
\label{Fig_S9}
\end{figure*}

For correspondence between theoretical calculations and experimental data, $U$ and $k_z$ have to be adapted.  
We firstly discuss the $U$ adaption.
Figure~\ref{Fig_S8} shows the calculated band structure with varying $U$ in comparison with ARPES data. 
Three characteristic trends with $U$ are marked by dashed arrows. They are well suited for a \MM{$U$} selection, since 
the band features \MM{there}  move continuously with  \MM{$U$, either downwards or upwards}. \MM{Moreover}, the features \MM{are apparent} in the experimental data. The other marked features   in Fig.~\ref{Fig_S8}a are \MM{additionally} used for \MM{optimizing the $U$ selection}.

Firstly, we adapt the calculated upper bands with strong S 3p$_{\rm z}$ character, that barely change with $U$, \MM{as expected, since the $U$ value is applied to the 3d states of Mn. 
These bands can be assigned}
to the flat band marked (i) in the ARPES data.
It is the highest energy valence band in DFT +U and the highest energy band below $E_{\rm F}$ observed in ARPES.
Subsequently, the relatively flat band (ii)  with strongest intensity close to $\overline{\rm M}$ matches most favorably to the calculations for $U=2.4\,$eV (see horizontal black line).
Comparing next the dispersing bands marked  iii-iv in the ARPES data, that have been previously attributed to a strong S 3p$_{\rm z}$ character, we find a very similar structure in the DFT+U data that shifts upwards with increasing $U$ (upper dashed arrow) and matches most favorably for $U=1.8$\,eV.
Here, the corresponding S 3p$_{\rm z}$-type band structure also expands in energy with increasing $U$ indicating increased dispersion. It is apparent that this dispersion is too weak (strong) at much lower (higher) $U$. We also find discrepancies at the selected $U$  as, e.g., the 'double-hump' structure is more distant in $k_\parallel$  for the experimental data than for the calculated band structure, but this could not be improved by changing $U$. We attribute this discrepancy to the fact that DFT+U is still an approximation.

Next, we compare the features v--vi of the ARPES data that have been previously attributed to a dominating Mn d character. The upper flat band (iv) shifts downwards with $U$ and matches best for \MM{$U=1.8\,$eV}. While its center at $\overline{\Gamma}$
is at the right energy for $U=1.8$\,eV, the downwards dispersion towards $\overline{\rm M}$ is better matched at $U=1.2$\,eV.
The lower flat band (vi) is identified with the lowest red Mn d band in the calculation that is at rather similar energy as in the experiment for $U=2.4$\,eV. These bands are shifting generally downwards with $U$, but also change their dispersion and hybridization with nearby S 3p bands. 

Lastly, the P 3p$_{\rm z}$ band (vii)  monotonically increases for increasing $U$, matching best for $U=0.6$\,eV. Here, 
however, the wrongly chosen $k_z$ for the comparison is relevant such that the correct match at $k_z=0/$\,\AA\, reveals $U=1.8$\,eV to be adequate (Fig.~\ref{Fig_S9}).

Obviously, there is a remaining ambiguity to attribute $U$, but most features match for $U=1.8$\,eV such that we use this value for other detailed comparisons. 

\FloatBarrier
\subsection{Relating $k_z$ to $h\nu$}
\label{sub:kzhnu}


With the selected $U$, we next adapt
$k_z$. Fig.~\ref{Fig_S9} compares the ARPES data with labeled features i--vi to the calculated band structure at different $k_z$.
Starting with the P 3p$_{\rm z}$ band (vi), the best match of the dispersion and the absolute energy is found for $k_z = 0.0 - 0.1/$\AA\,, as mentioned in subsection~\ref{sub:select_U}.
The band features with most pronounced changes by $k_z$ belong to the S 3p bands (i--iii).
Here, the energy at $\overline{\Gamma}$ (i) is reproduced for all $k_z$, but the 'double-hump' (ii) is only found with similar strength for $k_z = 0.0/$\AA.
The dispersive bands below (iii) are also quite similar in experiment and calculation for $k_z = 0.0/$\AA. 
For the Mn 3d bands (iv-v), $k_z=0.0-0.2/$\AA\, as well as $k_z=0.5/$\AA\, match reasonably. Here, one should keep in mind that the resonance condition preferentially excites the Mn 3d type states (red), if compared to the S 3p type states (green).
\MM{Summarizing}, $k_z=0.0/$\AA\, bears by far the strongest agreement between calculation and ARPES at $h\nu=50$\,eV.


In order to relate other photon energies  to $k_z$ values, we use the free electron model of final states with inner potential $V_0$ 
\begin{equation}
\label{eq:S4}
E_{\rm final}(\textbf{\textit{k}}) = \frac{\hbar^2\textbf{\textit k}^2}{2m}-V_0,   
\end{equation}
where $E_{\rm final}$ is the final state energy and $\textbf{\textit k}$ is the wave vector of the photoelectron inside the crystal. 

To determine $V_0$, we use the ARPES \MM{intensity at energy $E_\text{kin}^\text{A}$ for} $h\nu=50$\,eV and $\theta = 0^\circ$, i.e. at $k_z=0.0/$\AA\, and $k_\parallel=0.0/$\AA, such that an adequate rearrangement of eq.~(\ref{eq:S2}) with eq.(\ref{eq:S4}) reads \cite{Sobota2021}
\begin{equation}
V_0 = \frac{\hbar^2}{2m} (
n G_\perp)^2 - E_\text{kin}^\text{A} + \phi - \phi^\text{A}
\label{Eq_3}
\end{equation}
where $G_\perp = 0.9686/$\AA\, is the reciprocal lattice vector of MnPS$_3$ perpendicular to the surface \cite{Brec1986}, $n$ is an integer, $E_\text{kin}^\text{A}$ is the measured kinetic energy by the analyzer, and $\phi$ ($\phi^\text{A}$) is the work function of the sample (analyzer). \MM{The term $(n G_\perp)^2$ bears an  uncertainty in determining $V_0$  by the possible free choice of $n$.}

We use the Mn 3d band at $E-E_{\rm F}=-4.5$\,eV that matches the DFT+U calculation rather perfectly (feature \MM{iv} of Fig.~\ref{Fig_S9}). This results in $E_\text{kin}^\text{A} = h\nu - \phi^\text{A} + (E-E_\text{F})=40.9$\,eV. Inserting into eq.~(\ref{Eq_3}),
the lowest positive value of $V_0$ gets $V_0 \approx 17.1\,$eV for $n=4$ ($V_0 \approx 49.3\,$eV for $n=5$).
\MM{The value at $n=4$ is very reasonable}  for an inner potential \cite{zhang2022}, that we consequently use to estimate $k_z$ 
for all other photon energies according to
(e.g. eq.~(9) of \cite{Sobota2021})
\begin{equation}
\hbar k_z = \sqrt{2m(E_\text{kin}^\text{vac} \cos^2(\theta)+V_0)}
\label{Eq_2}
\end{equation}
with the kinetic energy in vacuum 
$$  E_\text{kin}^\text{vac} = E_\text{kin}^\text{A}+ \phi^\text{A} - \phi.$$

\FloatBarrier
\section{$k_z$ dispersion}
\label{sec:kzdisp}
 \begin{figure*}[t]
\centering
\includegraphics[width=0.75\textwidth]{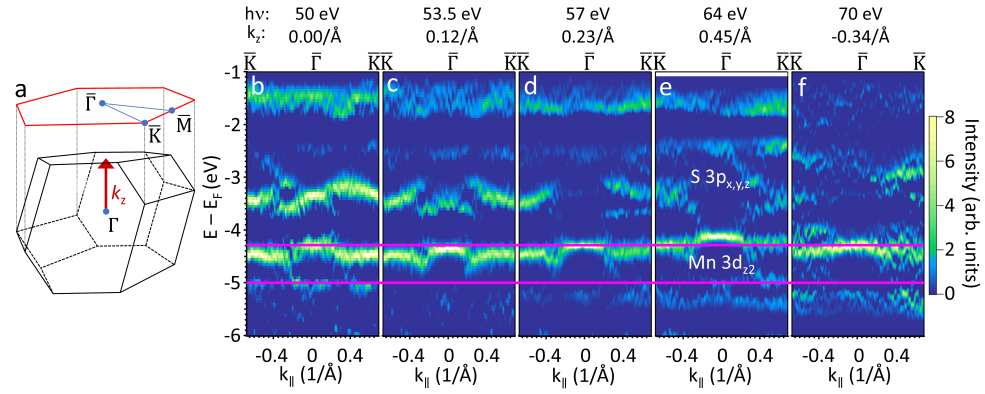}
\vspace{-0.5cm}
\caption{{\bf $k_z$ dispersion}
(a) 3D Brillouin zone of MnPS$_3$ (black) with the $k_z$ direction perpendicular to the cleavage plane indicated (red arrow). The Brillouin zone boundary is at \MM{$k_z=0.48/$\AA}. The 2D projection to the surface is drawn on top (red hexagon) with the main directions marked. 
(b)--(f) ARPES curvature intensity along $\overline{\rm K}\overline{\Gamma}\overline{\rm K}$ for various $h\nu$ as marked on top, $T=43$\,K, 62\,layers. The indicated $k_z$ values in the first Brillouin zone originate from a free electron final state model with deduced inner potential $V_0=17.1$\,eV (eq.~(\ref{Eq_2})), section~\ref{sec:U_kz_select}). 
The horizontal pink lines simplify the comparison between bands at different $k_z$. Orbitals are attributed according to \MM{Fig.~2}, main text. The increased noise for $h\nu=70$\,eV is likely caused by the fact that this $h\nu$ is far from the resonant Mn 3p\,$\rightarrow$\,3d excitation (Fig.~\ref{Fig_S6}) \cite{Kamata1997}.}
\label{Fig_S10}
\end{figure*}

Figure~\ref{Fig_S10} shows the $k_\parallel$ dispersion of the bands along $\overline{\rm K}\overline{\Gamma}\overline{\rm K}$ for varying $h\nu$.
We use the inner potential $V_0=17.1$\,eV  and eq.~(\ref{Eq_2}) (section~\ref{sec:U_kz_select}) to relate $h\nu$ to $k_z$ in the first Brillouin zone as marked on top. 
The $k_\parallel$ dispersion indeed changes with $k_z$. This includes the size of the band splitting around $\overline{\Gamma}$ of the Mn 3d$_{\rm z^2}$ band, which appears to increase between $h\nu=50$\,eV and $h\nu=64$\,eV (Brillouin zone boundary) as well as an upwards shift of the upper part of the Mn 3d band at $h\nu=64$\,eV (horizontal lines in Fig.~\ref{Fig_S10}b--f). Such an increased band splitting of the Mn 3d$_{\rm z^2}$ bands and an upshift of the maximum  is also observed in the DFT+U data around $k_z=0.3/${\AA} (Fig.~\ref{Fig_S9}). Hence, $k_z$ dispersion is found in both, ARPES data and DFT calculation albeit the interlayer exchange constant deduced from the measured magnon dispersion is negligible ($J_{\rm inter}\simeq 2\,\mu$eV \cite{Wildes1998}). 

%
\section{Orbital projection of the states from DFT+U}
\label{sec:orbital}

\begin{figure*}[htb]
\centering
\includegraphics[width=0.9\textwidth]{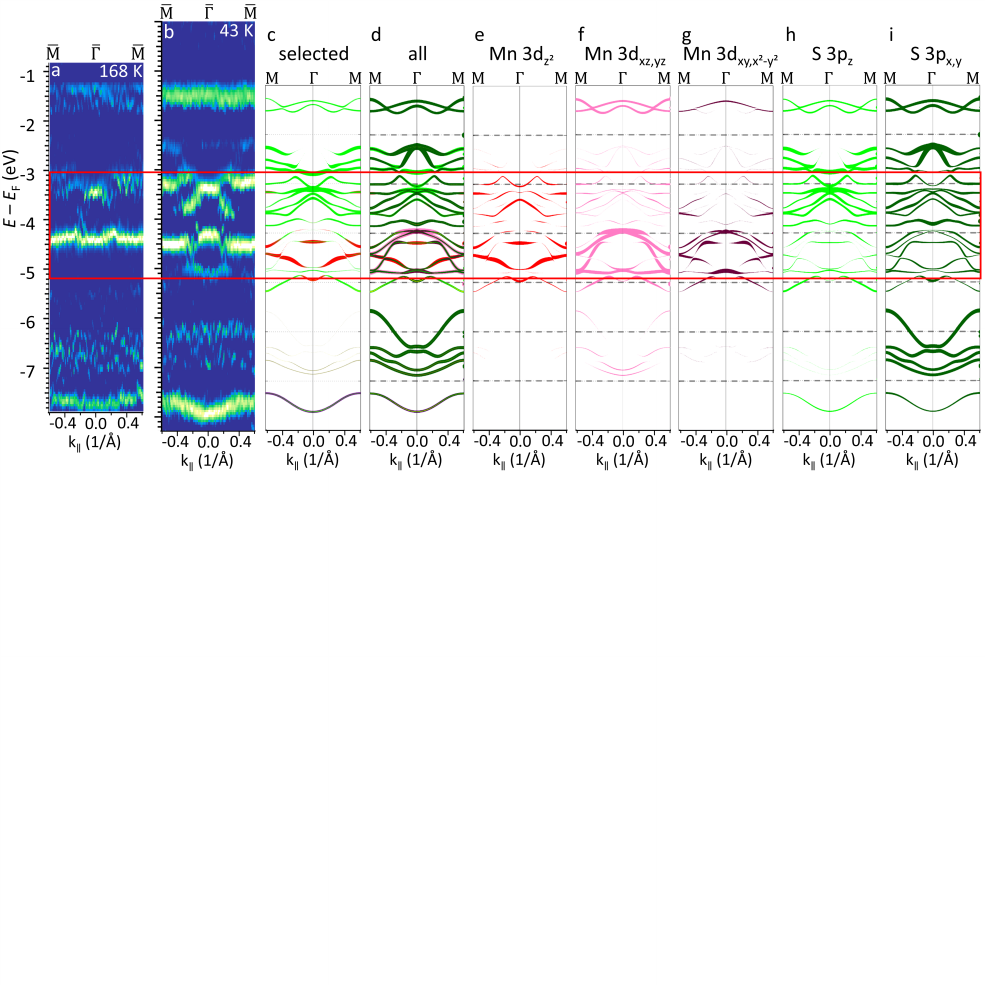}
\vspace{-8.8cm}
\caption{{\bf Orbital character of bands} (a) Curvature plot of ARPES data along $\overline{\rm M}\overline{\Gamma}\overline{\rm M}$, $T = 168$\,K ($T_{\rm N}=78$\,K), $h\nu =50$\,eV (same as \MM{Fig.~4c}, main text). (b) Same as (a) at  $T = 43$\,K (same as \MM{Fig.~2d}, main text). (c) DFT+U band structure along $\overline{\rm M}\overline{\Gamma}\overline{\rm M}$ ($U = 1.8$\,eV, $k_z = 0.0/$\AA) showing only bands with p$_{\rm z}$ (green) and d$_{\rm z^2}$ (red) orbital character. The size of the symbol corresponds to the strength of the displayed orbital character for each state. Only the stronger of the two contributions is displayed.  (d) DFT+U band structure ($U = 1.8$\,eV, $k_z = 0.0/$\AA) showing all orbital projections according to the color code of e--i. (e)--(i) Projections to single orbitals as marked on top. The red box highlights the bands with significant Mn 3d$_{\rm z^2}$ character.}
\label{Fig_S11}
\end{figure*}

Figure~\ref{Fig_S1} shows all relevant orbital contributions to the band structure separately, i.e. also the contributions that are barely probed in our ARPES experiments. It appears that the top valence band exhibits sizable contributions from S 3p$_{\rm xy}$ and \MM{weaker contributions} from the Mn d levels d$_{\rm xz}$ and d$_{\rm yz}$. The latter might explain that this band is also resonantly enhanced via the Mn 3p\,$\rightarrow$\,3d resonance (Fig.~\ref{Fig_S6}d) and that it shows significantly increased intensity below $T_{\rm N}$ (Fig.~\ref{Fig_S5}). Moreover, one observes
that most of the S 3p states are strongly mixed between p$_{\rm z}$ and p$_{\rm x,y}$. Only a few exceptions show a pure in-plane p$_{x,y}$ character, that interestingly always has negligible contributions from Mn 3d levels. Most importantly, the bands that we discussed in detail in the main text (red box in Fig.~\ref{Fig_S11}) are mixed  between all Mn 3d and all S 2p orbitals. This implies multifold interorbital interactions within the bands that change across $T_{\rm N}$ suggesting that both, direct exchange between the Mn atoms and superexchange across the S atoms contribute to the magnetism. Interestingly, within the probed energy window there are no states of pure Mn 3d character stressing \MM{again the relevance} of the S mediated exchange.  

%




\FloatBarrier
\section{Comparison between magnetic (AFM Neel)  and nonmagnetic (NM) case}
\label{sec:model}

\begin{figure*}[htb]
\centering
\includegraphics[width=\textwidth]{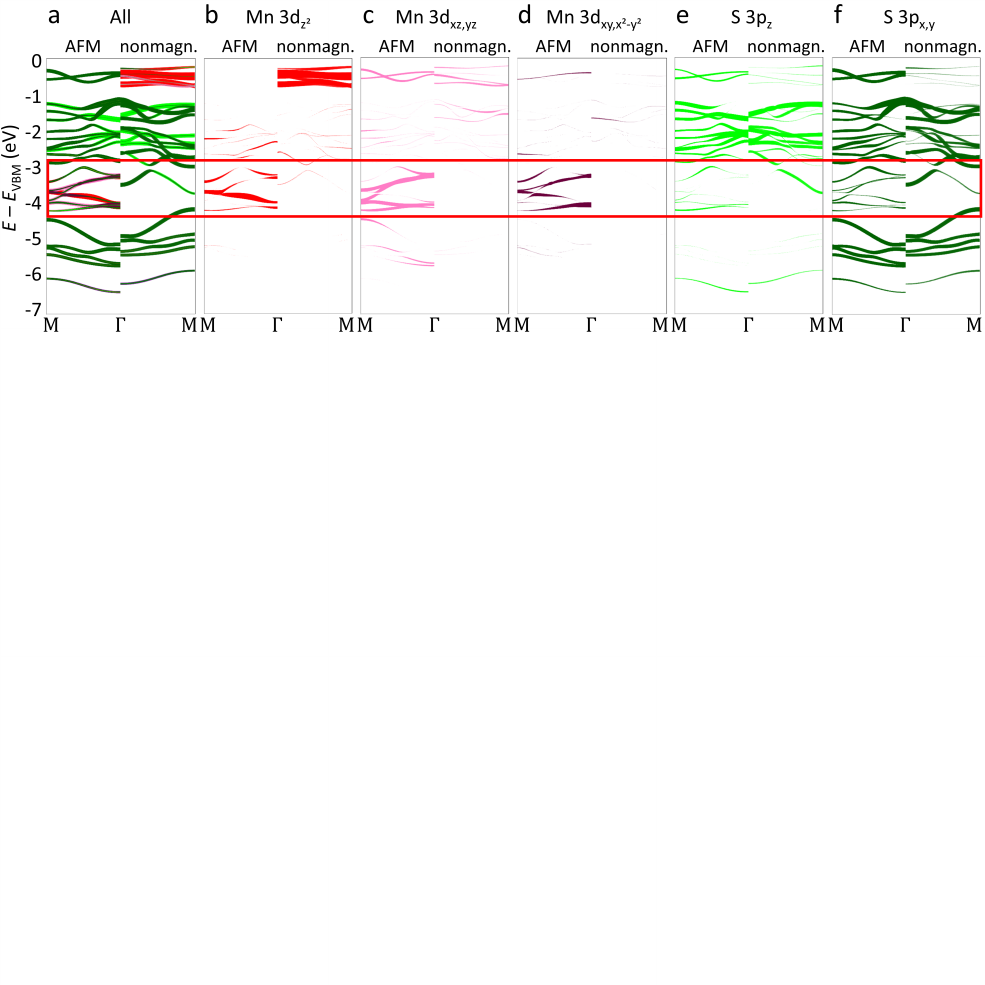}
\vspace{-12.4cm}
\caption{{\bf Orbital character of bands in magnetic (AFM Neel) and nonmagnetic calculation} (a) Calculated band structure with all bands using the orbital color code as  in b--f. The magnetic (nonmagnetic) band structure is displayed on the left (right). (b)-(f) Same as (a), but showing only the states with orbital chararacter as marked on top. The size of the symbols indicates the contribution of the corresponding orbital to the state. The red box highlights the states that change across $T_{\rm N}$ in the experiment.}
\label{Fig_S12}
\end{figure*}

\MM{Additional} nonmagnetic DFT calculation that neglect spin degrees of freedom  are presented and compared to the magnetic calculations obtained within DFT+U (Fig.~\ref{Fig_S12}). The resulting band structure exhibits bands of pure 3d$_{\rm z^2}$ character at $E-E_{\rm VBM}\simeq -0.5$\,eV ($E_{\rm VBM}$: valence band maximum). At slightly higher energies, one finds bands that originate purely from the other even 3d orbitals as expected for the trigonal anti-prismatic crystal field \MM{(not shown)}. These Mn 3d bands are all at higher energies than the occupied S 3p bands. Switching on the spin degrees of freedom (spin-polarized calculations), the S 3p bands are only moderately changed, while the Mn 3d$_{\rm z^2}$ bands split massively appearing now at $E-E_{\rm VBM}\simeq \pm 4$\,eV. This d level splitting also causes a large band gap of about 3\,eV \MM{similar to the experimentally observed one} for MnPS$_3$ \cite{Grasso1991,Du2015}. As \MM{also} visible in Fig.~\ref{Fig_S11}, the resulting occupied Mn 3d bands are far from having a pure orbital character. Instead, all bands within the red box ($E-E_{\rm F}\in \left[-3\,{\rm eV},-5\,{\rm eV}\right]$) are mixtures of all Mn 3d and  all S 3p orbitals. This is highlighted again in Fig.~\ref{Fig_S12} corroborating the strongly mixed character of Mn 3d levels at $E-E_{\rm VBM}\in \left[-3\,{\rm eV},-4.5\,{\rm eV}\right]$, but also that S 3p levels are split apart from their nonmagnetic energies in order to be involved in the exchange splitting of the Mn 3d levels. To a minor extent, that scenario also happens in the energy interval  $E-E_{\rm VBM}\in \left[-2\,{\rm eV},-3\,{\rm eV}\right]$, but there with a dominating S 3p character.
This matches the sketch in \MM{Fig.~3a--b}, main text, where two hybridized levels with preferential Mn 3d and S 3p character originate from the energy offset between the two original atomic levels.  

\FloatBarrier
\section{Influence of strain on the band dispersion for rough substrates}
\label{sec:strain}
 \begin{figure*}[htb]
\centering
\includegraphics[width=\textwidth]{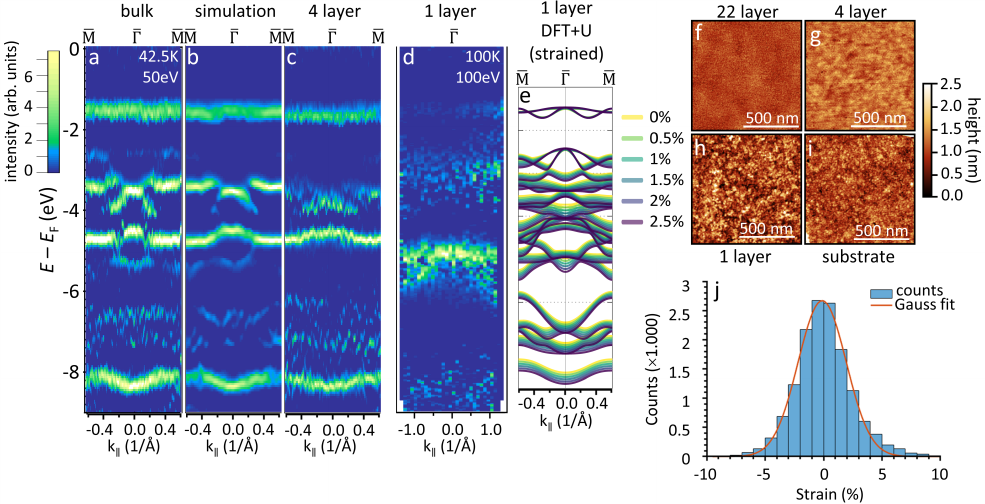}
\vspace{-0.7cm}
\caption{ {\bf Influence of strain \MM{on the} band structure} (a) Curvature plot of ARPES data along $\overline{\rm M}\overline{\Gamma}\overline{\rm M}$, $h\nu=50$\,eV, 22 layers, $T=43$\,K. 
(b) Artificial overlap of the data from (a) with multiple tilted surface normals ($k_\parallel=0/$\AA) according to a Gaussian distribution of tilt angles with a full width at half maximum of 2$^\circ$. This angular distribution features the measured surface orientations of the monolayer MnPS$_3$ displayed in h.
(c) Curvature plot of ARPES data along $\overline{\rm M}\overline{\Gamma}\overline{\rm M}$, $h\nu=50$\,eV, 4 layers, $T=43$\,K. 
(d) Curvature plot of ARPES data, $h\nu=100$\,eV, 1 layer, $T=100$\,K. The orientation of $k_\parallel$  could not be determined
due to a missing clear symmetry within the ($k_x$, $k_y$) data. The range on $k_\parallel$ is larger than in a--c.
(e) Calculated band structures along $\rm M\Gamma M$ for different isotropic tensile strains as marked, $U=1.2$\,eV, $k_z=0.05/$\,\AA. 
(f)--(h) Tapping mode AFMi images of MnPS$_3$ at different thicknesses as marked resulting in rms roughnesses: (f) $\sigma_{\rm rms}=1.3$\,\AA\,, 
 (g) $\sigma_{\rm rms}=1.8$\,\AA\,, (h) $\sigma_{\rm rms}=4.2$\,\AA. (i) Tapping mode AFMi image of the Au substrate, $\sigma_{\rm rms}=3.2$\,\AA.
(j) Strain distribution resulting from the analysis of the one-layer film \MM{shown in h}. 
A Gaussian fit (red line) with $\sigma=1$\,\% is added. 
}
\label{Fig_S13}
\end{figure*}

Figure~\ref{Fig_S13}a, c and d show ARPES data of MnPS$_3$ at different thicknesses obtained on a rougher substrate than for the sample of \MM{Fig.~1}, main text. The bands at a thickness of 4 layers (Fig.~\ref{Fig_S13}c) are less sharp than for the thicker film (Fig.~\ref{Fig_S13}a). This might be the reason that the energy splitting of the Mn d$_{\rm z^2}$ bands is not visible at 4 layers. There are also some minor changes in the dispersion, similar to the ones observed in the monolayer data of \MM{Fig. 1i}, main text. In contrast, the ARPES data of the rough monolayer in Fig.~\ref{Fig_S13}d changes significantly with respect to thicker layers. It appears much more blurry.  It was even not possible to determine the orientation of the Brillouin zone  due to missing clear rotational symmetries in the data. 

The strong blurring is most likely related to the corrugation of the thin MnPS$_3$ (Fig.~\ref{Fig_S13}h) on top of the  rough Au/SiO$_2$ substrate (Fig.~\ref{Fig_S13}i). The rms roughness $\sigma_{\rm rms}$ and the correlation length $\xi$ for the substrate and for the different flakes as well as for the ones presented in \MM{Fig.~1}, main text, are deduced from \MM{multiple} AFMi images and \MM{are} shown in Table~\ref{Tab_S2}. The substrate roughness is largely transferred to the monolayer, but already significantly reduced for the thickness of 4 layers and nearly absent at a thickness of 22 layers.
Thus, the ARPES spectrum of the thick MnPS$_3$ \MM{on the rough substrate} (Fig.~\ref{Fig_S13}a) is not \MM{very} different from the one obtained on the more smooth substrate (\MM{Fig.~1f}, main text).

\begin{table}[hb]
\centering
\begin{tabular}{|c | c  | c | c|}
 \hline
 material &  roughness &  correlation & Figure \\
 (thickness) & $\sigma_{\rm rms}$\,(nm)& length $\xi$\,(nm)& \\
 \hline
 Au/Ti/SiO$_2$/Si & 0.32 & 24 & \ref{Fig_S13}i \\ 
 \hline
 MnPS$_3$ (1 L) & 0.42 & 30 &\ref{Fig_S13}h \\
 \hline
  MnPS$_3$ (4 L) & 0.18 & 35 & \ref{Fig_S13}g\\
 \hline
MnPS$_3$ (22 L) & 0.13 & 69 & \ref{Fig_S13}f\\
 \hline\hline
 Au/Ti/SiO$_2$/Si & 0.2 & 45 & -- \\
 \hline
 MnPS$_3$ (1 L) & 0.3 & 35 & \MM{1e} \\
 \hline
MnPS$_3$ (14 L) & 0.24 & 63 & \MM{1d} \\
 \hline
 \end{tabular}
\caption{ Surface roughness and correlation length determined from the AFMi images shown in Fig.~\ref{Fig_S13} and \MM{Fig.~1}, main text. The AFMi image of the Au substrate for the MnPS$_3$ flakes of \MM{Fig.~1}, main text, is not shown.
}
\label{Tab_S2}
\end{table}

A possible explanation for the blurred ARPES of the monolayer 
is a local variation of the $\overline{\Gamma}$ point direction along the surface. To check this possibility, we use the spectrum of the thick film (Fig.~\ref{Fig_S13}a) and overlap correspondingly tilted spectra according to the measured angular distribution of the surface normals as deduced from the AFMi image (Fig.~\ref{Fig_S13}h). However, this effect barely changes the measured band structure (Fig.~\ref{Fig_S13}b) and, hence, cannot explain the blurring. Another possibility is the relatively strong strain within the outer sulfur layer due to the strong bending of the monolayer when adhering to the substrate. 
To estimate this strain $\epsilon$, the local radius of curvature $R(x,y)$ of the measured topography $z(x,y)$ (Fig.~\ref{Fig_S13}h) is determined with a resolution of 16\,nm using $R(x,y) = (1+z'(x,y)^2)^\frac{3}{2} / z''(x,y)$ \cite{gray2006}. 
The strain map then reads 
\begin{equation}
    \epsilon(x,y) = \frac{R(x,y)}{R(x,y)-d/2} - 1
\end{equation}
with $d$ being the thickness of the monolayer.

Figure~\ref{Fig_S13}j shows the resulting strain distribution exhibiting a standard deviation $\sigma = 1.0$\,\%. The four-layer film is still strained, but significantly less. Assuming the absence of mutual slips between the layers to obtain an upper bound of the strain, the strain distribution exhibits $\sigma = 0.45$\,\%. Interestingly, this $\sigma$ is similar to the value of the monolayer probed in Fig.~4a, main text ($\sigma = 0.39$\,\%).
DFT+U indeed reveals that isotropic tensile strain of up to 2.5\,\% shifts the bands downwards by 200--300\,meV (Fig.~\ref{Fig_S13}e). This explains the strong blurring of the spectroscopic data of the monolayer qualitatively, albeit the calculations can not be directly compared to the experimental data, since the latter are recorded above $T_{\rm N}$. The strong influence of strain on the band structure reveals that flat substrates are indispensable for ARPES measurements of thin layers, but also nicely evidences that strain can strongly change the band structure of MnPS$_3$.